\documentclass{aa}  

\usepackage{natbib}
\usepackage{amsmath}
\usepackage{graphicx}
\usepackage{svg}
\usepackage{txfonts}
\usepackage{ctable}
\usepackage{bbold}
\usepackage{placeins}
\usepackage{siunitx}
\usepackage[colorlinks]{hyperref}
\usepackage{xcolor}
\definecolor{dark-red}{rgb}{0.9,0.0,0.0}
\definecolor{dark-blue}{rgb}{0.15,0.15,0.9}
\definecolor{dark-green}{rgb}{0.15,0.8,0.15}
\definecolor{medium-blue}{rgb}{0,0,0.9}
\hypersetup{linkcolor={dark-blue},citecolor={dark-blue}, urlcolor={medium-blue}
}

\makeatletter
\renewcommand*\aa@pageof{, page \thepage{} of \pageref*{LastPage}} 
\makeatother

\newcommand\msun{\rm{M}_\mathrm{\sun}}

\newcommand\gaia{\textit{Gaia}}
\newcommand\gdrtwo{\gaia~DR2}
\newcommand\gedrthree{\gaia~EDR3}
\newcommand\gdrthree{\gaia~DR3}

\newcommand\secref[1]{Sect.~\ref{#1}}
\newcommand\figref[1]{Fig.~\ref{#1}}

\newcommand\equref[1]{Eq.~\eqref{#1}}
\newcommand\tabref[1]{Table~\ref{#1}}

\def\parallax{$\varpi$}
\def\parallaxerror{$\sigma_{\varpi}$}

\def\magrm{~{\rm mag}}

\bibpunct{(}{)}{;}{a}{}{,}

\usepackage{mathtools}
\newcounter{mysubequations}

\usepackage[normalem]{ulem}

\begin{document} 


   \title{Identification of new nearby white dwarfs using \textit{Gaia}~DR3}

   \author{Alex Golovin\inst{1}\thanks{Fellow of the International Max Planck Research School for Astronomy and Cosmic Physics at the University of Heidelberg (IMPRS-HD)}
        \and
        Sabine Reffert\inst{1}
        \and
        Akash Vani\inst{2}
        \and
        Ulrich Bastian\inst{2}
        \and
        Stefan Jordan\inst{2}
        \and
        Andreas Just\inst{2}}

   \institute{
   Landessternwarte,
   Zentrum f\"{u}r Astronomie der
       Universit\"{a}t Heidelberg, 
       K\"{o}nigstuhl 12, 69117 Heidelberg, Germany
   \and
   Astronomisches Rechen--Institut, Zentrum f\"{u}r Astronomie der
       Universit\"{a}t Heidelberg, 
       M\"{o}nchhofstr.~12--14, 69120 Heidelberg, Germany\\
       \email{agolovin@lsw.uni-heidelberg.de}}

  \date{Version: \today}

  \abstract{
A volume-complete sample of white dwarfs is essential for statistical studies of the white dwarf population.
The sample of nearby white dwarfs is the only one that allows the faint end of the luminosity function to be probed and thus is the only one that covers the entire range of white dwarf ages.
However, due to their intrinsic faintness, even nearby white dwarfs are difficult to identify.
}{
Our work focuses on improving the completeness and purity of the white dwarf census within 50~pc of the Sun. 
To accomplish this, we used \textit{Gaia} Data Release 3 (\gdrthree{}) to identify and characterise new and previously overlooked white dwarfs in the solar neighbourhood. We also identify objects with spurious astrometric solutions in \gdrthree{} but claimed as high-confidence white dwarfs in the \textit{Gaia} Catalogue of White Dwarfs (GCWD21) by Gentile Fusillo et al. (2021). 
}{
Based on the astrometry and photometry in \gdrthree{}, we identified new nearby white dwarfs and validated those that had been missed from recent white dwarf catalogues despite being previously documented. To ensure the reliability of their astrometric solutions, we used a cut on just two parameters from \gdrthree{}: the amplitude of the image parameter determination goodness-of-fit and the parallax-over-error ratio. 
In addition, we imposed photometric signal-to-noise requirements to ensure the reliable identification of white dwarfs when using the colour-magnitude diagram.
}{
We have identified nine previously unreported white dwarfs within the local population of 50~pc, and validated 21 previously reported white dwarfs missing from the GCWD21 and other recent volume-limited white dwarf samples. A few of these objects belong to the rare class of ultra-cool white dwarfs. Four white dwarfs in our sample have an effective temperature of $T_{\rm eff}\leq4000$~K within the $1\sigma$ interval, and two of them have an absolute magnitude of $M_G > 16.0$~mag.
The identified white dwarfs are predominantly located in crowded fields, such as near the Galactic plane or in the foreground of the Large Magellanic Cloud.
We also find that 19 of these white dwarfs have common proper motion companions with angular separations ranging from $1.1\arcsec$ to $7.1\arcsec$ and brightness differences between the components of up to 9.8 magnitudes. One of these systems is a triple system consisting of a white dwarf and two K~dwarfs, while another is a double white dwarf system. 
The identified white dwarfs represent a 1.3\% improvement in the completeness of the 50~pc sample, resulting in a new total of 2265 known white dwarfs located within 50~pc of the Sun.
We have identified 103 contaminants among the 2338 high-confidence white dwarfs in the 50~pc subsample of the GCWD21 and have found that their astrometric solutions in \gdrthree{} are spurious, improving the purity by 4.4\%.
}{}


   \maketitle

   \authorrunning{A.\,Golovin\,et\,al.}
 \titlerunning{New nearby white dwarfs}

\section{Introduction}
\label{sec:intro}

The vast majority of stars in the solar neighbourhood will end their evolutionary path as white dwarfs.
Statistical studies of the white dwarf population provide insights into the stellar formation history of the solar neighbourhood \citep{Yuan_1992A&A...261..105Y, Rowell_2013MNRAS.434.1549R, Isern_2019ApJ...878L..11I}. The luminosity function of white dwarfs allows us to infer the age of the local stellar population \citep{Schmidt_1959ApJ...129..243S, Liebert_1979ApJ...233..226L, Liebert_1988ApJ...332..891L, Yuan_1992A&A...261..105Y, Harris_2006AJ....131..571H, Catalan_2008MNRAS.387.1693C} and can even help in the search for evidence of dark matter, by constraining the properties of hypothetical dark matter particles, such as axions \citep{Isern_2008ApJ...682L.109I}.

However, due to the intrinsic faintness of white dwarfs, the sample of nearby white dwarfs in the solar neighbourhood is the only sample that contains the full range of white dwarf ages.
If the sample is incomplete, the derived luminosity function may be biased.
Discoveries of nearby white dwarfs can have a significant impact on the statistical properties of the luminosity function, its shape, and in particular the position of the cutoff at the faint end, which is defined by the age of the oldest white dwarfs in the solar neighbourhood. Besides increasing the completeness of the sample, such discoveries may improve the accuracy of the luminosity function, thus providing a better understanding of our Galaxy and its history.

Furthermore, the high completeness of the local white dwarf census is important for selecting suitable targets for studies of individual objects. For instance, \citet{McGill_2018MNRAS.478L..29M} predicted an astrometric microlensing event caused by the nearby white dwarf \object{GJ~440} (= \object{LAWD~37}) by using \textit{Gaia} Early Data Release 3 \citep[EDR3;][]{gaiaedr3_summary, gaiaedr3_astromertry}. Subsequently, they confirmed this event using the \textit{Hubble} Space Telescope,
which allowed them to directly obtain the mass of \object{GJ~440} through gravitational microlensing \citep{McGill_2023MNRAS.520..259M}. This was the first time such a measurement had been made for a single star other than the Sun from a predicted astrometric microlensing event.

When hunting for white dwarfs that may have been overlooked in previous studies, it is reasonable to expect that they are likely hidden in the densest part of the Galactic plane, which is often avoided by surveys due to high source density, such as 
RAVE \citep{RAVE_2020AJ....160...82S}, GALAH \citep{GALAH_2015MNRAS.449.2604D}, and SEGUE \citep{SEGUE_2009AJ....137.4377Y}. White dwarfs can also go unnoticed when they are paired with a considerably brighter companion in a binary system, which outshines them and makes them challenging to detect \citep{Ferrario_2012MNRAS.426.2500F}.

The {\it Gaia} Catalogue of White Dwarfs \citep[hereafter referred to as the GCWD21]{gentilefusillo21_2021MNRAS.508.3877G} is based on \gedrthree{}
and lists 359~073 high-confidence white dwarfs with a probability, $P_{\mathrm{WD}}$, of being a white dwarf greater than 0.75. The selection criteria used to identify these white dwarfs are complex and described in Eqs.~1--21 of \cite{gentilefusillo21_2021MNRAS.508.3877G}. These selection criteria include cuts on the photometric excess factor, which is based on photometry alone \citep{Evans_gaia_dr2_photom_2018A&A...616A...4E, lindegren18}, and a conservative cut on the re-normalised unit weight error (RUWE), whose re-normalisation factor is also a function of the $G$-band magnitude and the $BP-RP$ colour \citep{lindegren18b}.

The completeness of the local white dwarf census can be further improved by applying selection criteria based solely on astrometry rather than employing cuts that involve both astrometry and photometry in \gdrthree{}.
When constructing the {Fifth Catalogue of Nearby Stars} (CNS5), \citet{golovin23} demonstrated that a cut on the amplitude of angular variation of the image parameter determination goodness-of-fit (IPD GoF; \texttt{ipd\_gof\_harmonic\_amplitude} in \gedrthree{} and \gdrthree{}) effectively eliminates sources with spurious astrometric solutions. This cut (\equref{eq:ipd_gof_harm_ampl_cut} of this paper), when applied to nearby sources, outperforms the commonly used cut on RUWE.
The IPD GoF cut is based solely on astrometry and is independent of the object's magnitudes. This approach allows the recovery of objects that have reliable parallaxes but with peculiarities in their photometry, hence increasing sample completeness.
The cut was validated and discussed in detail in \citet{golovin23}.
More recently, Vani et al. (in preparation) demonstrated that this approach is effective even for distances beyond 25~pc.
On the other hand, 50~pc is the distance limit at which parallaxes of white dwarfs are measured with a relative uncertainty of 5\% or better, given that the 95th percentile of the \gaia{} parallax uncertainty for white dwarfs in the GCWD21 is 1~mas. The GCWD21 contains 2338 white dwarfs that are possibly located within 50~pc of the Sun.

In this paper we report the identification of nine new white dwarfs in the 50~pc local population using \gdrthree{}.
The white dwarfs that have been identified (referred to as `new white dwarfs' hereafter) were not included in the GCWD21. Additionally, they are not listed in the Montreal White Dwarf Database\footnote{\url{https://www.montrealwhitedwarfdatabase.org/home.html}} \citep{MWDD_2017ASPC..509....3D} and were omitted in previous white dwarf studies that relied on \gdrthree{}. Therefore, these newly identified white dwarfs are also missing from the recent volume-limited white dwarf samples \citep[e.g.][]{Fleury_2022MNRAS.511.5984F, Torres_2022MNRAS.511.5462T, Cukanovaite_2023MNRAS.522.1643C, JimenezEstebaan_2023MNRAS.518.5106J, Vincent_2023MNRAS.521..760V, OBRien_2023MNRAS.518.3055O, Torres_2023A&A...677A.159T}.
In addition, our selection validated 21 white dwarf candidates that are missing from the GCWD21 and the samples mentioned above but designated as white dwarfs in \citet{El_Badry_inflation_2021MNRAS.506.2269E} or in \citet{OBRien_2023MNRAS.518.3055O} or listed in the \textit{Gaia} Catalogue of Nearby Stars \citep[GCNS;][]{gaiaedr3_gcns} with a probability of being a white dwarf greater than 0.75.

It is noteworthy that most of these objects have been previously identified as nearby stars. Specifically, 27 of them are included in the GCNS, and four of them are listed in the CNS5 catalogue.
However, they have not been identified as white dwarfs in these catalogues, with the exception of three objects with higher white dwarf probabilities in the GCNS.

Finally, among the 2338 high-confidence white dwarfs in the 50~pc subsample of the GCWD21, we have identified 103 contaminants whose astrometric solutions in \gdrthree{} are most likely spurious. We argue that these objects should be omitted from the 50~pc sample.

This paper is organised as follows.
In \secref{sec:data_selection} we describe our selection criteria.
Section~\ref{sec:new_wds} presents newly identified white dwarfs in the 50~pc local population. 
The stellar parameters of the identified white dwarfs are reported in \secref{sec:stellar_parameters}.
In \secref{sec:contaminants} we focus on improving the purity of the local white dwarf census.
Section~\ref{sec:discussion} provides our conclusions and a summary, and in
Appendix \ref{sec:dsc} we discuss the discrete source classifier (DSC) of the white dwarfs in \gdrthree{}.

\section{Data selection in \gdrthree{}}
\label{sec:data_selection}

To identify new white dwarfs, we constructed a primary sample by retrieving all the objects that are possibly located within 50~pc of the Sun from \gdrthree{}:
\begin{equation}
    \label{eq:plx_cut}
    \varpi_{\rm DR3}+3~\sigma_\varpi{}_{\rm DR3} \geq 20~\text{mas}, 
\end{equation}
where \parallax{$_{\rm DR3}$} denotes the measured parallax in the \gdrthree{} catalogue and \parallaxerror{$_{\rm DR3}$} is its uncertainty.

In \gdrthree{}, the amplitude of angular variation of the IPD GoF indicates the degree of asymmetry in the image. This parameter allows us to distinguish single stars from sources with asymmetric images, such as poorly resolved binaries or mismatched sources (two or more different sources assigned to the same \texttt{source\_id}, resulting in a spurious catalogue entry).

\begin{figure}
    \centering
    \includegraphics[width=0.49\textwidth]{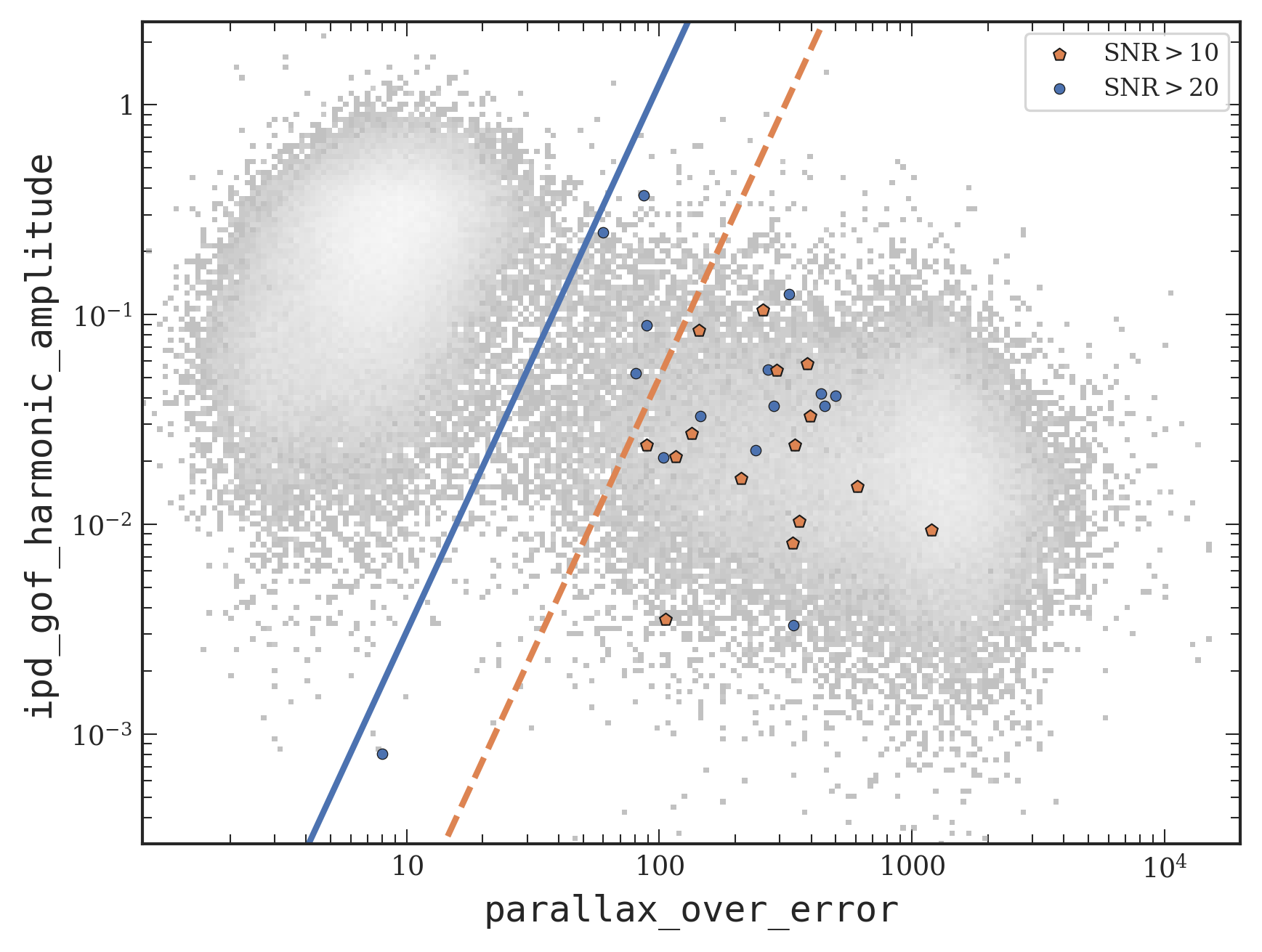}
    \caption{Amplitude of angular variation in the IPD GoF as a function of $|\varpi/\sigma_\varpi|$ for white dwarfs in this work (blue and orange symbols). The blue line corresponds to \equref{eq:ipd_gof_harm_ampl_cut}; the orange dashed line corresponds to the cut defined in \equref{eq:phot_flux_over_error_low}-(iv) to select objects with lower S/N in photometry (orange pentagons). For comparison, the 50~pc \gdrthree{} sample is shown in the background (grey symbols).}
    \label{fig:ipd_gof_ha_cuts}
\end{figure}

We eliminated spurious sources or those with unreliable astrometry by applying a cut on the amplitude of the IPD GoF in \gdrthree{} originally introduced by \citet{golovin23}.  
For the reader's convenience, we restate this condition here:
\begin{equation}
    \label{eq:ipd_gof_harm_ampl_cut}
    A_{\rm GoF}<10^{-5.12}(\varpi_{\rm DR3}/\sigma_\varpi{}_{\rm DR3})^{2.61},
\end{equation}
where $A_{\rm GoF}$ is the amplitude of the IPD GoF (\texttt{ipd\_gof\_harmonic\_amplitude}). 
The cut parameters are chosen to separate the two groups formed by sources with reliable and spurious astrometric solutions in the ($A_{\rm GoF}, \varpi_{\rm DR3}/\sigma_\varpi{}_{\rm DR3}$) parameter space (\figref{fig:ipd_gof_ha_cuts}).
This selection criterion has been validated in \citet{golovin23} against the machine learning classifier from \citet{Rybizki_fidelity_2021arXiv210111641R}, the classifier from \citet{gaiaedr3_gcns}, which was used to generate the GCNS, and the negative parallax sample.

To ensure the reliable identification of white dwarfs based on their position in the colour-magnitude diagram (CMD), we required that the $G$-, $BP$-, and $RP$-band photometry have a sufficiently high signal-to-noise ratio:
\begin{equation}
    \label{eq:phot_flux_over_error}
    \left.\begin{aligned}
    \text{(i)}\quad&\texttt{phot\_g\_mean\_flux\_over\_error} >20 \\
    \text{(ii)}\quad&\texttt{phot\_bp\_mean\_flux\_over\_error} >20 \\
    \text{(iii)}\quad&\texttt{phot\_rp\_mean\_flux\_over\_error} >20
    \end{aligned} \quad \right\} 
.\end{equation}

Furthermore, we augmented our sample by including objects that have photometry with a lower signal-to-noise ratio. However, for these objects, we imposed a more conservative requirement on the IPD~GoF amplitude in \gdrthree{}:
\begin{equation}
    \label{eq:phot_flux_over_error_low}
    \left.\begin{aligned}
    \text{(i)}\quad&\texttt{phot\_g\_mean\_flux\_over\_error} > 10 \\
    \text{(ii)}\quad&\texttt{phot\_bp\_mean\_flux\_over\_error} > 10 \\
    \text{(iii)}\quad&\texttt{phot\_rp\_mean\_flux\_over\_error} > 10 \\
    \text{(iv)}\quad&A_{\rm GoF}<10^{-6.52}(\varpi_{\rm DR3}/\sigma_\varpi{}_{\rm DR3})^{2.61}
    \end{aligned} \quad \right\} 
.\end{equation}

It should be noted that sources with a flux-over-error ratio < 10 are affected by the low flux threshold issue in \gaia{}, resulting in a clipped distribution of their fluxes in epoch photometry at the faint end, which leads to an overestimated mean flux \citep{gaiaedr3_photometry,gaiaedr3_validation}. This issue predominantly affects the $BP$ band, as the majority of sources have lower fluxes in $BP$ compared to the $RP$ band. Consequently, the faint end of the main sequence on the CMD bends towards bluer colours, becoming statistically indistinguishable from the faint end of the white dwarf sequence.

In this study, we opted to use the $(G, BP-RP)$ parameter space
to mitigate the impact of blending and contamination in photometry on our analysis.
Although the $G-RP$ colour is often favoured for its typically higher signal-to-noise ratio as compared to the $BP-RP$ colour \citep[e.g.][]{Scholz_2020A&A...637A..45S, Kaltenegger_2021Natur.594..505K, Reyle_2023arXiv230202810R, JimenezEstebaan_2023MNRAS.518.5106J},
it is prone to blending of faint sources, particularly in crowded fields. In \gaia{}, blending primarily affects the $BP$ and $RP$ fluxes rather than the $G$ band because of the very different window sizes used in the photometric instrument and the astrometric field CCD \citep{gaiaedr3_documentation_ch5_2021gdr3.reptE...5B}. Consequently, blended objects appear redder than expected in the CMD when the $G-RP$ colour is used (see \citealt{golovin23} for further details).

The CMD provides a reliable means of identifying white dwarfs in the solar neighbourhood, given the distinct position of their sequence and the availability of extremely precise astrometry from \gdrthree{}.
We employed the following cut to select objects located in the white dwarf region on the CMD:
\begin{equation}
    \label{eq:wd_locus_cut}
    \left.\begin{aligned}
    \text{(i)}\quad&M_G > 10 \magrm+2.5\,(BP-RP)\\
    \text{(ii)}\quad&BP-RP < 1.9 \magrm
    \end{aligned} \quad \right\}
,\end{equation}
where $M_G$ represents the object's absolute magnitude in the $G$ band, calculated as $M_G = \texttt{phot\_g\_mean\_mag}+5\log_{10}(\varpi_{\rm DR3}~\text{[mas]}/100)$.
This cut is shown by the dotted line in \figref{fig:wd_HRD_cuts}.

As a final step, we cross-matched our resulting sample, consisting of 2250 objects, with the GCWD21 catalogue. Subsequently, we excluded 
2220 objects that are already listed in the GCWD21 catalogue.
This left 30 objects. These objects are examined in more detail in the following section.

It is important to emphasise that our selection criteria solely employ the \gdrthree{} content that was already available in \gedrthree{}. Consequently, the selection process using either \gedrthree{}\footnote{Even though in this particular case the resulting samples will be the same, for sources with 6-p astrometric solutions in \gedrthree{} a correction to the $G$-band photometry has to be applied \citep{gaiaedr3_photometry}. In \gdrthree{} these corrections are already included.} or \gdrthree{} leads to the same resulting samples.

\section{White dwarf sample}
\label{sec:new_wds}

\subsection{New and validated white dwarfs within 50~pc}
\label{sec:sample}

\begin{figure}
    \centering
    \includegraphics[width=0.49\textwidth]{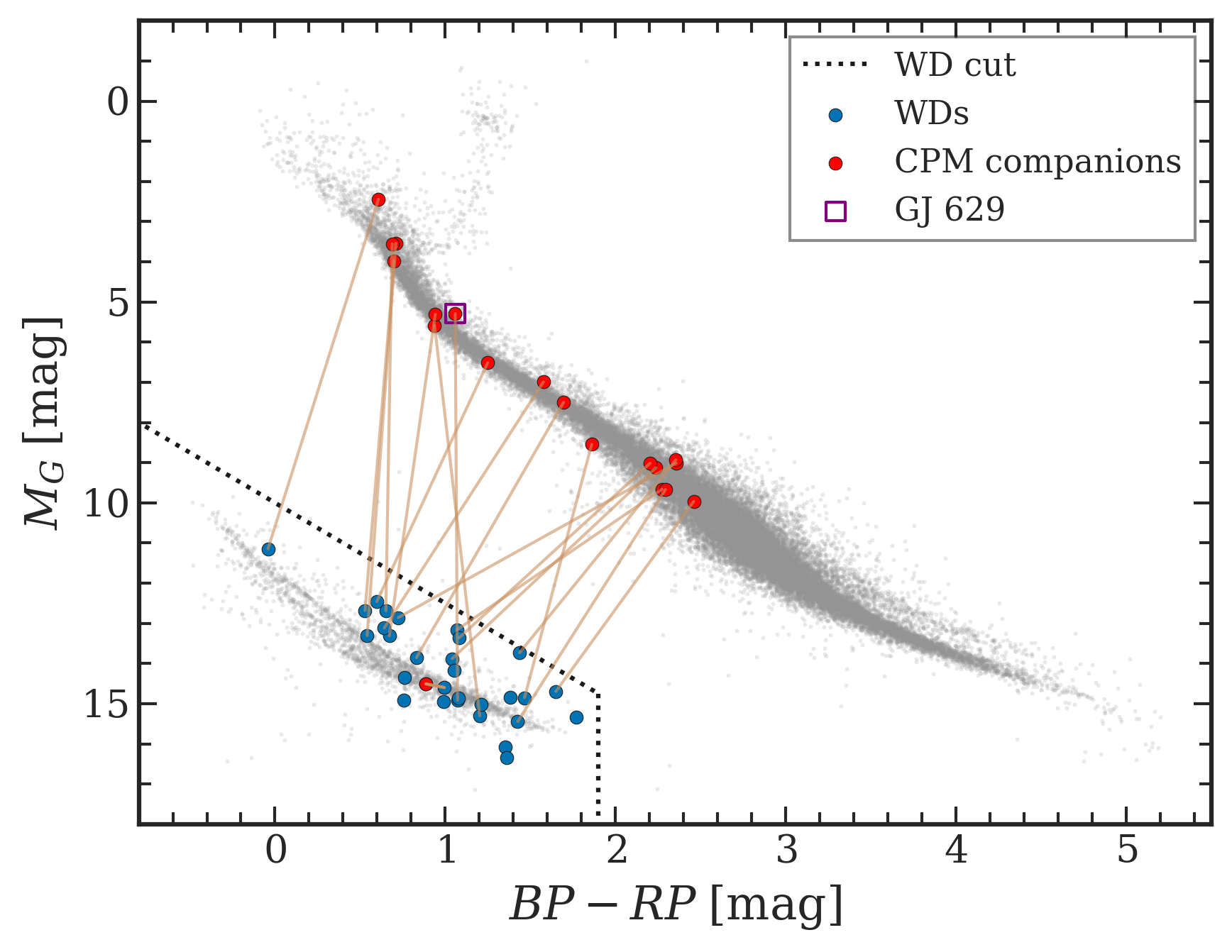}
    \caption{\gdrthree{} CMD for the identified white dwarfs (blue symbols) and their CPM companions (red symbols). 
    The adopted cut on the white dwarf locus is indicated by the dotted line. 
    The solid orange lines connect white dwarfs to their corresponding CPM companions. 
    Note that one of the pairs is a double white dwarf system (red symbol below the dotted line).
     For comparison, the CMD for the 50~pc \gdrthree{} sample is shown in the background (grey symbols). 
     The square symbol indicates \object{GJ~629}, which is itself a spectroscopic binary (see \secref{sec:cpm} for details).}
    \label{fig:wd_HRD_cuts}
\end{figure}

We identified 30 new or so far overlooked white dwarfs within 50~pc with exquisite astrometry in \gdrthree{}, following the criteria outlined in \secref{sec:data_selection}.
In \figref{fig:wd_HRD_cuts} we show the location of these white dwarfs on the CMD, as well as their common proper motion (CPM) companions (see \secref{sec:cpm}).
Their astrometric and photometric properties are summarised in \tabref{tab:new_wds}.
In total, 21 of these white dwarfs have been previously documented as such at least once, but are missing from the GCWD21 and recent volume-limited white dwarf samples.
The remaining nine objects in our list are new white dwarfs that were not previously reported as members of this class.

According to the DSC of white dwarfs in \gdrthree{}, all these 30 objects have probabilities lower than 0.5 for being categorised as white dwarfs. However, prior research has shown the shortcomings of this classifier, highlighting its limited purity and suboptimal performance \citep{GDR3_Creevey_ApsisI_2022arXiv220605864C, GDR3_documentation_ch11_2022gdr3.reptE..11U}. Notably, the sudden decline in inferred probabilities coincides with $BP-RP = 0.6$~mag, corresponding to a flux ratio of $I_{RP}/I_{BP} = 1$ (see Appendix \ref{sec:dsc} for more details).

\begin{figure}
    \centering
    \includegraphics[width=0.49\textwidth]{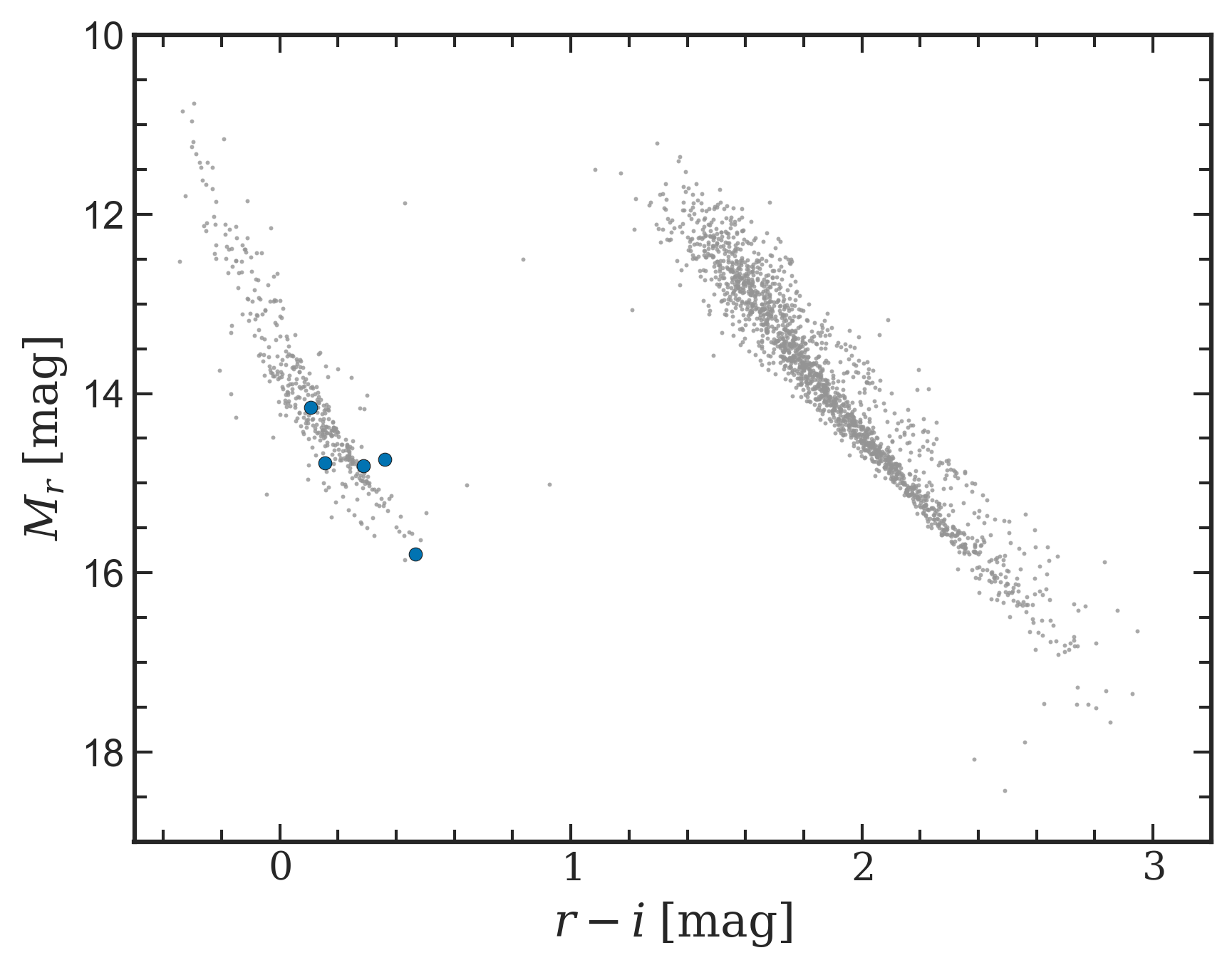} 
    \caption{PS1 CMD of the five counterparts for the identified white dwarfs. For illustrative purposes, the CMD for the 50~pc sample is displayed in the background (grey points), containing only objects that have at least 10 detections in $r$ and $i$ PS1 bands, and magnitude uncertainties below 0.01~mag.}
    \label{fig:wd_HRD_PS1}
\end{figure}

After cross-matching our list with Pan-STARRS1 \citep[PS1;][]{Chambers_2016arXiv161205560C}, we found that five objects in our list have a counterpart in PS1. Combining photometry from PS1 with extremely accurate astrometry from \gdrthree{} further supports that these objects are indeed white dwarfs.
The position of these objects on the CMD perfectly overlaps with the white dwarf cooling sequence, as shown in \figref{fig:wd_HRD_PS1}.
Notably, one of these objects (\object{\gdrthree{}~4314988445029569920} = PS1~122052927423559882) even appears to be one of the reddest white dwarfs in the solar neighbourhood. Although not the reddest in terms of $BP-RP$ colour, this white dwarf is also the faintest in our sample in terms of absolute $G$-band magnitude. We estimated its effective temperature to be about $T_{\mathrm{eff}} \sim 3700$~K by fitting the Montreal models to \gdrthree{} and PS1 photometry (see \secref{sec:stellar_parameters} for details).
This suggests that this object is potentially one of the oldest and coldest white dwarfs in the local population, placing it in the rare class of ultra-cool white dwarfs.

\begin{sidewaystable*}
\caption{New and validated white dwarfs within 50 pc and their parameters from \gdrthree{}. For counterparts in Pan-STARRS1, the PS1\_id is given. The last column lists references where the object has been reported as a white dwarf.}
    \label{tab:new_wds}
    \scriptsize
    \centering
    \begin{tabular}{l S S S S S S S S S S S S l c}
\hline \hline
  \multicolumn{1}{l}{\gdrthree{}} &
  \multicolumn{1}{c}{$\alpha_{\rm J2016.0}$} &
  \multicolumn{1}{c}{$\delta_{\rm J2016.0}$} &
  \multicolumn{1}{c}{$\varpi$} &
  \multicolumn{1}{c}{$\sigma_{\varpi}$} &
  \multicolumn{1}{c}{$\mu_{\alpha*}$} & 
  \multicolumn{1}{c}{$\sigma_{\mu_{\alpha*}}$} &
  \multicolumn{1}{c}{$\mu_{\delta}$} &
  \multicolumn{1}{c}{$\sigma_{\mu_{\delta}}$} &
  \multicolumn{1}{c}{$G$} &
  \multicolumn{1}{c}{$\sigma_G$} &
  \multicolumn{1}{c}{$BP-RP$} &
  \multicolumn{1}{c}{$\sigma_{BP-RP}$} &
  \multicolumn{1}{l}{PS1\_id} & 
  \multicolumn{1}{c}{Reference\footnote{1 - \citet{El_Badry_inflation_2021MNRAS.506.2269E}, 2 - \citet{OBRien_2023MNRAS.518.3055O}, 3 - \citet{gaiaedr3_gcns}, 4 - \citet{gentilefusillo19_2019MNRAS.482.4570G}}} \\
  \texttt{source\_id} & {[deg]} & {[deg]} & {[mas]} & {[mas]} & {[mas yr$^{-1}$]} & {[mas yr$^{-1}$]} & {[mas yr$^{-1}$]} & {[mas yr$^{-1}$]} & {[mag]} & {[mag]} & {[mag]} & {[mag]} & & \\
\hline
  6013647666939138688 & 232.34488 & -35.87226 & 56.76 & 0.95 & 2.66 & 0.37 & 2.01 & 0.44 & 17.3027 & 0.0036 & 1.354 & 0.011 &  & This work \\
  4651329704762754944 & 79.00237 & -72.23388 & 44.802 & 0.090 & -61.54 & 0.10 & 812.23 & 0.12 & 14.60519 & 0.00093 & 0.724 & 0.049 &  & 1 \\
  6017724140666000896 & 247.87332 & -39.01564& 44.46 & 0.11 & -417.61 & 0.15 & -308.31 & 0.15 & 16.6801 & 0.0057 & 1.073 & 0.077 &  & 1      \\
  2983256662868370048 & 79.99882 & -15.83773 & 42.279 & 0.035 & 182.685 & 0.031 & 216.203 & 0.032 & 14.3316 & 0.0016 & 0.60 & 0.15 &  & 1 \\
  1938960722332184704 & 355.69590 & 46.15057 & 37.08 & 0.43 & -2.25 & 0.23 & -9.11 & 0.39 & 17.0716 & 0.0016 & 0.756 & 0.047 & 163383556957661286 & This work               \\
  2078105327586616832 & 295.44066 & 43.74985 & 36.73 & 0.11 & 61.28 & 0.15 & 72.55 & 0.18 & 17.4825 & 0.0036 & 1.20 & 0.12 &  & 1          \\
  6119354336882826752 & 214.60295 & -36.55506 & 35.965 & 0.083 & -17.988 & 0.096 & -19.642 & 0.094 & 16.9170 & 0.0017 & 1.650 & 0.065 &  & This work  \\
  2831490694928280576 & 346.63271 & 19.91053 & 34.020 & 0.10 & 281.90 & 0.15 & -1.409 & 0.098 & 15.0260 & 0.0025 & 0.529 & 0.061 &  & 1  \\
  5956907713001974656 & 268.21937 & -41.99552 & 31.512 & 0.097 & 145.58 & 0.11 & -172.515 & 0.010 & 15.8141 & 0.0043 & 0.541 & 0.040 &  & 1  \\
  6665685378201412992 & 299.06923 & -52.97232 & 31.319 & 0.081 & 141.86 & 0.12 & -73.72 & 0.11 & 16.3813 & 0.0029 & 0.83 & 0.10 &  & 1, 2         \\
  205635204411687168 & 76.71714 & 44.44491 & 29.585 & 0.065 & 13.402 & 0.072 & 6.695 & 0.062 & 16.5293 & 0.0013 & 1.041 & 0.057 &  & 1  \\
  2651214734078859648 & 346.61426 & -0.49425 & 28.33 & 0.20 & -383.12 & 0.15 & -244.34 & 0.13 & 16.4721 & 0.0020 & 1.44 & 0.11 &  & This work        \\
  4110515669211359744 & 262.19720 & -24.15883 & 26.89 & 0.30 & -265.23 & 0.36 & -293.65 & 0.24 & 17.7038 & 0.0021 & 1.383 & 0.020 & 79012621975390142 & This work  \\
  5463514273884166016 & 150.15300 & -30.68801 & 26.277 & 0.090 & 77.660 & 0.067 & 37.642 & 0.094 & 17.0753 & 0.0028 & 1.054 & 0.075 &  & This work         \\
  4648527839871194880 & 87.24230 & -75.12733 & 24.988 & 0.088 & -23.094 & 0.094 & 425.34 & 0.12 & 18.0312 & 0.0013 & 1.210 & 0.033 &  & 2, 3, 4  \\
  6566046912935892352 & 325.76201 & -43.49541 & 24.83 & 0.12 & 204.021 & 0.098 & -66.24 & 0.14 & 15.7188 & 0.0030 & 0.65 & 0.10 &  & 1   \\
  6063480282704764928 & 198.77439 & -56.54848 & 24.77 & 0.31 & -31.24 & 0.19 & -259.62 & 0.23 & 18.3635 & 0.0018 & 1.771 & 0.015 &  & This work  \\
  1355203232910297600 & 257.96584 & 43.26073 & 24.528 & 0.040 & -112.197 & 0.048 & -184.413 & 0.047 & 16.22051 & 0.00083 & 1.068 & 0.099 &  & 1          \\
  2533660345315705344 & 18.70369 & -0.97166 & 23.702 & 0.092 & -22.41 & 0.11 & 203.772 & 0.073 & 14.2693 & 0.0029 & -0.04 & 0.10 &  & 1      \\
  4663902104803506816 & 76.72861 & -64.81502 & 22.970 & 0.096 & 306.58 & 0.13 & 108.81 & 0.12 & 18.0639 & 0.0015 & 1.076 & 0.023 &  & 3  \\
  3462945170562185088 & 183.23736 & -34.08703 & 22.11 & 0.21 & 100.03 & 0.22 & -164.90 & 0.21 & 18.7238 & 0.0066 & 1.43 & 0.11 &  & 1  \\
  6080038412403680256 & 197.61442 & -52.99006 & 21.882 & 0.064 & -100.018 & 0.069 & 13.532 & 0.058 & 16.6650 & 0.0013 & 1.082 & 0.069 &  & 1  \\
  2021862490380335232 & 294.58601 & 25.69982 & 21.65 & 0.19 & 58.58 & 0.28 & -1.76 & 0.26 & 17.6669 & 0.0077 & 0.760 & 0.074 &  & This work  \\
  6039626481000400128 & 238.17214 & -31.86765 & 21.20 & 0.15 & -125.28 & 0.16 & 26.29 & 0.10 & 17.8826 & 0.0045 & 0.887 & 0.020 &  & 1  \\
  6039626481009257472 & 238.17232 & -31.86738 & 21.14 & 0.16 & -132.03 & 0.17 & 41.05 & 0.11 & 17.9793 & 0.0019 & 0.995 & 0.096 &  & 1  \\
  4837326390227065344 & 63.93907 & -45.97118 & 20.710 & 0.077 & 245.026 & 0.076 & -129.57 & 0.12 & 16.7347 & 0.0024 & 0.671 & 0.064 &  & 1  \\
  1323632298413108096 & 241.81026 & 34.02664 & 20.504 & 0.057 & 96.269 & 0.065 & -5.02 & 0.11 & 16.5644 & 0.0015 & 0.642 & 0.089 &  & 1 \\
  4468278954501360128 & 272.03420 & 1.87819 & 19.87 & 0.22 & -22.35 & 0.24 & 36.11 & 0.23 & 18.3603 & 0.0031 & 1.466 & 0.069 & 110252720342434261 & 1 \\
  625602672887745152 & 155.17523 & 20.46312 & 19.73 & 0.19 & -160.20 & 0.21 & -188.34 & 0.22 & 18.4834 & 0.0035 & 0.991 & 0.062 & 132551551754976643 & 3 \\
  4314988445029569920 & 292.74236 & 11.71608 & 19.2 & 2.4 & 11.8 & 1.6 & -16.1 & 1.5 & 19.9190 & 0.0094 & 1.362 & 0.043 & 122052927423559882 & This work \\
\hline\end{tabular}
\end{sidewaystable*}

\begin{sidewaystable*}
\caption{CPM companions to the white dwarfs identified in this work.}
    \label{tab:cpm_companions}
    \centering
    \begin{tabular}{l l S l S l S l S l l l S}
\hline \hline
  \multicolumn{1}{c}{WD} &
  \multicolumn{1}{c}{CPM companion} &
  \multicolumn{1}{c}{$\varpi$} &
  \multicolumn{1}{c}{$\sigma_{\varpi}$} &
  \multicolumn{1}{c}{$\mu_{\alpha*}$} &
  \multicolumn{1}{c}{$\sigma_{\mu_{\alpha*}}$} &
  \multicolumn{1}{c}{$\mu_{\delta}$} &
  \multicolumn{1}{c}{$\sigma_{\mu_{\delta}}$} &
  \multicolumn{1}{c}{$G$} &
  \multicolumn{1}{c}{$BP-RP$} &
  \multicolumn{1}{c}{SpT} &
  \multicolumn{1}{c}{$\rho$} &
  \multicolumn{1}{c}{$s$} \\
  \gdrthree{} \texttt{source\_id} & \gdrthree{} \texttt{source\_id} & [mas] & [mas] & [mas yr$^{-1}$] & [mas yr$^{-1}$] & [mas yr$^{-1}$] & [mas yr$^{-1}$] & [mag] & [mag] & & [\arcsec] & [AU] \\
\hline
  4651329704762754944 & 4651329704762754176 & 44.856 & 0.016 & -46.932 & 0.018 & 825.909 & 0.022 & 10.755 & 2.357 &  M1.5V  &   2.98  &  66.5  \\
  6017724140666000896 & 6017724140678769024 & 44.373 & 0.049 & -429.581 & 0.054 & -330.311 & 0.044 & 7.057 & 1.059 &  G9V  &   7.14  &  160.8  \\
  2983256662868370048 & 2983256662869643776 & 42.116 & 0.016 & 173.704 & 0.014 & 207.709 & 0.014 & 8.394 & 1.250 &  K4V  &   3.65  &  86.5  \\
  2078105327586616832 & 2078105327586616704 & 36.644 & 0.018 & 50.702 & 0.021 & 75.325 & 0.019 & 7.762 & 0.937 &   K0V &   6.54  &  178.3  \\
  6119354336882826752 & 6119354336882826880 & 36.189 & 0.023 & -34.259 & 0.027 & -22.779 & 0.028 & 12.173 & 2.460 &  M3V  &   2.55  &  70.9  \\
  2831490694928280576 & 2831490694929214464 & 34.115 & 0.019 & 286.890 & 0.023 & 5.057 & 0.017 & 6.314 & 0.700 &  F8V  &   5.24  &  154.1  \\
  6665685378201412992 & 6665685343840128384 & 31.494 & 0.015 & 134.355 & 0.014 & -74.846 & 0.012 & 10.009 & 1.696 &  K7V  &   6.83  &  217.17  \\
  5956907713001974656 & 5956907713041688960 & 31.361 & 0.026 & 140.745 & 0.028 & -190.891 & 0.018 & 6.061 & 0.710 &  F6V  &   4.90  &  156.6  \\
  205635204411687168 & 205635200112139520 & 29.550 & 0.019 & 15.824 & 0.022 & 24.176 & 0.018 & 11.778 & 2.237 &  M2V  &   2.85  &  96.5  \\
  2651214734078859648 & 2651214734078859776 & 28.378 & 0.024 & -376.420 & 0.024 & -238.486 & 0.020 & 11.667 & 2.355 &  M1.5V  &   2.21  &  78.1  \\
  6566046912935892352 & 6566046809856626944 & 25.168 & 0.023 & 206.609 & 0.018 & -62.312 & 0.021 & 6.554 & 0.688 &  F6V  &   5.62  &  226.05  \\
  1355203232910297600 & 1355203232910297728 & 24.557 & 0.014 & -101.075 & 0.016 & -185.807 & 0.016 & 12.713 & 2.273 &  M2.5V  &   2.15  &  87.7  \\
  2533660345315705344 & 2533660345315782528 & 23.539 & 0.087 & -14.263 & 0.099 & 205.414 & 0.060 & 5.582 & 0.608 &   F0V &   5.99  &  253.5  \\
  3462945170562185088 & 3462945170563600640 & 21.894 & 0.020 & 107.608 & 0.021 & -166.659 & 0.021 & 12.972 & 2.294 &  M2.5V  &  3.52  &  159.5  \\
  6080038412403680256 & 6080038408112210176 & 21.837 & 0.017 & -98.057 & 0.015 & 23.998 & 0.014 & 12.318 & 2.204 &   M1.5V &   2.59  &  118.6  \\
  6039626481009257472 & 6039626481000400128 & 21.20 & 0.15 & -125.28 & 0.16 & 26.29 & 0.10 & 17.883 & 0.887 &  --  &   1.13  &  53.5  \\
  4837326390227065344 & 4837326385930314368 & 20.707 & 0.012 & 242.804 & 0.012 & -123.338 & 0.017 & 8.728 & 0.942 &   G9V &   4.65  &  224.8  \\
  1323632298413108096 & 1323632508865678080 & 20.551 & 0.011 & 95.297 & 0.010 & -2.102 & 0.013 & 10.432 & 1.577 & K6V &   4.55  &  222.0  \\
  4468278954501360128 & 4468275995268525056 & 20.397 & 0.014 & -16.585 & 0.014 & 40.233 & 0.013 & 12.001 & 1.860 &   M0.5V &   5.48  &  275.7  \\
\hline\end{tabular}

\end{sidewaystable*}

\subsection{Common proper motion companions}
\label{sec:cpm}

For 19 white dwarfs in our list, we identified CPM
companions in \gdrthree{} by selecting pairs with statistically identical parallaxes, proper motions consistent
with a Keplerian orbit, and projected separations of less than 1~pc (see \citealt{El_Badry_inflation_2021MNRAS.506.2269E} for details). The identified binaries have angular separations~$\rho$ ranging from 1.1\arcsec to 7.1\arcsec and projected separations~$s$ between 54 and 276~AU.
To estimate the spectral types of the main-sequence CPM companions, we used their absolute magnitudes $M_G$ and relied on the empirical $M_G$ and spectral type sequence from \citet{pecaut13}\footnote{updates available at \url{https://www.pas.rochester.edu/~emamajek/EEM_dwarf_UBVIJHK_colors_Teff.txt}}.
We list the astrometric parameters of the CPM companions from \gdrthree{} along with their spectral types and angular and projected separations in \tabref{tab:cpm_companions}. Their positions on the CMD are shown in \figref{fig:wd_HRD_cuts}, denoted by red symbols.

One of the identified CPM pairs turned out to be a double white dwarf system, as the CPM companion to \object{\gdrthree{}~6039626481000400128} is another white dwarf from our list (\object{\gdrthree{}~6039626481009257472}).
Assuming a circular orbit and that the projected separation of this system is equal to its true semi-major axis, this corresponds to an orbital period of $P \sim 375~\text{yr}$.

In addition to providing single-star astrometric solutions, \gdrthree{} also incorporates non-single-star (NSS) solutions for astrometric and spectroscopic binaries \citep{GDR3_Halbwachs_2022arXiv220605726H, GDR3_Holl_2022arXiv220605439H}.
Notably, in this data release only unresolved binaries were considered for the NSS processing \citep{GDR3_documentation_ch7_2022gdr3.reptE...7P}.

Of all the white dwarfs and their CPM companions identified in this work, only one object, namely \object{GJ~629} (= \object{\gdrthree{}~6017724140678769024}), which is the CPM companion of \object{\gdrthree{}~6017724140666000896}, has an entry in the NSS tables of \gdrthree{} indicating that \object{GJ~629} is a spectroscopic binary itself. This suggests that \object{\gdrthree{}~6017724140666000896} and \object{GJ~629} constitute a triple system.
Although it has been demonstrated that some of the listed NSS solutions in \gdrthree{} may be spurious and not all objects with NSS-solutions represent genuine binaries \citep{Spaeth_2023RNAAS...7...12S}, as well as that scan-angle-dependent signals in \gdrthree{} can lead to spurious periods \citep{GDR3_Holl_spurious_2022arXiv221211971H}, it is important to note that it had been already established before \gdrthree{} that \object{GJ~629} is a spectroscopic binary, consisting of two early-K dwarfs \citep{Bopp_1970MNRAS.147..355B, Raghavan_2010ApJS..190....1R, Pourbaix_SB9_2004A&A...424..727P, Fuhrmann_2017ApJ...836..139F}. Furthermore, the orbital period of $P=31.8$~d obtained from the NSS-solution in \gdrthree{} is consistent with previously reported values \citep{Bopp_1970MNRAS.147..355B, Raghavan_2010ApJS..190....1R}.
Remarkably, \object{GJ~629} is also the only main-sequence companion that lies photometrically in the main-sequence binary band of \figref{fig:wd_HRD_cuts}.

Regarding the remaining objects in our list, their astrometric solutions are consistent with a single-star model, and there is no evidence supporting the existence of additional components orbiting these objects.

\section{Stellar parameters}
\label{sec:stellar_parameters}

\begin{figure*}
    \centering
    \includegraphics[width=0.49\textwidth]{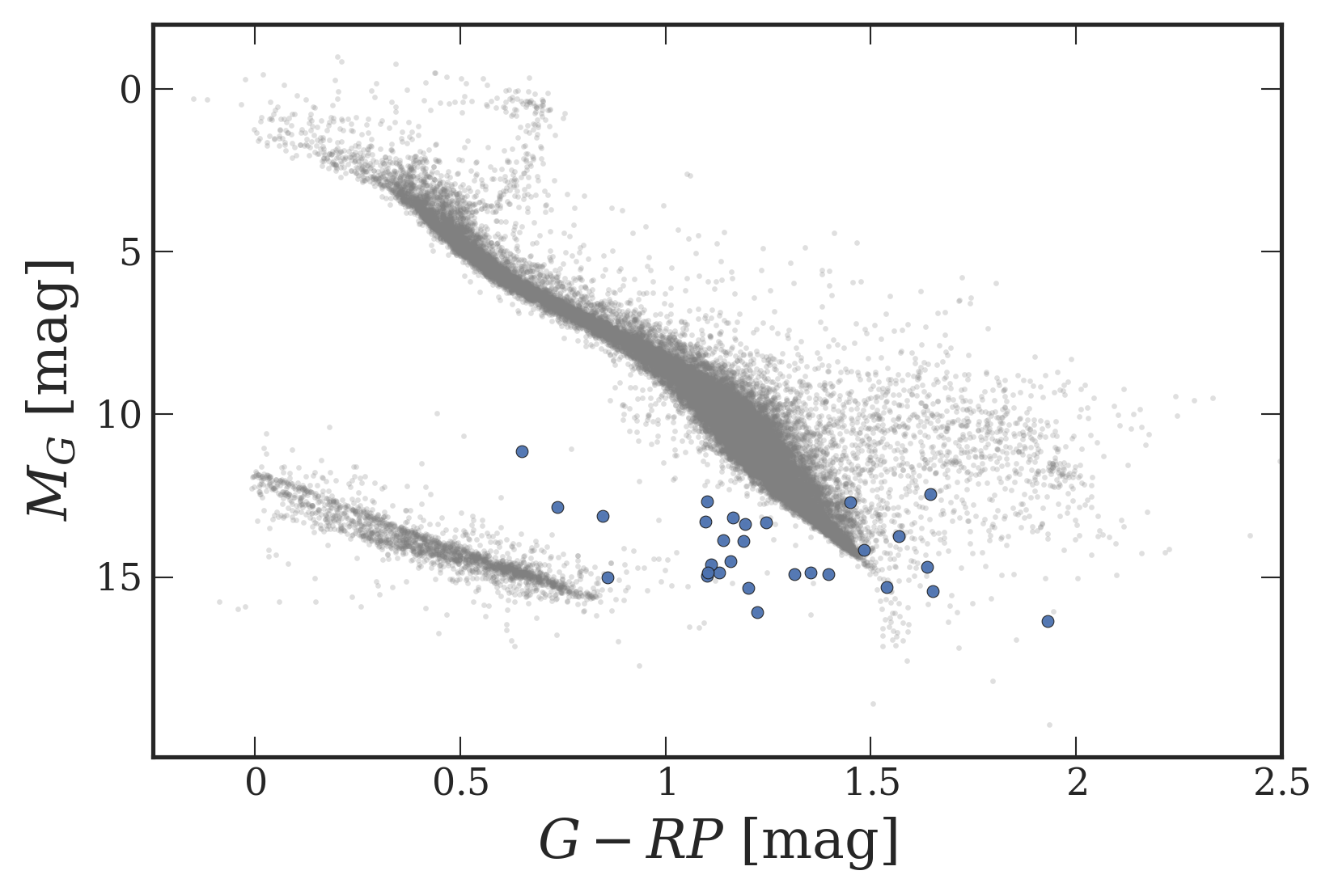}
    \includegraphics[width=0.49\textwidth]{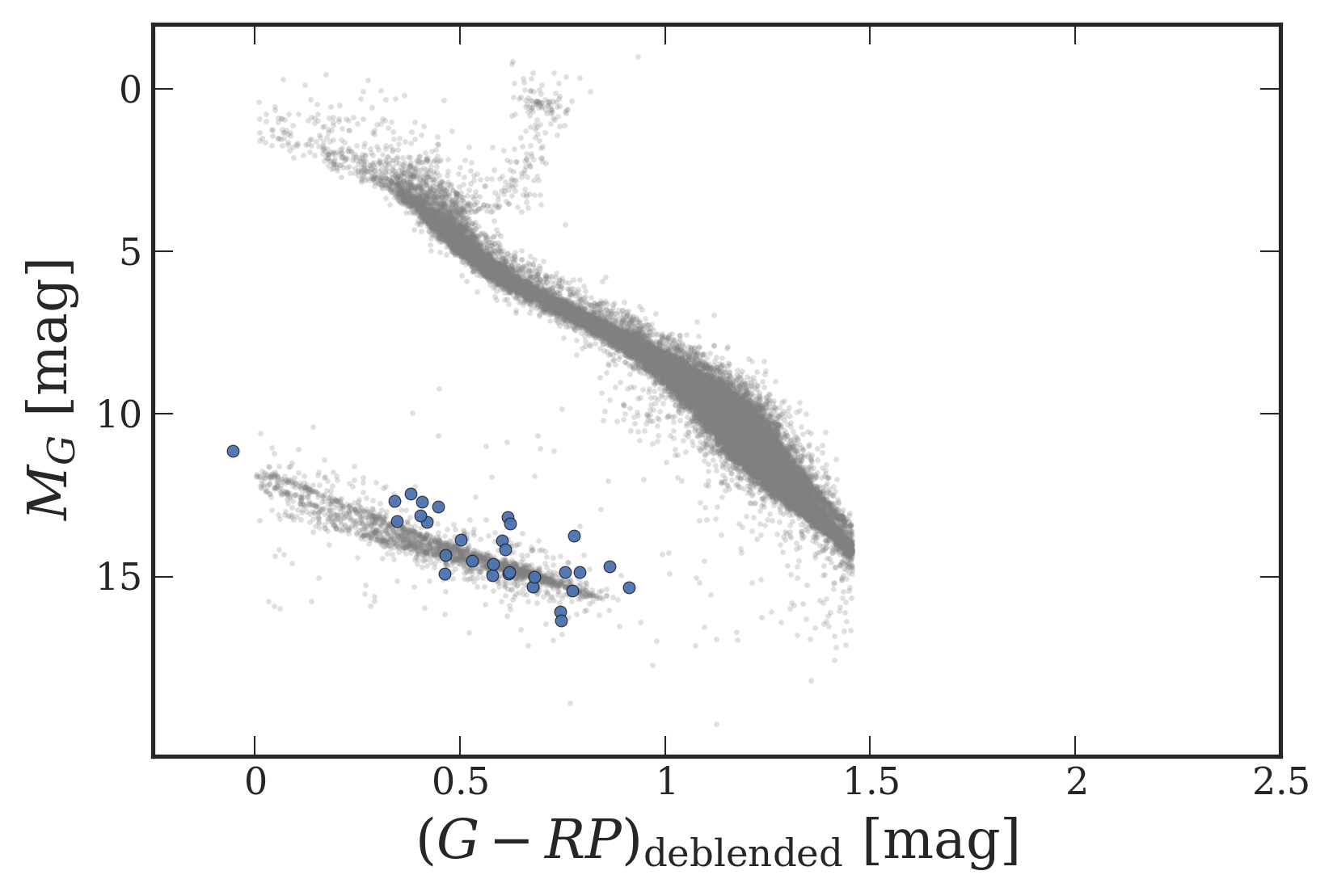}
    \caption{Difference between the measured and de-blended $G-RP$ colours for the identified white dwarfs and their location on the CMD. For comparison, the 50~pc \gdrthree{} sample is shown in the background (grey points); only objects with $BP-RP$ colours in the range $0.0~{\rm mag} < BP-RP < 4.25~{\rm mag}$ (the applicability range of the de-blending correction) are plotted. \textit{Left:} CMD using the published $G-RP$ colour. \textit{Right:} CMD for the same sample, but using de-blended $G-RP$ colour.}
    \label{fig:hrd_delblending}
\end{figure*}

\begin{figure}
    \centering
    \includegraphics[width=0.49\textwidth]{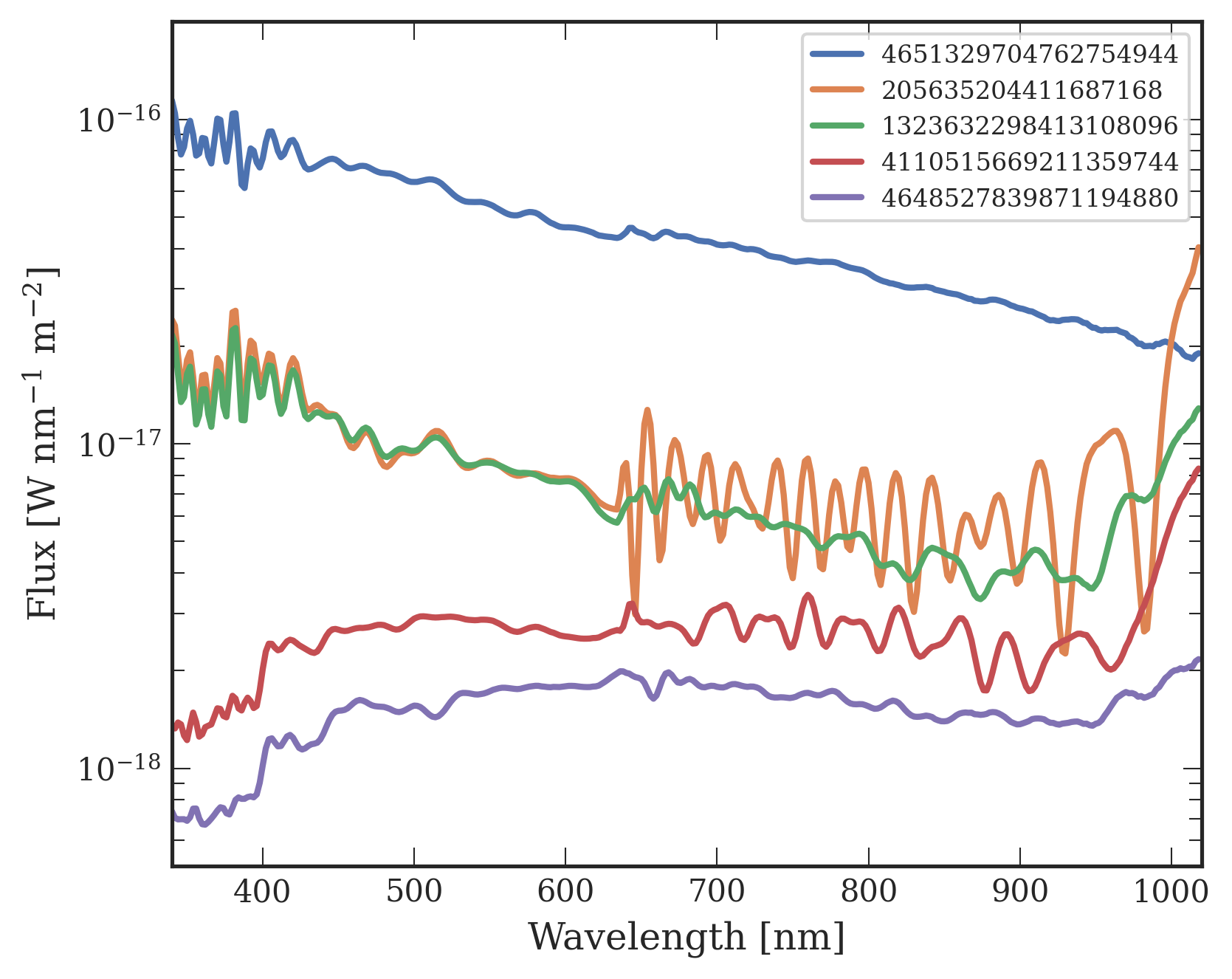} 
    \caption{Calibrated mean XP spectra for five white dwarfs in our sample.
The corresponding \gdrthree{} source identifiers are indicated in the legend.
}
    \label{fig:xp_spec}
\end{figure}

\begin{figure*}
    \centering
    \includegraphics[width=0.24\textwidth]{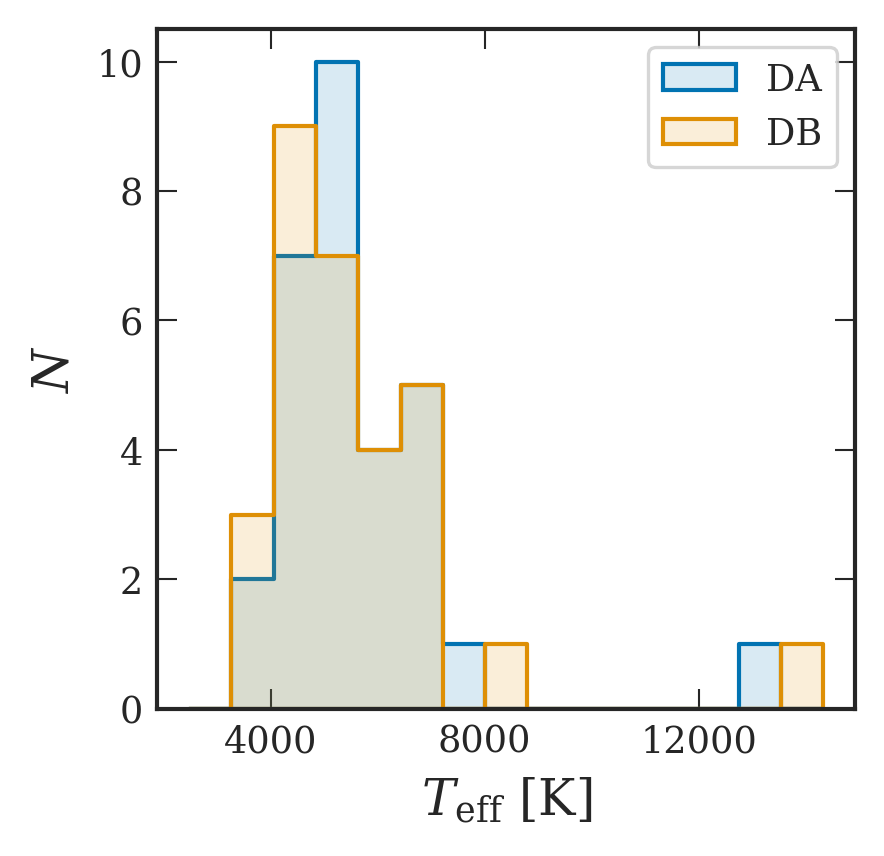}
    \includegraphics[width=0.24\textwidth]{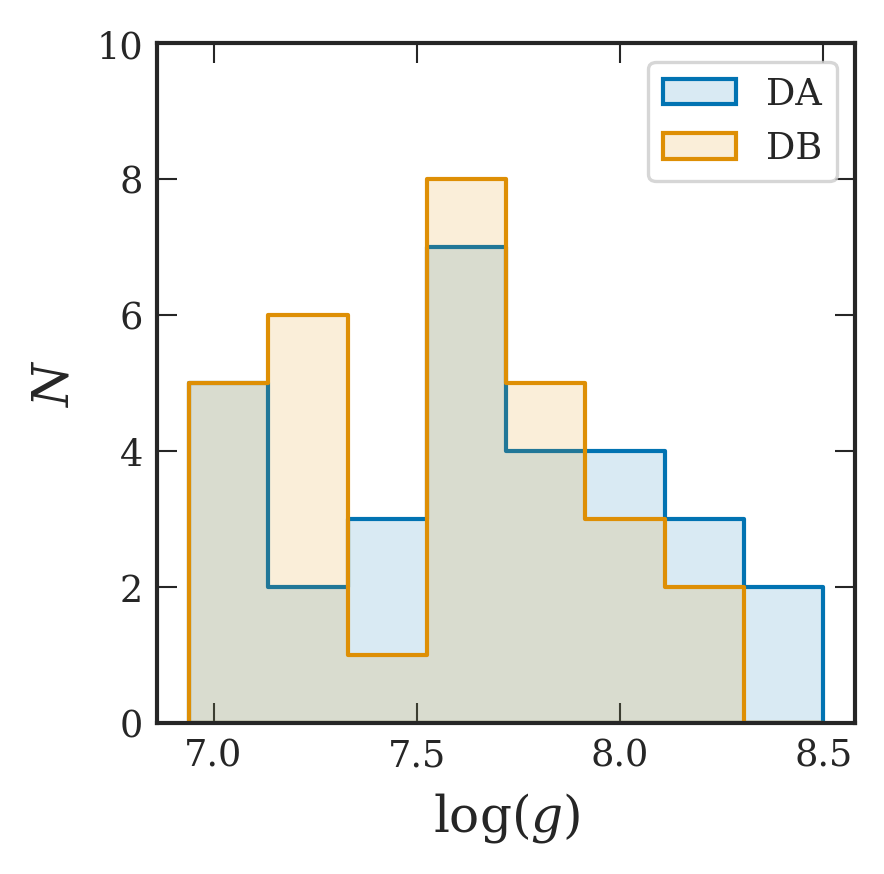}
    \includegraphics[width=0.24\textwidth]{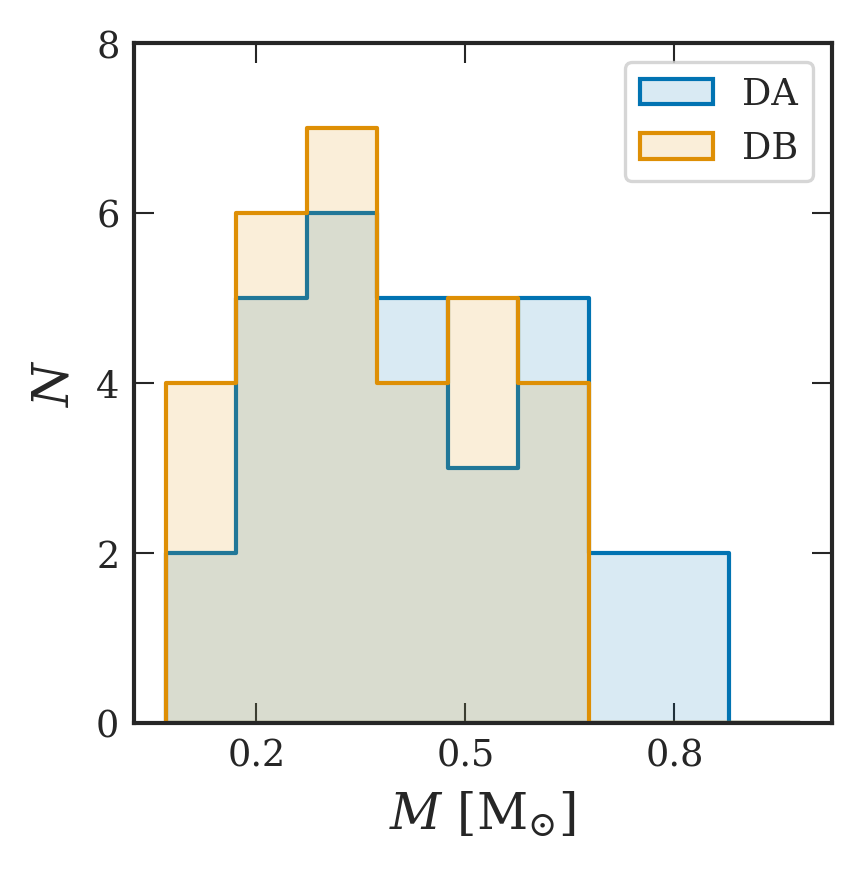}
    \includegraphics[width=0.24\textwidth]{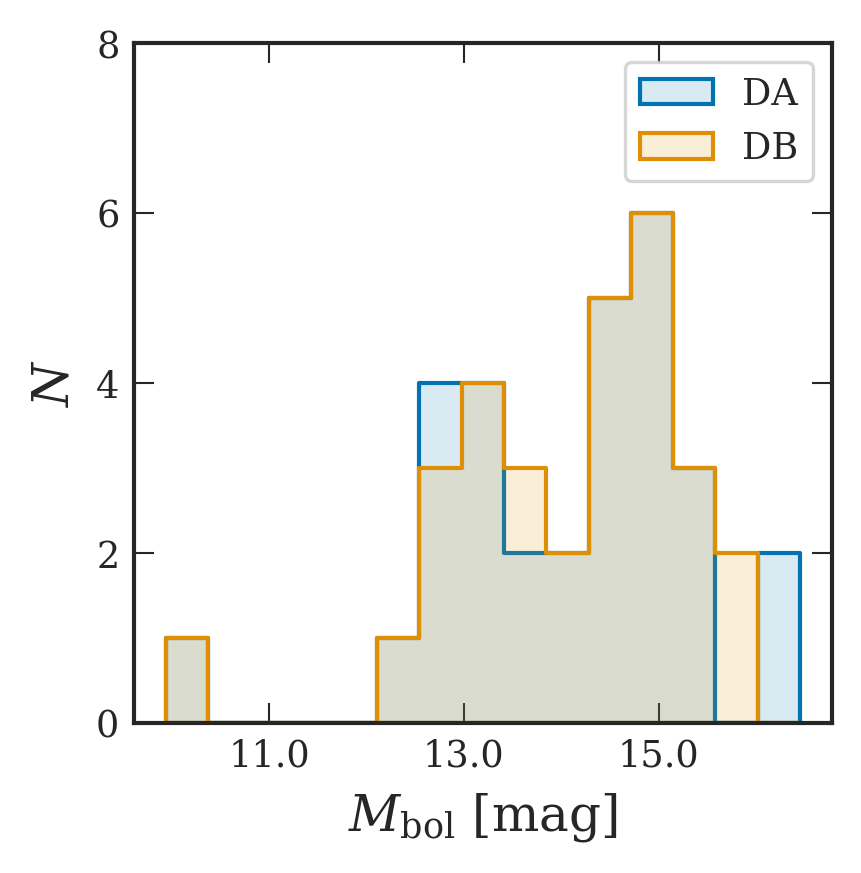}
    \caption{Comparison of the distributions of stellar parameters derived in this work from DA and DB models, showing that there is substantial overlap between the two solutions. The DA solutions tend to have slightly higher values for both masses and surface gravities than the DB solutions, although in most cases the difference is not statistically significant.}
    \label{fig:stellar_parameters_hist}
\end{figure*}

The ultra-precise parallaxes available from \gdrthree{}, along with multi-colour photometry, enable us to estimate the stellar parameters of the identified white dwarfs.
This is accomplished by photometric fitting of the Montreal DA (pure hydrogen) and DB (pure helium) models\footnote{\url{https://www.astro.umontreal.ca/~bergeron/CoolingModels/}} \citep{Bedard_Montreal_models_2020ApJ...901...93B} to the spectral energy distribution of a white dwarf.
The photometric fitting was performed with the \texttt{WDPhotTools} package \citep{Lam_WDPhotTools_2022RASTI...1...81L} utilising the Markov chain Monte Carlo method to sample the posterior distribution of the inferred parameters, such as effective temperature $T_{\rm eff}$, surface gravity $\log(g)$, mass $M$, and bolometric magnitude $M_{\rm bol}$.

The input parameters required by the \texttt{WDPhotTools} for the photometric fitting include the type of atmosphere 
and fluxes in different photometric bands along with their associated uncertainties.
As the atmospheric compositions are unknown, we obtained both DA and DB solutions for each white dwarf in our sample.
Optionally, to allow for a much more reliable photometric fit, it can be supplemented with the distance, which in our case is derived from the \textit{Gaia}~DR3 parallax measurements, and the interstellar reddening.
We did not apply any de-reddening corrections to the photometry, as it is reasonable to assume that the interstellar absorption within 50~pc is negligible. Thus, there is no degeneracy between $T_{\rm eff}$ and reddening -- another advantage of nearby white dwarfs.

To infer the stellar parameters for objects with \gaia{} XP spectra, we used these spectra to derive synthetic photometry. For the remaining objects, we used $G$, $BP$, and $RP$ magnitudes from \gdrthree{}, supplemented with PS1 photometry when available.
In \gaia{}, the $G$ and $BP/RP$ magnitudes are obtained from two different instruments with different CCD window sizes.
We note that although the blending can bias the measured values of the $BP$ and $RP$ fluxes, it results in only negligible changes in their $BP-RP$ colour. Furthermore, the $G$ band is less prone to blending due to the smaller window size of the \gaia{} astrometric field.
In contrast to our selection criterion in \equref{eq:wd_locus_cut}, where we use only the $G$-band magnitude and the $BP-RP$ colour, here we also consider the $BP$ and $RP$ magnitudes separately to derive the stellar parameters.
To minimise the impact of the blending on our analysis, we derived the de-blended $G-RP$ colour, as fully described in Appendix B of \citet{golovin23}, and used this value to correct the $BP$ and $RP$ magnitudes. The $G$-band magnitude and $BP-RP$ colour were kept fixed.
This correction is only applied to objects that do not have \gaia{} XP spectra. 
\figref{fig:hrd_delblending} displays CMDs before and after de-blending, illustrating the inconsistency in photometry and the power of the de-blending correction.
Omitting de-blending would significantly impact the derived stellar parameters, resulting in, for instance, significantly underestimated masses of white dwarfs.

XP spectra in \gdrthree{} are available for five white dwarfs in our sample. These spectra are published in a `continuous' form, where the coefficients of 55 basis functions, which are Gauss-Hermite polynomials, are provided.
In \figref{fig:xp_spec}, we present the mean sampled spectra obtained from the coefficients of the Gauss-Hermite polynomials. These spectra are then calibrated from internally calibrated mean spectra to the absolute system using the \texttt{GaiaXPy} package. Sampling is performed from 340~nm to 1020~nm in steps of 2~nm.
We note that the wiggles observed in the resulting spectra are not physical, but rather artefacts resulting from correlated errors of the polynomial coefficients.

Using these spectra, we computed synthetic photometry in various photometric systems, including \gdrthree{}, Johnson-Kron-Cousins (JKC), PS1, and SDSS. However, we excluded the $U_{\rm JKC}$ and $u_{\rm SDSS}$ bands due to imperfections in the internal calibration process, as the XP spectra do not cover the bluest part of the transmission curves of these two bands.

Furthermore, it has been shown that the uncertainties of the synthetic fluxes are often underestimated and, to address this, we applied the empirical corrections to the uncertainties, following \citet{GDR3_Montegriffo_XP_synth_phot_2022arXiv220606215G, GDR3_Montegriffo_XP_calibration_2022arXiv220606205M}.
Finally, we employed the derived spectral energy distribution to estimate the stellar parameters of the white dwarfs, considering only fluxes with a flux-over-error ratio greater than 10.

For each white dwarf, we report in \tabref{tab:param} the DA and DB solutions for the inferred stellar parameters and their uncertainties. 
The distribution of the stellar parameters from both solutions derived in this work is shown in \figref{fig:stellar_parameters_hist}.
There is a significant overlap between the distributions of the two solutions. While the derived masses and surface gravities generally tend to be slightly higher in the DA solutions than in the DB solutions, it is worth noting that this difference is not statistically significant for 90\% of our sample.

The Montreal models, which we used to obtain the stellar parameters, assume a carbon-oxygen core for both DA and DB atmospheres.
While we found a considerable number of white dwarfs with apparently low masses in our sample, it is crucial to note that such extremely low-mass white dwarfs are rare and are characterised by a helium-core composition \citep{Althaus_2013A&A...557A..19A}.
Given that the main-sequence lifetime of low-mass stars can surpass the age of the Universe, the existence of extremely low-mass white dwarfs today suggests a formation process beyond single-star evolution, excluding cases of intense mass loss in high-metallicity stars \citep{Kilic_2007ApJ...671..761K}. Instead, the formation of extremely low-mass white dwarfs is anticipated to occur through binary evolution, involving one or more instances of common envelope evolution, as described in \citet{Li_2019ApJ...871..148L}. The outcome of this binary evolution is manifested as a population of compact binary systems that have undergone significant changes, housing low-mass helium-core white dwarfs alongside their evolved companions.
Importantly, there is currently no robust observational evidence supporting the existence of helium-core DB white dwarfs \citep{Bergeron_2011ApJ...737...28B, Brown_2020ApJ...889...49B, Kosakowski_2020ApJ...894...53K}.
Furthermore, \citet{Sarna_2000MNRAS.316...84S} have performed comprehensive calculations to elucidate the binary evolution for the formation of low-mass helium white dwarfs with stellar masses $<0.25~\msun$. Their results show that after the detachment of the Roche lobe, the helium cores are enveloped by a substantial hydrogen layer with a mass ranging from 0.01 to $0.06~\msun$.  Therefore, only low-mass DA white dwarfs are formed.
Assuming that the mass values in our analysis are correct and not an artefact of photometric excess, we argue that the identified low-mass white dwarfs in our sample are most likely non-DBs.

\citet{Bergeron_2019ApJ...876...67B} noticed that there is a significant population of white dwarfs with apparently low masses ($< 0.4~{\rm M_\sun}$) positioned slightly above the densest part of the white dwarf sequence on the CMD. These objects were attributed by \citet{Bergeron_2019ApJ...876...67B} to unresolved double degenerate binaries. 
However, as can be seen in \figref{fig:wd_HRD_cuts}, the majority of the white dwarfs in our sample within this region are actually found in binaries with a main-sequence component.
Given their small angular separation ($\rho\leq 7.1\arcsec$) and the extreme difference in magnitude and sometimes also in colour between the components, it is plausible that the contamination of photometry, especially in the $BP$ and $RP$ bands, by a bright object in the field of view may also account for this effect.
This conclusion gains further support from the fact that the white dwarfs in our sample with the unrealistically low estimated masses have the reddest CPM companions ($BP-RP > 2.0$~mag) and the smallest angular separations ($\rho < 3\arcsec$). Consequently, this raises concerns about the accuracy of the photometric masses in the presence of a bright, albeit well-resolved, companion in the vicinity of a white dwarf, and these masses should be interpreted with caution, as they may not be physically meaningful and can potentially cause a spurious excess in the white dwarf mass distribution at the low-mass end.

\section{Improving the purity of the local white dwarf census: Contaminants in the 50~pc subsample of the GCWD21}
\label{sec:contaminants}

\begin{figure}
    \centering
    \includegraphics[width=0.5\textwidth]{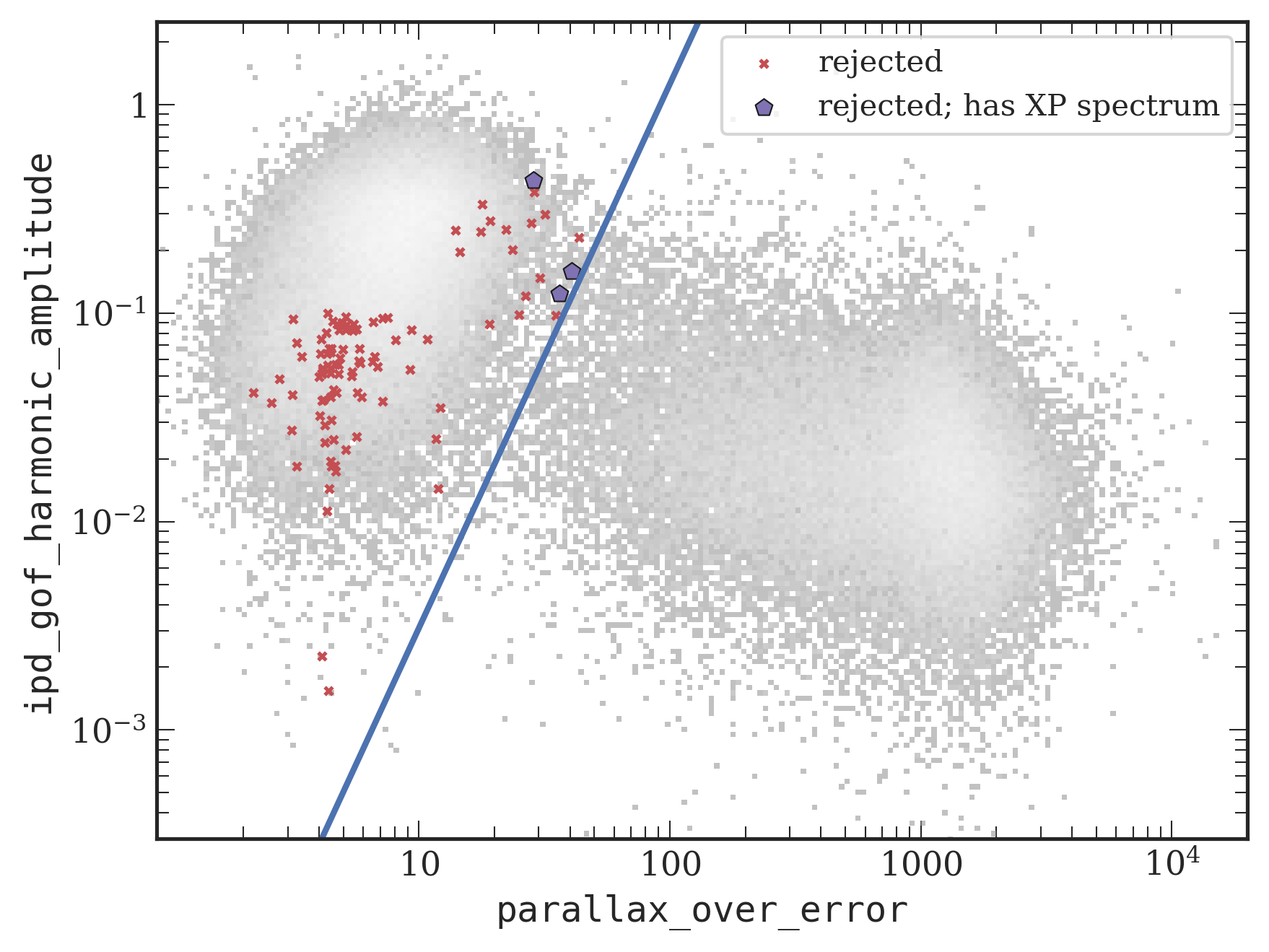} 
    \caption{Amplitude of the IPD GoF as a function of $|\varpi/\sigma_\varpi|$ for the contaminants
    overplotted on the \gdrthree{} 50~pc sample. Purple pentagon symbols indicate three contaminants with available XP spectra. The solid line corresponds to the cut described by \equref{eq:ipd_gof_harm_ampl_cut}.}
    \label{fig:gcswd21_rej_ipd_gof}
\end{figure}

The cut on the amplitude of the IPD~GoF can be used to significantly improve not only the completeness but also the purity of the local white dwarf sample. 
This is achieved by eliminating the sources -- hereafter referred to as contaminants -- whose astrometric solutions fail to meet the criterion defined in \equref{eq:ipd_gof_harm_ampl_cut} (see \figref{fig:gcswd21_rej_ipd_gof}).
By doing so we were able to identify a total of 103 contaminants among the 2338 high-confidence white dwarfs ($P_{\mathrm{WD}}>0.75$) in the 50~pc subset of the GCWD21.

All identified contaminants are located at nominal distances greater than 25~pc.
This is not entirely surprising, given that the total fraction of contaminants in \gdrthree{} is smaller for the 25~pc volume than the 50~pc volume. When considering the full range of absolute magnitudes and colours of the CMD, $11.78\%$ of sources within 25~pc have spurious astrometry in \gdrthree{}, whereas, within 50~pc, the fraction of sources with spurious astrometry increases to $50.84\%$  (see Figs.\,1 and A.2. in \citealt{golovin23}).

\begin{figure}
    \centering
    \includegraphics[width=0.5\textwidth]{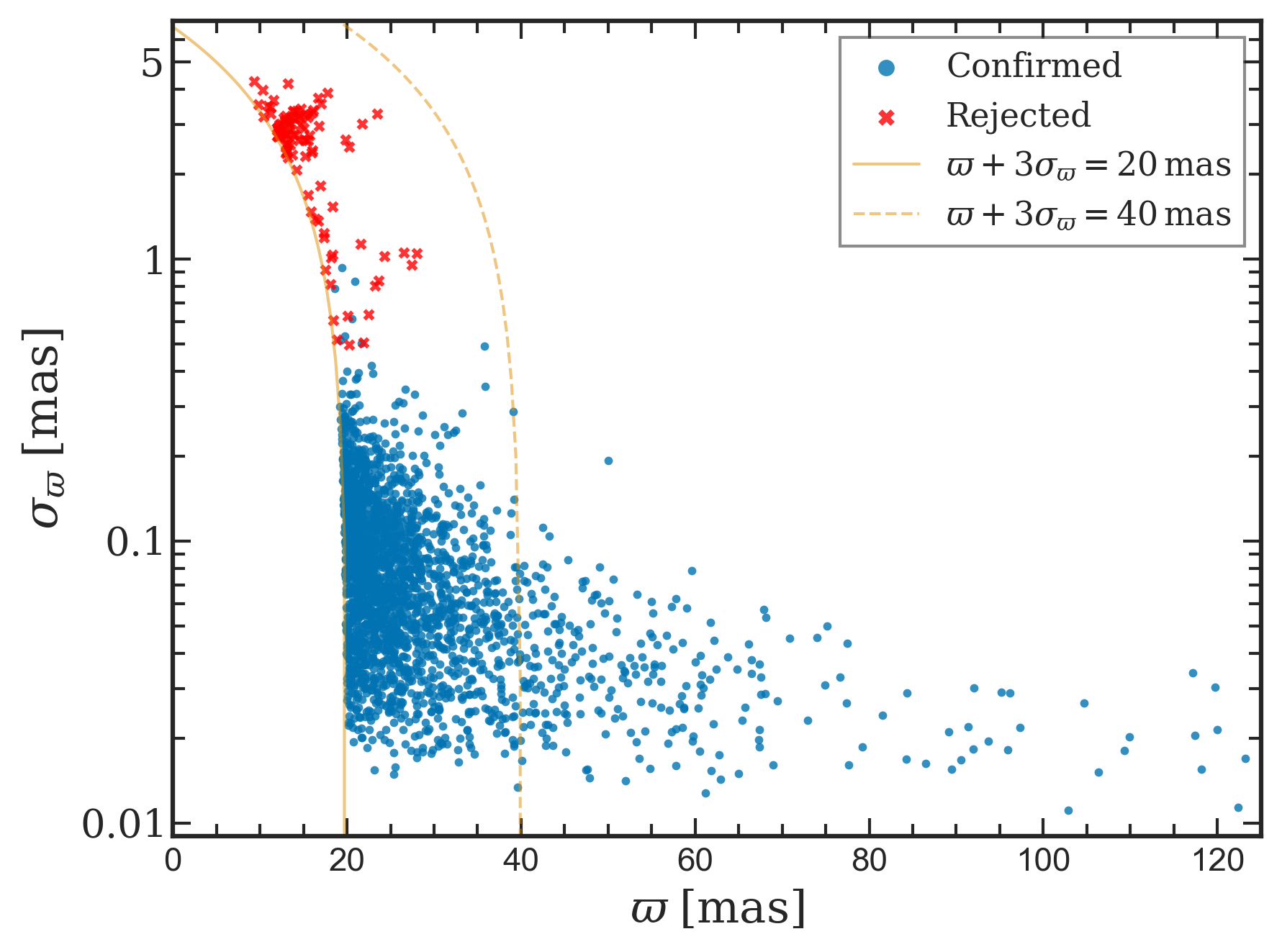} 
    \caption{Parallax uncertainty vs\ parallax for the contaminants (red symbols) and confirmed nearby white dwarfs (blue symbols). The dashed and the solid lines represent the 25~pc and 50~pc thresholds, respectively. The seven nearest confirmed white dwarfs are outside of the frame.}
    \label{fig:rejected_test}
\end{figure}

As shown in \figref{fig:rejected_test}, the contaminants exhibit significantly larger absolute parallax uncertainties than confirmed white dwarfs, and they form two distinct populations in the ($\sigma_\varpi, \varpi$) parameter space.
Parallax uncertainties of contaminants range from 0.50~mas to 4.28~mas,
whereas parallax uncertainties of confirmed white dwarfs range from 0.01~mas to 0.93~mas.
Among the 2235 confirmed white dwarfs, only 7 have parallax uncertainties above 0.50~mas, the minimum value observed for contaminants.

We note the presence of 9 sources with $\varpi/\sigma_\varpi < 4$. However, since they have $\mu/\sigma_\mu > 10$, they all fulfil one of the conditions listed in Eqs. 6, 12, or 19 in \citet{gentilefusillo21_2021MNRAS.508.3877G}.
On the other hand,
we found out that 3\,115 of all 359\,073 high-confidence white dwarfs in the GCWD21 ($0.9\%$ of the catalogue) have both $\varpi/\sigma_\varpi < 4$ and $\mu/\sigma_\mu \leq 10$.
These sources are erroneously retained and will be removed from the catalogue in the next release based on \gdrthree{} (Gentile Fusillo, private communication).

\begin{figure}
    \centering
    \includegraphics[width=0.5\textwidth]{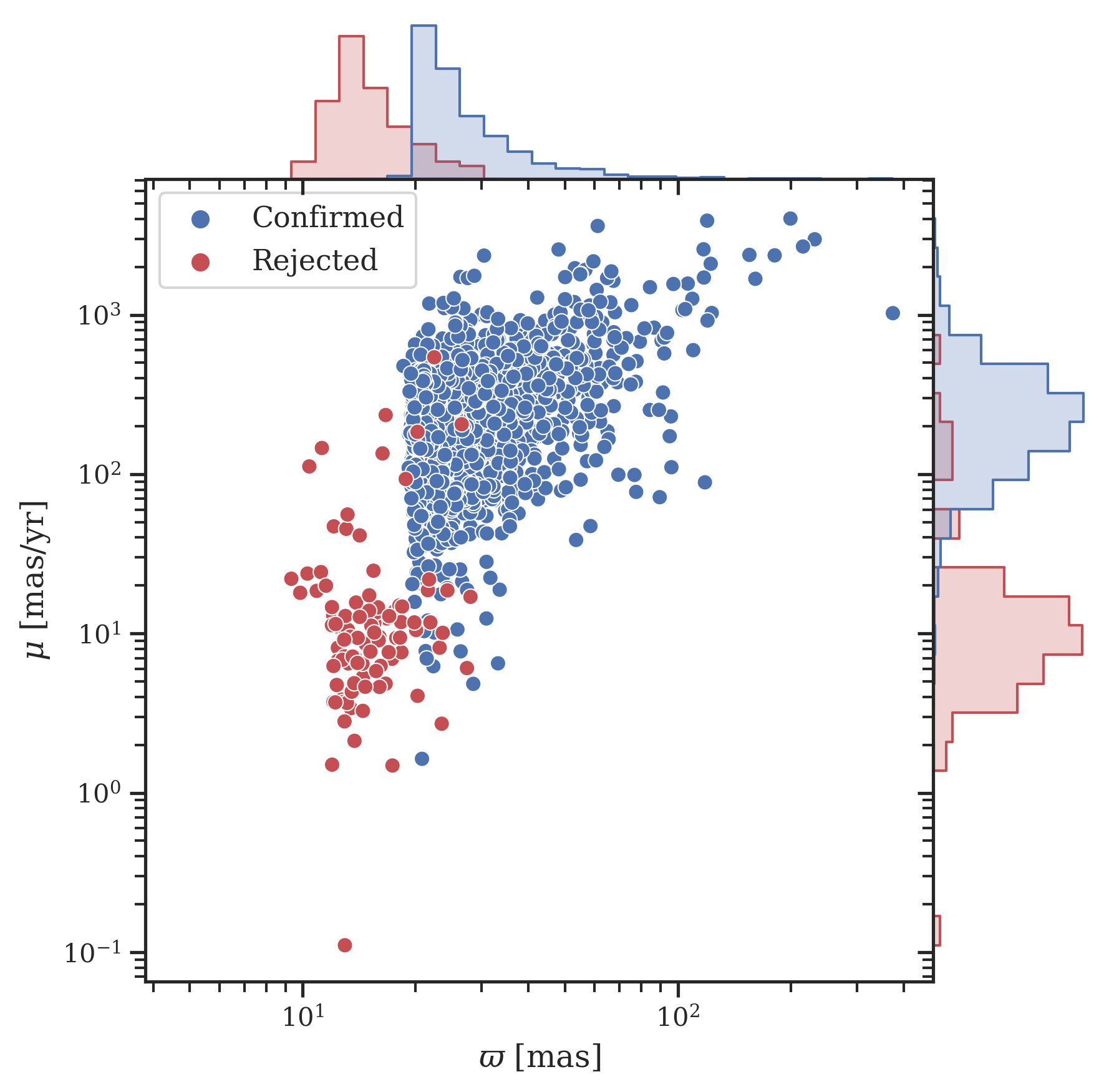} 
    \caption{Distributions of the total proper motions and parallaxes of the confirmed white dwarfs (blue symbols) and contaminants (red symbols) within 50~pc in the GCWD21.}
    \label{fig:ppm}
\end{figure}

\begin{figure}
    \centering
    \includegraphics[width=0.5\textwidth]{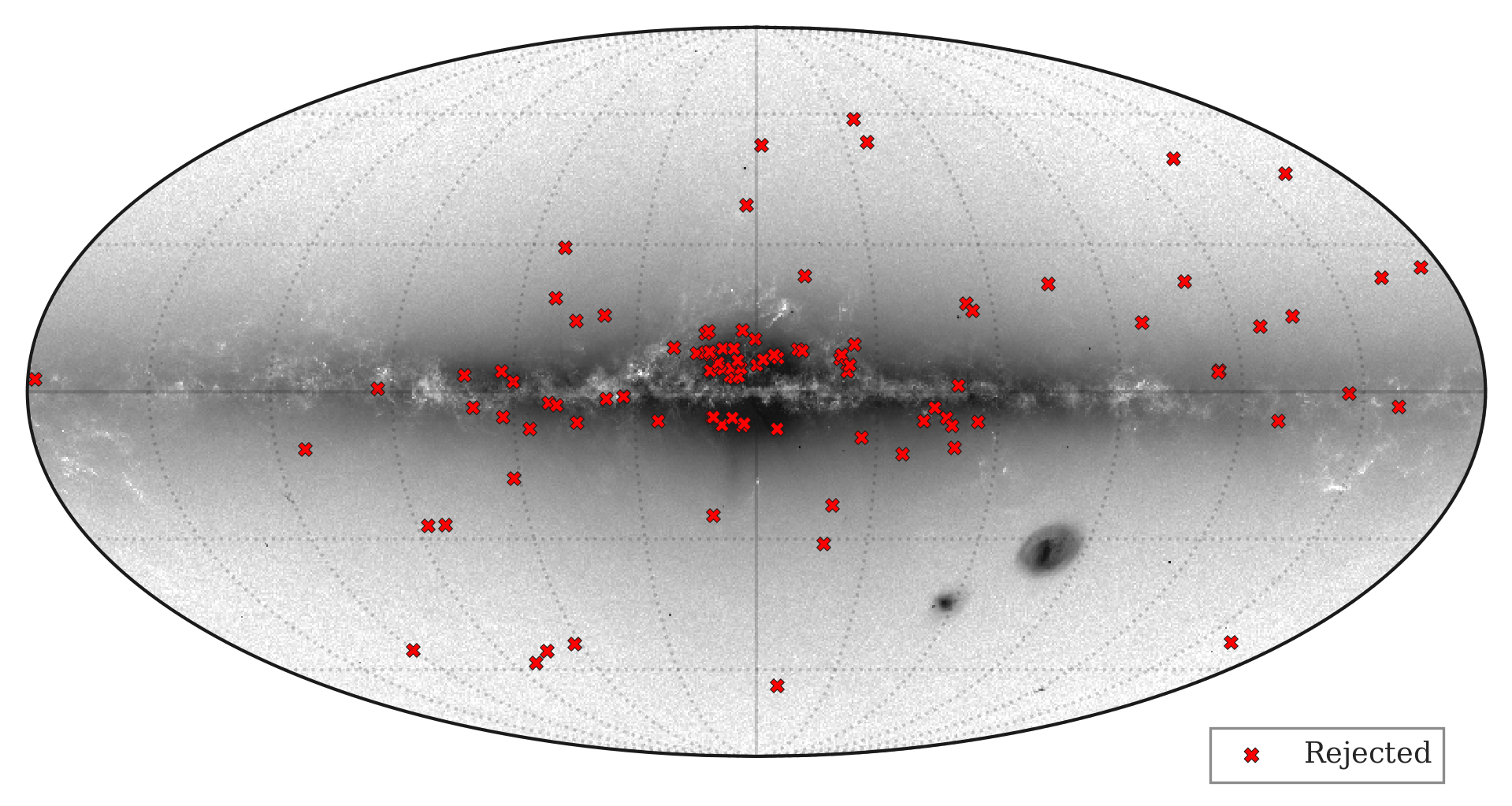} 
    \caption{Sky distribution of the contaminants in Galactic coordinates, with $l = b = 0$ at the centre. The majority of contaminants are located in the vicinity of the Galactic bulge.}
    \label{fig:skymap_rejected}
\end{figure}

Furthermore, the contaminants typically have much lower proper motion values than the confirmed white dwarfs, and their distributions differ significantly, as shown in \figref{fig:ppm}.
The majority of contaminants are located close to the Galactic bulge (\figref{fig:skymap_rejected}). 

Of the 103 contaminants, only three have an XP spectrum available in \gdrthree{}.
However, \citet{GDR3_documentation_ch14_2022gdr3.reptE..14F} conclude that the XP spectra alone are not sufficient to correctly distinguish white dwarfs from other types of stellar sources.
Nor can the reliability of an astrometric solution be inferred from these spectra.
On the other hand, it is noteworthy that these three objects have particularly large RUWE values that range from 7.12 to 10.73.
In fact, these are the objects with the largest RUWE values among all 103 contaminants.
Furthermore, \citet{OBRien_2023MNRAS.518.3055O} conducted spectroscopic follow-up observations of two contaminants from our sample: \object{\textit{Gaia}~DR3~5681511931765351936}, which has the largest amplitude of the IPD GoF as illustrated in \figref{fig:gcswd21_rej_ipd_gof},
and \object{\textit{Gaia}~DR3~4124294027400835584}. Their observations
revealed that both contaminants are main-sequence stars.

\begin{figure}
    \centering
    \includegraphics[width=0.5\textwidth]{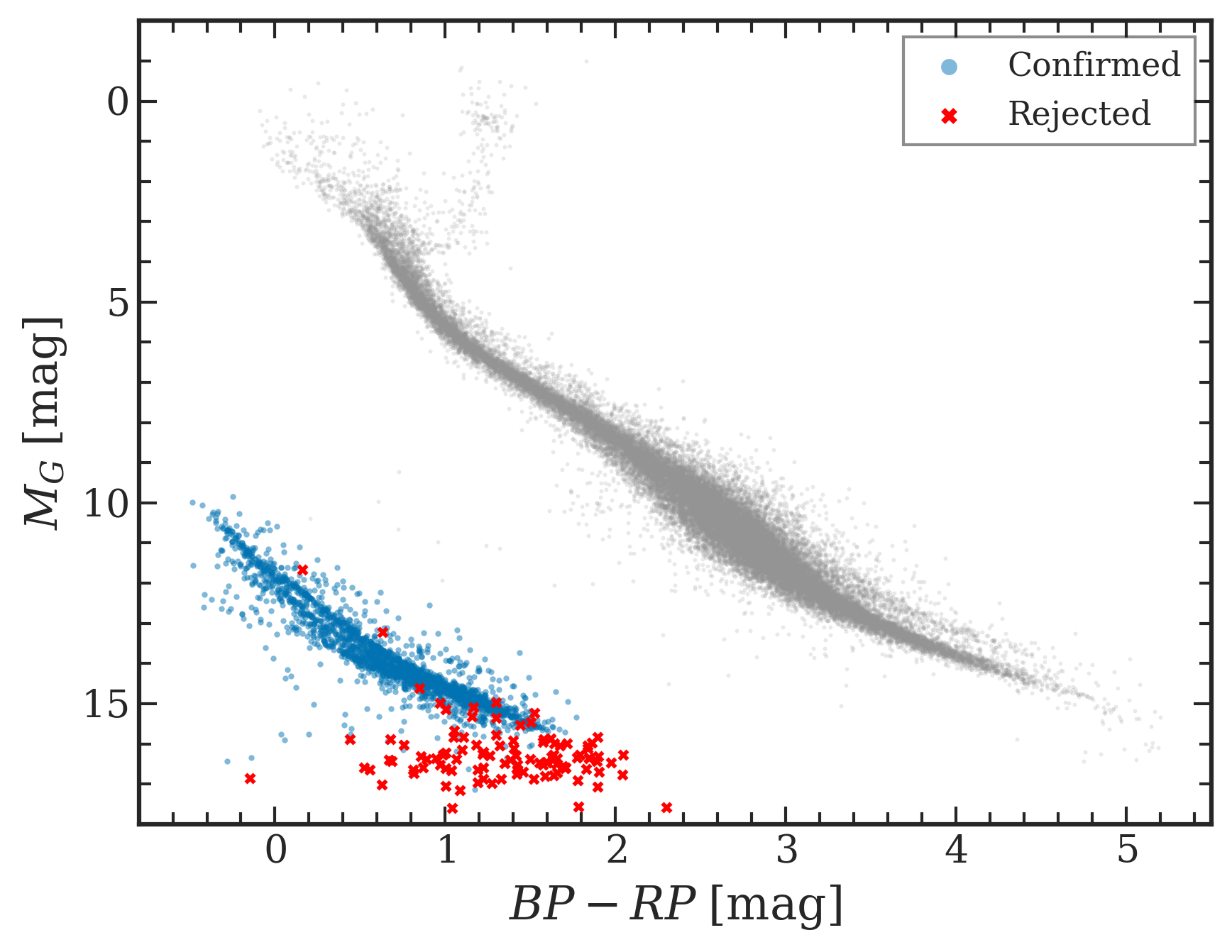} 
    \caption{\gdrthree{} CMD of the contaminants (red symbols) as compared against confirmed white dwarfs within 50~pc in the GCWD21 (blue symbols). For illustrative purposes, the CMD for the 50~pc \gdrthree{} sample is shown in the background (grey points).}
    \label{fig:HRD_rejected}
\end{figure}

Finally, the position of the identified contaminants on the CMD (\figref{fig:HRD_rejected}) coincides with the location of the sources classified as spurious in \citet[see their Fig.~18]{Rybizki_fidelity_2021arXiv210111641R}. Specifically, 97 out of 103 contaminants have an astrometric fidelity (\texttt{fidelity\_v2}) below 0.55 in \citet{Rybizki_fidelity_2021arXiv210111641R}, supporting that their astrometric solutions are likely to be spurious.
We note that the position of the identified contaminants on the CMD is also inconsistent with them being brown dwarfs.
We argue that their astrometric solutions in \gdrthree{} are indeed spurious and that these contaminants should be excluded from the 50~pc sample, thereby improving the sample purity by 4.4\%.

\section{Discussion and conclusions}
\label{sec:discussion}

\begin{figure}
    \centering
    \includegraphics[width=0.5\textwidth]{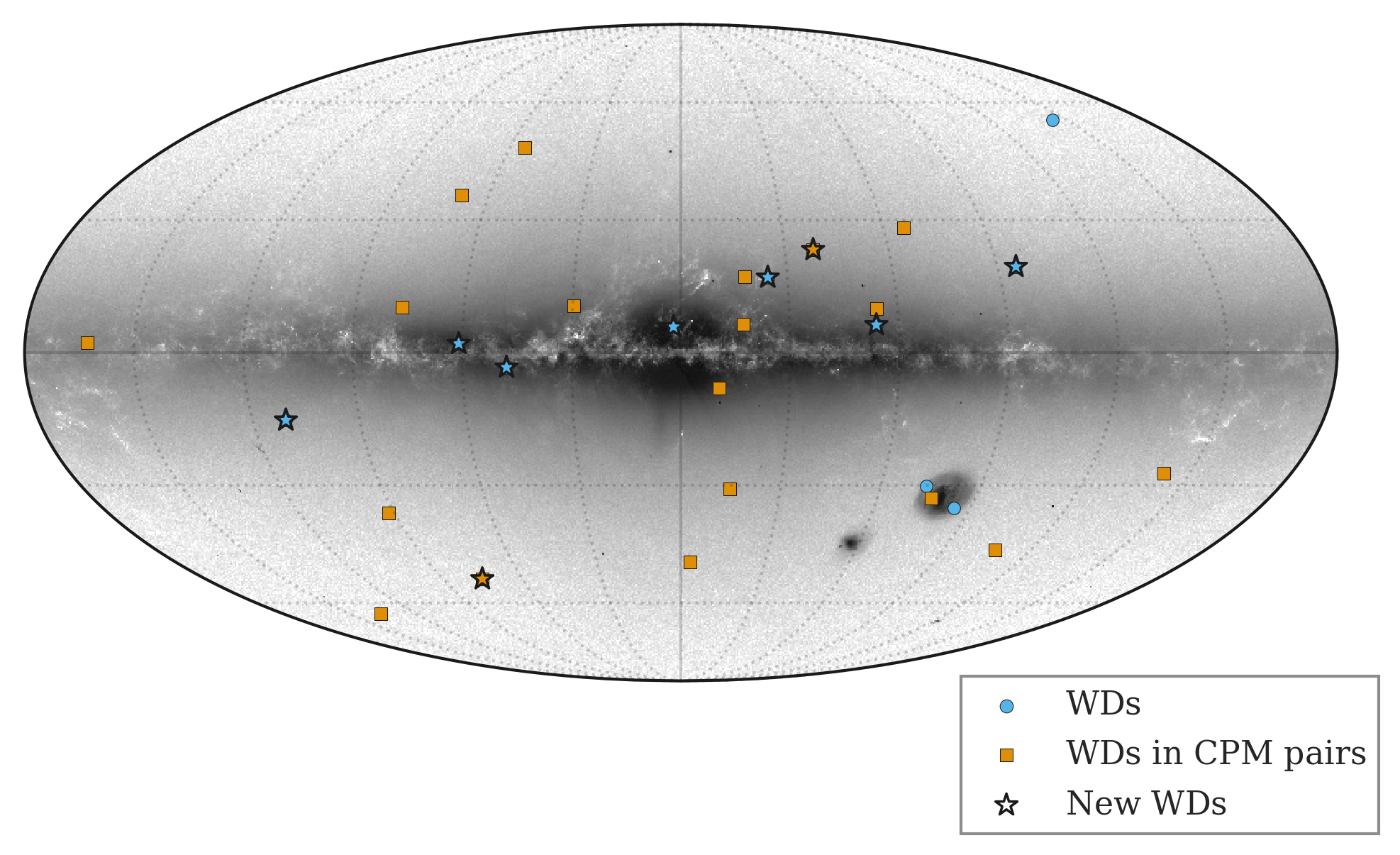} 
    \caption{Map of the identified white dwarfs (blue and orange symbols) in Galactic coordinates, with $l = b = 0$ at the centre. Orange symbols represent objects with a CPM companion.
    A star symbol denotes each of the nine new white dwarfs presented in this paper.
    For comparison, the source density map for \gdrthree{} is shown in the background.
    }
    \label{fig:wd_skymap}
\end{figure}

\begin{figure}
    \centering
    \includegraphics[width=0.5\textwidth]{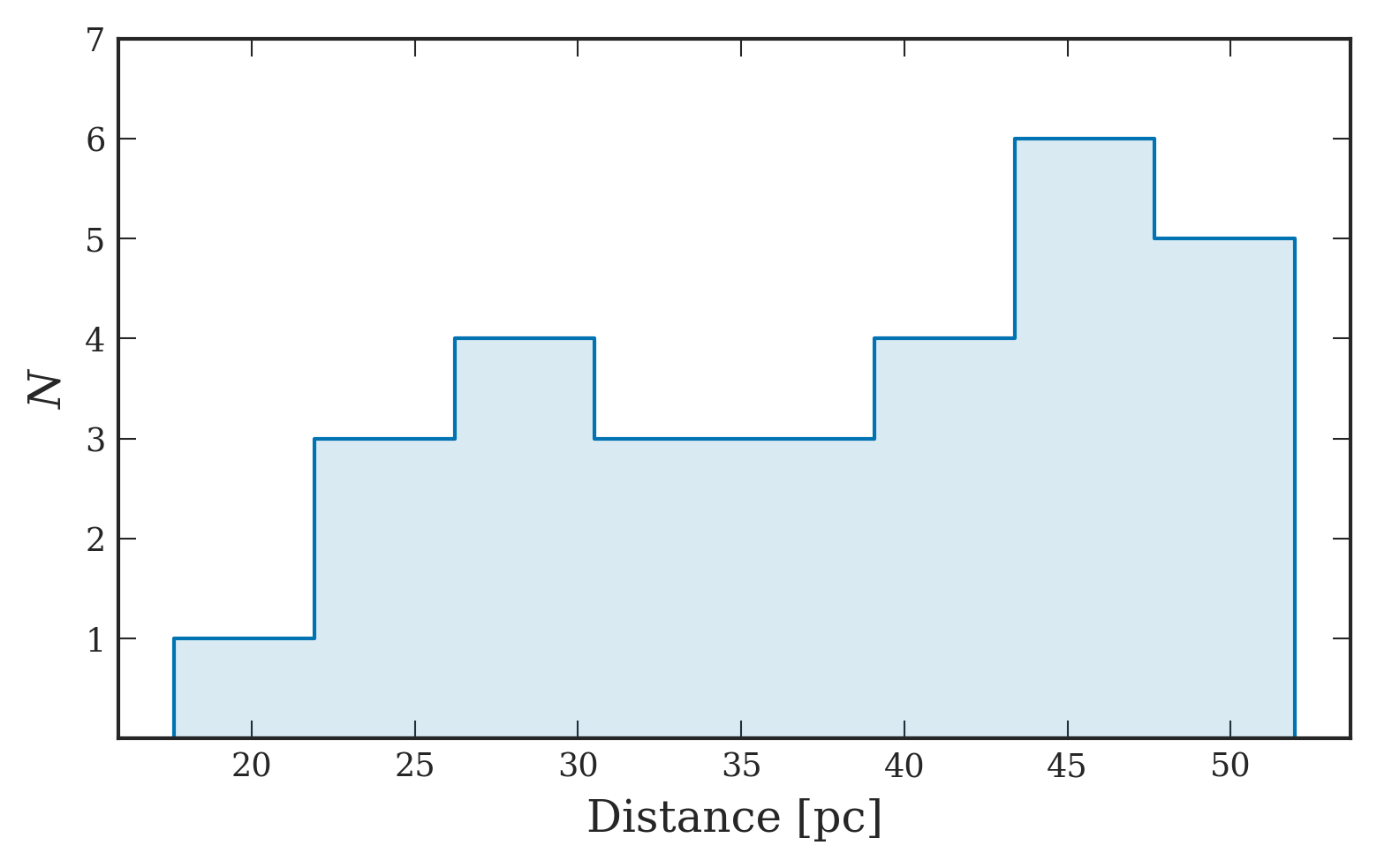}
    \caption{Distribution of distances of the white dwarfs identified in this study.}
    \label{fig:dist_hist}
\end{figure}

We have identified nine new white dwarfs and validated 21 previously reported white dwarf candidates within the 50~pc local population based on the astrometry and photometry from \gdrthree{}.
To eliminate sources with spurious astrometry, we used only two key parameters from \gdrthree{}, namely \texttt{ipd\_gof\_harmonic\_amplitude} and \texttt{parallax\_over\_error}. 
Notably, we find that 20 white dwarfs, or two-thirds of our sample, have a CPM companion.
At the same time, the identified white dwarfs are preferentially located close to the Galactic plane or in crowded fields.
\figref{fig:wd_skymap} shows the distribution of these white dwarfs in galactic coordinates. 
Of the 30 white dwarfs identified in this work, 14 are located within the region defined by $|b < 30 \degr|$ and $(0\degr < l < 90\degr$ or $270\degr < l < 360\degr)$. In this region, which covers one-quarter of the entire sky, a mere 7.5 stars would be statistically expected if they were uniformly distributed.
In addition, three white dwarfs from our sample are located in the foreground of the Large Magellanic Cloud.
The distribution of distances of the identified white dwarfs is shown in \figref{fig:dist_hist}.

In our sample, we have identified a total of nine white dwarfs that, to the best of our knowledge, have not been previously reported as members of this class. Among these, one white dwarf (\object{\gdrthree{}~6013647666939138688}) is within 25~pc. Additionally, two of the new white dwarfs have main-sequence CPM companions. While PS1 photometry is available for three of the new white dwarfs, further supporting their classification as white dwarfs, it also indicates that one of them may be one of the reddest white dwarfs in the solar neighbourhood.

The majority of the white dwarfs identified in this paper have been reported as nearby stars in the past. 
Among them, all four white dwarfs within 25~pc are by construction part of the CNS5, having been included during the catalogue compilation using the same selection criterion as in \equref{eq:ipd_gof_harm_ampl_cut}; 
however, no photometric signal-to-noise ratio threshold was enforced in the CNS5.
The CNS5 catalogue, on the other hand, does not explicitly classify white dwarfs.
Of the 30 identified white dwarfs, the GCNS lists 27 as nearby stars and rejects three, one of which is within 25~pc.

As noted in the introduction to this paper, our selection validates 21 white dwarfs that have been previously documented as such at least once. 
However, these particular white dwarfs are missing from the GCWD21 and other recent volume-limited white dwarf samples. We argue that these objects are indeed white dwarfs with reliable \gdrthree{} astrometry and should therefore be included in such catalogues.

Within our sample, 18 objects have been classified as white dwarfs according to \citet{El_Badry_inflation_2021MNRAS.506.2269E}, who compiled a catalogue of binary stars using \gedrthree{} and categorised the components as either white dwarfs or main-sequence stars.
However, it is worth noting that two objects from our sample (\object{\textit{Gaia}~DR3~6119354336882826752} and \object{\textit{Gaia}~DR3~2651214734078859648}) are also present in the catalogue but have been misclassified as main-sequence stars due to an overly steep colour-magnitude cut employed in the study.
\citet{El_Badry_inflation_2021MNRAS.506.2269E} designated any object not falling under the category of a white dwarf as a main-sequence star, which also included giants and subgiants.
The quality cuts used in their study solely required a parallax-over-error ratio greater than 5, a parallax uncertainty of less than 2~mas, and a valid (non-null) value for the $G$-band magnitude.

The GCNS provides a parameter that indicates the probability of an object being a white dwarf.
We consider an object with a probability above 0.75 in the GCNS to be a reported white dwarf and do not count it as new in this paper.
Remarkably, only three of the 27 catalogued white dwarfs have probabilities above 0.75, and only one of them 
has a probability above the 0.9 threshold.
This discrepancy is most likely due to the effect of blending, which was not taken into account when the probabilities were derived.

Noteworthy is that our selection has
rediscovered a white dwarf, \object{\textit{Gaia}~DR3~4648527839871194880}, previously reported by \citet{gentilefusillo19_2019MNRAS.482.4570G} and with a 0.97 white dwarf probability in the GCNS but currently missing from the GCWD21 catalogue. Recently, this white dwarf was spectroscopically confirmed by \citet{OBRien_2023MNRAS.518.3055O}, who classified it as a magnetic DZH white dwarf. Using photometry from \gdrtwo{}, they also estimated its effective temperature to be $T_{\rm eff} = 4720 \pm 170$~K, with a surface gravity of $\log(g) = 7.9 \pm 0.1$~dex.
Our independent estimate, derived from \gaia{} XP spectroscopy and assuming a pure-hydrogen atmosphere, gives $T_{\rm eff} = 4666 \pm 22$~K and $\log(g) = 7.81 \pm 0.02$~dex, in agreement with their values. The derivation of stellar parameters that are more realistic for a DZH white dwarf would require spectroscopy with a higher resolution and a higher signal-to-noise ratio to fit synthetic magnetic spectra to it \citep[see e.g.][]{Kuelebi_2009A&A...506.1341K, Hardy_2023MNRAS.520.6111H}.

Source density in \gdrthree{} correlates only weakly with the true source density on the sky, and consequently, it is a poor predictor of completeness, since one must also account for the \gaia{} selection function, as highlighted by \citet{Cantat-Gaudin_2023A&A...669A..55C}. However, fields with a higher concentration of sources are more prone to blending and contamination, which affect the $BP$ and $RP$ bands even more strongly than the $G$ band, leading to discrepancies in the photometry. Within our sample, all but eight sources have at least one other source in \gdrthree{} within a radius of $4\arcsec$. In the most extreme case (\object{\gdrthree{}~4314988445029569920}), there are five other sources within this radius.

Furthermore, it is important to keep in mind that in \gaia{}, both telescopes project images onto the same CCD.
Consequently, when one of the telescopes is pointed towards a crowded field, it inevitably influences the completeness and the selection function of the other field. 
Moreover, this can introduce peculiarities in the photometry of sources located within that field, even if it is not crowded. 
These peculiarities are often quantified through the photometric excess factor (\texttt{phot\_bp\_rp\_excess\_factor} in \gdrthree{}; \citealt{Evans_gaia_dr2_photom_2018A&A...616A...4E, lindegren18, gaiaedr3_photometry}). This factor is the dominant reason for the absence of the white dwarfs identified in this work from both the GCWD21 and the volume-limited samples listed in the introduction. All the white dwarfs in our sample have a highly significant flux excess in \gdrthree,{} with $C^*/\sigma_{C^*} > 37$, where $C^*$ is the corrected flux excess factor and $\sigma_{C^*}$ is its dispersion for a sample of isolated stellar sources with good quality \gaia{} photometry, as defined in \citet{gaiaedr3_photometry}. In \citet{JimenezEstebaan_2023MNRAS.518.5106J}, a $3\sigma_{C^*}$ cut is applied, while \citet{gentilefusillo21_2021MNRAS.508.3877G} use a $5\sigma_{C^*}$ cut for sources in high-density fields and a $|C^*| < 0.6$ cut for sources at high Galactic latitudes ($|b| < 25 \deg$) or low in low-density fields ($\tt density < 400$)\footnote{The \texttt{density} parameter is defined in their paper as the total number of sources in \gedrthree{} within 50~arcsec$^2$ pixels.}.

The use of cuts on the RUWE is another reason why these white dwarfs have been rejected in prior studies. Within our sample, ten objects have $\rm RUWE > 1.4$, while 20 objects have $\rm RUWE > 1.1$, which is the threshold used in the GCWD21.

Besides improving the overall completeness of the local white dwarf census, the identified white dwarfs are important objects for constraining the statistical properties of the solar neighbourhood.
For example, four white dwarfs in our sample have an effective temperature of $T_{\rm eff}\leq4000$~K within the $1\sigma$ interval in either DA or DB solutions, and two of them have an absolute magnitude of $M_G > 16.0$~mag. Such ultra-cool white dwarfs are extremely rare: only 24 white dwarfs within 50~pc with $M_G > 16.0$~mag are listed in the GCWD21  (the contaminants are excluded).
These objects define the faint end of the white dwarf luminosity function from which the age of the local stellar population can be inferred \citep{Schmidt_1959ApJ...129..243S}.

In our sample, we have identified a total of 19 white dwarfs within binary or triple systems. Of these, one is a triple system in which the white dwarf (\object{\gdrthree{}~6017724140666000896}) has a CPM companion (\object{GJ~629}) that is itself a spectroscopic binary. 
Additionally, one is a double white dwarf system (\object{\gdrthree{}~6039626481000400128} and \object{\gdrthree{}~6039626481009257472}). The remaining binaries in our sample are Sirius-like systems, characterised by the presence of a significantly brighter main-sequence companion. Within our sample, the difference in brightness between the components of these systems ranges from 3.5 to 9.8 magnitudes in the $G$ band.

One of the CPM companions of the white dwarfs validated in our study, \object{GJ~3344}, had been previously targeted in the spectroscopic search for undetected white dwarfs in binaries with main-sequence companions.
\citet{Bar_2017ApJ...850...34B} selected 101 M dwarfs from the CNS3 catalogue \citep{cns3}, limiting their sample to objects with strong near-UV emission, as detected by the GALEX survey \citep{Martin_GALEX_2005ApJ...619L...1M}. This near-UV emission can indicate the presence of an unresolved white dwarf companion.
Despite their efforts, \citet{Bar_2017ApJ...850...34B} could not detect any white dwarfs around these 101 M dwarfs. The authors concluded that either the undetected white dwarfs are located outside the $5\arcsec$ slit used for observations or the effective temperature of the undetected white dwarfs is below the detection limit, estimated to be $8500\sim10000$~K for \object{GJ~3344}.
For \object{GJ~3344}, both of these explanations could be valid, as (i) the targets in \citet{Bar_2017ApJ...850...34B} were positioned in the centre of the $5\arcsec$ slit, while we found the white dwarf companion at a separation of $\rho = 2.98\arcsec$, and (ii) our estimation of its effective temperature from \gaia{} XP spectra suggests a value of about $T_{\mathrm{eff}} \sim 6700$~K.

Although the addition of these previously overlooked white dwarfs to the 50~pc sample has undoubtedly improved its completeness, the overall sample size has shrunk from 2338 to 2265 high-confidence white dwarfs. This reduction is a consequence of the improved sample purity achieved through the identification and exclusion of 103 sources with spurious astrometric solutions within the 50~pc subsample of the GCWD21. 
The distribution of the astrometric and photometric properties of these contaminants, along with their position in the sky, indicates that their astrometric solutions in \gdrthree{} are unreliable and that applying the cut on the amplitude of angular variation of the IPD GoF can significantly improve both the completeness and purity of the local white dwarf sample. 

Furthermore, four white dwarfs identified in this study (three of them validated and one new) are located within 25~pc of the Sun. This is a significant addition ($1.6\%$) to the 256 white dwarfs in the 25~pc sample of the GCWD21.

The identification of these new nearby white dwarfs is yet another step forward towards a volume-complete sample of white dwarfs.

\begin{sidewaystable*}

\caption{Stellar parameters for the white dwarfs listed in \tabref{tab:new_wds}}
    \label{tab:param}
    \centering
    \begin{tabular}{l| l l l l l l l l| l l l l l l l l}
    \hline \hline
    & \multicolumn{1}{l}{DA solution} & & & & & & & & \multicolumn{1}{l}{DB solution} \\
    \multicolumn{1}{l|}{\gdrthree{}}&
    \multicolumn{1}{c}{$T_\mathrm{eff}$} &
    \multicolumn{1}{c}{$\sigma_{T_\mathrm{eff}}$} &
    \multicolumn{1}{c}{$\log(g)$} &
    \multicolumn{1}{c}{$\sigma_{\log(g)}$} &
    \multicolumn{1}{c}{$M$} &
    \multicolumn{1}{c}{$\sigma_{M}$} &
    \multicolumn{1}{c}{$M_\mathrm{bol}$} &
    \multicolumn{1}{c|}{$\sigma_{M_\mathrm{bol}}$} &
    \multicolumn{1}{c}{$T_\mathrm{eff}$} &
    \multicolumn{1}{c}{$\sigma_{T_\mathrm{eff}}$} &
    \multicolumn{1}{c}{$\log(g)$} &
    \multicolumn{1}{c}{$\sigma_{\log(g)}$} &
    \multicolumn{1}{c}{$M$} &
    \multicolumn{1}{c}{$\sigma_{M}$} &
    \multicolumn{1}{c}{$M_\mathrm{bol}$} &
    \multicolumn{1}{c}{$\sigma_{M_\mathrm{bol}}$} \\
    \texttt{source\_id} & [K] & [K] & [dex] & [dex] & [$\msun$] & [$\msun$] & [mag] & [mag] & [K] & [K] & [dex] & [dex] & [$\msun$] & [$\msun$] & [mag] & [mag] \\
    \hline
    6013647666939138688     & 4326 & 180 & 8.4 & 0.1 & 0.83 & 0.09 & 16.11 & 0.04 &  4057 & 131 & 8.15 & 0.1 & 0.67 & 0.08 & 16.01 & 0.04 \\
    4651329704762754944     & 6734 & 55 & 7.28 & 0.03 & 0.27 & 0.01 & 12.62 & 0.01 & 6723 & 55 & 7.25 & 0.03 & 0.24 & 0.01 & 12.67 & 0.01 \\
    6017724140666000896     & 5038 & 245 & 8.08 & 0.17 & 0.64 & 0.12 & 14.96 & 0.06 & 4717 & 266 & 7.86 & 0.20 & 0.49 & 0.10 & 14.97 & 0.02 \\
    2983256662868370048     & 7935 & 1789 & 7.62 & 0.51 & 0.31 & 0.16 & 12.39 & 0.12 & 8194 & 1389 & 7.69 & 0.36 & 0.29 & 0.16 & 12.43 & 0.14 \\
    1938960722332184704     & 5633 & 106 & 8.24 & 0.05 & 0.74 & 0.04 & 14.71 & 0.01 & 5368 & 130 & 8.08 & 0.07 & 0.64 & 0.03 & 14.70 & 0.01 \\
    2078105327586616832     & 4702 & 398 & 8.1 & 0.3 & 0.66 & 0.17 & 15.33 & 0.06 & 4436 & 411 & 7.9 & 0.3 & 0.53 & 0.17 & 15.32 & 0.04 \\
    6119354336882826752     & 4167 & 47 & 7.04 & 0.06 & 0.17 & 0.02 & 14.62 & 0.01 &  4091 & 33 & 7.03 & 0.04 & 0.16 & 0.01 & 14.71 & 0.01 \\
    2831490694928280576     & 6960 & 387 & 7.42 & 0.16 & 0.32 & 0.06 & 12.65 & 0.04 & 6841 & 375 & 7.3 & 0.2 & 0.28 & 0.06 & 12.70 & 0.04 \\
    5956907713001974656     & 6857 & 188 & 7.86 & 0.07 & 0.51 & 0.05 & 13.28 & 0.02 & 6707 & 189 & 7.78 & 0.08 & 0.45 & 0.05 & 13.32 & 0.02 \\
    6665685378201412992     & 5714 & 325 & 7.7 & 0.2 & 0.44 & 0.1 & 13.90 & 0.01 & 5480 & 350 & 7.55 &  0.2 & 0.36 & 0.11 & 13.93 & 0.01 \\
    205635204411687168      & 6015 & 53 & 7.74 & 0.03 & 0.43 & 0.01 & 13.70 & 0.01 & 5812 & 59 & 7.60 & 0.04 & 0.36 & 0.02 & 13.73 & 0.01 \\
    2651214734078859648     & 5008 & 75 & 7.05 & 0.08 & 0.18 & 0.02 & 13.75 & 0.02 & 4979 & 70 & 7.04 & 0.07 & 0.17 & 0.02 & 13.84 & 0.02 \\
    4110515669211359744     & 4867 & 42 & 7.69 & 0.04 & 0.41 & 0.02 & 14.58 & 0.01 & 4494 & 42 & 7.39 & 0.04 & 0.28 & 0.02 & 14.64 & 0.01 \\
    5463514273884166016     & 5087 & 66 & 7.52 & 0.05 & 0.34 & 0.02 & 14.19 & 0.01 & 4788 & 70 & 7.27 & 0.06 & 0.24 & 0.02 & 14.24 & 0.02 \\
    4648527839871194880     & 4666 & 22 & 7.81 & 0.02 & 0.47 & 0.01 & 14.91 & 0.01 & 4323 & 21 & 7.54 & 0.02 & 0.33 & 0.01 & 14.96 & 0.01 \\
    6566046912935892352     & 6596 & 527 & 7.27 & 0.24 & 0.26 & 0.09 & 12.68 & 0.05 & 6606 & 467 & 7.23 & 0.23 & 0.24 & 0.08 & 12.73 & 0.05 \\
    6063480282704764928     & 3474 & 68 & 7.06 & 0.08 & 0.17 & 0.02 & 15.44 & 0.03 & 3626 & 28 & 7.03 & 0.04 & 0.16 & 0.01 & 15.25 & 0.02 \\
    1355203232910297600     & 5586 & 96 & 7.05 & 0.07 & 0.19 & 0.02 & 13.20 & 0.01 & 5645 & 91 & 7.04 & 0.07 & 0.18 & 0.02 & 13.25 & 0.01 \\
    2533660345315705344     & 13333 & 2526 & 7.7 & 0.2 & 0.45 & 0.12 & 10.14 & 0.38 & 13964 & 2714 & 7.8 & 0.3 & 0.49 & 0.18 & 10.06 & 0.36 \\
    4663902104803506816     & 5040 & 72 & 8.04 & 0.05 & 0.61 & 0.03 & 14.90 & 0.02 & 4706 & 77 & 7.82 & 0.06 & 0.46 & 0.03 & 14.92 & 0.01 \\
    3462945170562185088     & 4056 & 399 & 7.7 & 0.3 & 0.45 & 0.15 & 15.44 & 0.07 & 3914 & 253 & 7.53 & 0.27 & 0.35 & 0.13 & 15.39 & 0.04 \\
    6080038412403680256     & 5370 & 80 & 7.05 & 0.07 & 0.19 & 0.02 & 13.40 & 0.01 & 5398 & 66 & 7.04 & 0.05 & 0.17 & 0.02 & 13.46 & 0.01 \\
    2021862490380335232     & 5969 & 338 & 8.22 & 0.16 & 0.73 & 0.11 & 14.41 & 0.03 & 5737 & 370 & 8.09 & 0.19 & 0.62 & 0.11 & 14.41 & 0.03 \\
    6039626481000400128     & 5554 & 102 & 8.12 & 0.06 & 0.67 & 0.04 & 14.59 & 0.01 & 5277 & 116 & 7.95 & 0.07 & 0.54 & 0.04 & 14.59 & 0.01 \\
    6039626481009257472     & 5241 & 308 & 7.9 & 0.2 & 0.55 & 0.1 & 14.55 & 0.05 &  4923 & 334 & 7.7 & 0.25 & 0.4 & 0.1 & 14.58 & 0.02 \\
    4837326390227065344     & 6286 & 308 & 7.6 & 0.15 & 0.39 & 0.07 & 13.33 & 0.02 & 6106 & 318 & 7.5 & 0.18 & 0.32 & 0.07 & 13.37 & 0.02 \\
    1323632298413108096     & 6856 & 53 & 7.59 & 0.02 & 0.38 & 0.01 & 12.93 & 0.01 & 6755 & 53 & 7.53 & 0.03 & 0.34 & 0.01 & 12.98 & 0.01 \\
    4468278954501360128     & 4522 & 72 & 7.54 & 0.08 & 0.34 & 0.03 & 14.74 & 0.02 & 4182 & 71 & 7.23 & 0.09 & 0.22 & 0.03 & 14.79 & 0.02 \\
    625602672887745152      & 4384 & 94 & 7.51& 0.09 & 0.33 & 0.04 & 14.85 & 0.01 & 4121 & 79 & 7.26 & 0.09 & 0.24 & 0.03 & 14.89 & 0.01 \\
    4314988445029569920     & 3677 & 572 & 8.0 & 0.5 & 0.6 & 0.2 & 16.25 & 0.2 &  3646 & 407 & 7.7 & 0.4 & 0.5 & 0.2 & 16.0 & 0.15 \\
         \hline
    \end{tabular}
\end{sidewaystable*}

\begin{acknowledgements}
We would like to thank the anonymous referee for a thorough review and many helpful comments, which have improved this manuscript.

Part of this work was supported by the International Max Planck Research School for Astronomy and Cosmic Physics at the University of Heidelberg, IMPRS-HD, Germany.
A.G. and A.J. gratefully acknowledge funding from the Deutsche Forschungsgemeinschaft (DFG, German Research Foundation) -- Project-ID 138713538 -- SFB 881 (``The Milky Way System'', subproject A06).

This work has made use of:
TOPCAT \citep{topcat_2005ASPC..347...29T, topcat_2019ASPC..523...43T}, a GUI analysis package for working with tabular data in astronomy;
Astropy, a community-developed core Python package for Astronomy \citep{astropy2018};
Scipy, a set of open source scientific and numerical tools for Python \citep{scipy_2020NatMe..17..261V};
WDPhotTools \citep{Lam_WDPhotTools_2022RASTI...1...81L}, a white dwarf photometric toolkit in Python;
GaiaXPy \citep{GDR3_Montegriffo_XP_synth_phot_2022arXiv220606215G}, a Python package, developed and maintained by members of the \gaia{} Data Processing and Analysis Consortium (DPAC), and in particular, Coordination Unit 5 (CU5), and the Data Processing Centre located at the Institute of Astronomy, Cambridge, UK (DPCI).
the VizieR catalogue access tool and the SIMBAD database operated at CDS, Strasbourg, France;
the National Aeronautics and Space Administration (NASA) Astrophysics Data System (ADS).

This work has made use of
data from the European Space Agency (ESA) mission {\it Gaia} (\url{https://www.cosmos.esa.int/gaia}), processed by the {\it Gaia}
Data Processing and Analysis Consortium (DPAC, \url{https://www.cosmos.esa.int/web/gaia/dpac/consortium}). Funding for the DPAC has been provided by national institutions, in particular, the institutions participating in the {\it Gaia} Multilateral Agreement.

The Pan-STARRS1 Surveys (PS1) have been made possible through contributions of the Institute for Astronomy, the University of Hawaii, the Pan-STARRS Project Office, the Max-Planck Society and its participating institutes, the Max Planck Institute for Astronomy, Heidelberg and the Max Planck Institute for Extraterrestrial Physics, Garching, The Johns Hopkins University, Durham University, the University of Edinburgh, Queen's University Belfast, the Harvard-Smithsonian Center for Astrophysics, the Las Cumbres Observatory Global Telescope Network Incorporated, the National Central University of Taiwan, the Space Telescope Science Institute, the National Aeronautics and Space Administration under Grant No. NNX08AR22G issued through the Planetary Science Division of the NASA Science Mission Directorate, the National Science Foundation under Grant No. AST-1238877, the University of Maryland, and Eotvos Lorand University (ELTE).
\end{acknowledgements}

\bibliographystyle{aa}
\bibliography{refs}

\begin{thebibliography}{76}
\expandafter\ifx\csname natexlab\endcsname\relax\def\natexlab#1{#1}\fi

\bibitem[{{Althaus} {et~al.}(2013){Althaus}, {Miller Bertolami}, \& {C{\'o}rsico}}]{Althaus_2013A&A...557A..19A}
{Althaus}, L.~G., {Miller Bertolami}, M.~M., \& {C{\'o}rsico}, A.~H. 2013, \aap, 557, A19

\bibitem[{{Astropy Collaboration} {et~al.}(2018){Astropy Collaboration}, {Price-Whelan}, {Sip{\H o}cz}, {G{\"u}nther}, {Lim}, {Crawford}, {Conseil}, {Shupe}, {Craig}, {Dencheva}, {Ginsburg}, {VanderPlas}, {Bradley}, {P{\'e}rez-Su{\'a}rez}, {de Val-Borro}, {Aldcroft}, {Cruz}, {Robitaille}, {Tollerud}, {Ardelean}, {Babej}, {Bach}, {Bachetti}, {Bakanov}, {Bamford}, {Barentsen}, {Barmby}, {Baumbach}, {Berry}, {Biscani}, {Boquien}, {Bostroem}, {Bouma}, {Brammer}, {Bray}, {Breytenbach}, {Buddelmeijer}, {Burke}, {Calderone}, {Cano Rodr{\'{\i}}guez}, {Cara}, {Cardoso}, {Cheedella}, {Copin}, {Corrales}, {Crichton}, {D'Avella}, {Deil}, {Depagne}, {Dietrich}, {Donath}, {Droettboom}, {Earl}, {Erben}, {Fabbro}, {Ferreira}, {Finethy}, {Fox}, {Garrison}, {Gibbons}, {Goldstein}, {Gommers}, {Greco}, {Greenfield}, {Groener}, {Grollier}, {Hagen}, {Hirst}, {Homeier}, {Horton}, {Hosseinzadeh}, {Hu}, {Hunkeler}, {Ivezi{\'c}}, {Jain}, {Jenness}, {Kanarek}, {Kendrew}, {Kern}, {Kerzendorf}, {Khvalko}, {King}, {Kirkby}, {Kulkarni},
  {Kumar}, {Lee}, {Lenz}, {Littlefair}, {Ma}, {Macleod}, {Mastropietro}, {McCully}, {Montagnac}, {Morris}, {Mueller}, {Mumford}, {Muna}, {Murphy}, {Nelson}, {Nguyen}, {Ninan}, {N{\"o}the}, {Ogaz}, {Oh}, {Parejko}, {Parley}, {Pascual}, {Patil}, {Patil}, {Plunkett}, {Prochaska}, {Rastogi}, {Reddy Janga}, {Sabater}, {Sakurikar}, {Seifert}, {Sherbert}, {Sherwood-Taylor}, {Shih}, {Sick}, {Silbiger}, {Singanamalla}, {Singer}, {Sladen}, {Sooley}, {Sornarajah}, {Streicher}, {Teuben}, {Thomas}, {Tremblay}, {Turner}, {Terr{\'o}n}, {van Kerkwijk}, {de la Vega}, {Watkins}, {Weaver}, {Whitmore}, {Woillez}, {Zabalza}, \& {Astropy Contributors}}]{astropy2018}
{Astropy Collaboration}, {Price-Whelan}, A.~M., {Sip{\H o}cz}, B.~M., {et~al.} 2018, \aj, 156, 123

\bibitem[{{Bar} {et~al.}(2017){Bar}, {Vreeswijk}, {Gal-Yam}, {Ofek}, \& {Nelemans}}]{Bar_2017ApJ...850...34B}
{Bar}, I., {Vreeswijk}, P., {Gal-Yam}, A., {Ofek}, E.~O., \& {Nelemans}, G. 2017, \apj, 850, 34

\bibitem[{{B{\'e}dard} {et~al.}(2020){B{\'e}dard}, {Bergeron}, {Brassard}, \& {Fontaine}}]{Bedard_Montreal_models_2020ApJ...901...93B}
{B{\'e}dard}, A., {Bergeron}, P., {Brassard}, P., \& {Fontaine}, G. 2020, \apj, 901, 93

\bibitem[{{Bergeron} {et~al.}(2019){Bergeron}, {Dufour}, {Fontaine}, {Coutu}, {Blouin}, {Genest-Beaulieu}, {B{\'e}dard}, \& {Rolland}}]{Bergeron_2019ApJ...876...67B}
{Bergeron}, P., {Dufour}, P., {Fontaine}, G., {et~al.} 2019, \apj, 876, 67

\bibitem[{{Bergeron} {et~al.}(2011){Bergeron}, {Wesemael}, {Dufour}, {Beauchamp}, {Hunter}, {Saffer}, {Gianninas}, {Ruiz}, {Limoges}, {Dufour}, {Fontaine}, \& {Liebert}}]{Bergeron_2011ApJ...737...28B}
{Bergeron}, P., {Wesemael}, F., {Dufour}, P., {et~al.} 2011, \apj, 737, 28

\bibitem[{{Bopp} {et~al.}(1970){Bopp}, {Evans}, {Laing}, \& {Deeming}}]{Bopp_1970MNRAS.147..355B}
{Bopp}, B.~W., {Evans}, D.~S., {Laing}, J.~D., \& {Deeming}, T.~J. 1970, \mnras, 147, 355

\bibitem[{{Brown} {et~al.}(2020){Brown}, {Kilic}, {Kosakowski}, {Andrews}, {Heinke}, {Ag{\"u}eros}, {Camilo}, {Gianninas}, {Hermes}, \& {Kenyon}}]{Brown_2020ApJ...889...49B}
{Brown}, W.~R., {Kilic}, M., {Kosakowski}, A., {et~al.} 2020, \apj, 889, 49

\bibitem[{{Busso} {et~al.}(2021){Busso}, {Cacciari}, {Bellazzini}, {Carrasco}, {De Angeli}, {Evans}, {Fabricius}, {Jordi}, {Montegriffo}, {Pancino}, {Rainer}, \& {Sanna}}]{gaiaedr3_documentation_ch5_2021gdr3.reptE...5B}
{Busso}, G., {Cacciari}, C., {Bellazzini}, M., {et~al.} 2021, {Gaia EDR3 documentation Chapter 5: Photometric data}, Gaia EDR3 documentation

\bibitem[{{Cantat-Gaudin} {et~al.}(2023){Cantat-Gaudin}, {Fouesneau}, {Rix}, {Brown}, {Castro-Ginard}, {Kostrzewa-Rutkowska}, {Drimmel}, {Hogg}, {Casey}, {Khanna}, {Oh}, {Price-Whelan}, {Belokurov}, {Saydjari}, \& {Green}}]{Cantat-Gaudin_2023A&A...669A..55C}
{Cantat-Gaudin}, T., {Fouesneau}, M., {Rix}, H.-W., {et~al.} 2023, \aap, 669, A55

\bibitem[{{Catal{\'a}n} {et~al.}(2008){Catal{\'a}n}, {Isern}, {Garc{\'\i}a-Berro}, \& {Ribas}}]{Catalan_2008MNRAS.387.1693C}
{Catal{\'a}n}, S., {Isern}, J., {Garc{\'\i}a-Berro}, E., \& {Ribas}, I. 2008, \mnras, 387, 1693

\bibitem[{{Chambers} {et~al.}(2016){Chambers}, {Magnier}, {Metcalfe}, {Flewelling}, {Huber}, {Waters}, {Denneau}, {Draper}, {Farrow}, {Finkbeiner}, {Holmberg}, {Koppenhoefer}, {Price}, {Rest}, {Saglia}, {Schlafly}, {Smartt}, {Sweeney}, {Wainscoat}, {Burgett}, {Chastel}, {Grav}, {Heasley}, {Hodapp}, {Jedicke}, {Kaiser}, {Kudritzki}, {Luppino}, {Lupton}, {Monet}, {Morgan}, {Onaka}, {Shiao}, {Stubbs}, {Tonry}, {White}, {Ba{\~n}ados}, {Bell}, {Bender}, {Bernard}, {Boegner}, {Boffi}, {Botticella}, {Calamida}, {Casertano}, {Chen}, {Chen}, {Cole}, {Deacon}, {Frenk}, {Fitzsimmons}, {Gezari}, {Gibbs}, {Goessl}, {Goggia}, {Gourgue}, {Goldman}, {Grant}, {Grebel}, {Hambly}, {Hasinger}, {Heavens}, {Heckman}, {Henderson}, {Henning}, {Holman}, {Hopp}, {Ip}, {Isani}, {Jackson}, {Keyes}, {Koekemoer}, {Kotak}, {Le}, {Liska}, {Long}, {Lucey}, {Liu}, {Martin}, {Masci}, {McLean}, {Mindel}, {Misra}, {Morganson}, {Murphy}, {Obaika}, {Narayan}, {Nieto-Santisteban}, {Norberg}, {Peacock}, {Pier}, {Postman}, {Primak}, {Rae}, {Rai},
  {Riess}, {Riffeser}, {Rix}, {R{\"o}ser}, {Russel}, {Rutz}, {Schilbach}, {Schultz}, {Scolnic}, {Strolger}, {Szalay}, {Seitz}, {Small}, {Smith}, {Soderblom}, {Taylor}, {Thomson}, {Taylor}, {Thakar}, {Thiel}, {Thilker}, {Unger}, {Urata}, {Valenti}, {Wagner}, {Walder}, {Walter}, {Watters}, {Werner}, {Wood-Vasey}, \& {Wyse}}]{Chambers_2016arXiv161205560C}
{Chambers}, K.~C., {Magnier}, E.~A., {Metcalfe}, N., {et~al.} 2016, arXiv e-prints, arXiv:1612.05560

\bibitem[{{Creevey} {et~al.}(2023){Creevey}, {Sordo}, {Pailler}, {Fr{\'e}mat}, {Heiter}, {Th{\'e}venin}, {Andrae}, {Fouesneau}, {Lobel}, {Bailer-Jones}, {Garabato}, {Bellas-Velidis}, {Brugaletta}, {Lorca}, {Ordenovic}, {Palicio}, {Sarro}, {Delchambre}, {Drimmel}, {Rybizki}, {Torralba Elipe}, {Korn}, {Recio-Blanco}, {Schultheis}, {De Angeli}, {Montegriffo}, {Abreu Aramburu}, {Accart}, {{\'A}lvarez}, {Bakker}, {Brouillet}, {Burlacu}, {Carballo}, {Casamiquela}, {Chiavassa}, {Contursi}, {Cooper}, {Dafonte}, {Dapergolas}, {de Laverny}, {Dharmawardena}, {Edvardsson}, {Le Fustec}, {Garc{\'\i}a-Lario}, {Garc{\'\i}a-Torres}, {Gomez}, {Gonz{\'a}lez-Santamar{\'\i}a}, {Hatzidimitriou}, {Jean-Antoine Piccolo}, {Kontiza}, {Kordopatis}, {Lanzafame}, {Lebreton}, {Licata}, {Lindstr{\o}m}, {Livanou}, {Magdaleno Romeo}, {Manteiga}, {Marocco}, {Marshall}, {Mary}, {Nicolas}, {Pallas-Quintela}, {Panem}, {Pichon}, {Poggio}, {Riclet}, {Robin}, {Santove{\~n}a}, {Silvelo}, {Slezak}, {Smart}, {Soubiran}, {S{\"u}veges}, {Ulla},
  {Utrilla}, {Vallenari}, {Zhao}, {Zorec}, {Barrado}, {Bijaoui}, {Bouret}, {Blomme}, {Brott}, {Cassisi}, {Kochukhov}, {Martayan}, {Shulyak}, \& {Silvester}}]{GDR3_Creevey_ApsisI_2022arXiv220605864C}
{Creevey}, O.~L., {Sordo}, R., {Pailler}, F., {et~al.} 2023, \aap, 674, A26

\bibitem[{{Cukanovaite} {et~al.}(2023){Cukanovaite}, {Tremblay}, {Toonen}, {Temmink}, {Manser}, {O'Brien}, \& {McCleery}}]{Cukanovaite_2023MNRAS.522.1643C}
{Cukanovaite}, E., {Tremblay}, P.~E., {Toonen}, S., {et~al.} 2023, \mnras, 522, 1643

\bibitem[{{De Silva} {et~al.}(2015){De Silva}, {Freeman}, {Bland-Hawthorn}, {Martell}, {de Boer}, {Asplund}, {Keller}, {Sharma}, {Zucker}, {Zwitter}, {Anguiano}, {Bacigalupo}, {Bayliss}, {Beavis}, {Bergemann}, {Campbell}, {Cannon}, {Carollo}, {Casagrande}, {Casey}, {Da Costa}, {D'Orazi}, {Dotter}, {Duong}, {Heger}, {Ireland}, {Kafle}, {Kos}, {Lattanzio}, {Lewis}, {Lin}, {Lind}, {Munari}, {Nataf}, {O'Toole}, {Parker}, {Reid}, {Schlesinger}, {Sheinis}, {Simpson}, {Stello}, {Ting}, {Traven}, {Watson}, {Wittenmyer}, {Yong}, \& {{\v{Z}}erjal}}]{GALAH_2015MNRAS.449.2604D}
{De Silva}, G.~M., {Freeman}, K.~C., {Bland-Hawthorn}, J., {et~al.} 2015, \mnras, 449, 2604

\bibitem[{{Dufour} {et~al.}(2017){Dufour}, {Blouin}, {Coutu}, {Fortin-Archambault}, {Thibeault}, {Bergeron}, \& {Fontaine}}]{MWDD_2017ASPC..509....3D}
{Dufour}, P., {Blouin}, S., {Coutu}, S., {et~al.} 2017, in Astronomical Society of the Pacific Conference Series, Vol. 509, 20th European White Dwarf Workshop, ed. P.~E. {Tremblay}, B.~{Gaensicke}, \& T.~{Marsh}, 3

\bibitem[{{El-Badry} {et~al.}(2021){El-Badry}, {Rix}, \& {Heintz}}]{El_Badry_inflation_2021MNRAS.506.2269E}
{El-Badry}, K., {Rix}, H.-W., \& {Heintz}, T.~M. 2021, \mnras, 506, 2269

\bibitem[{{Evans} {et~al.}(2018){Evans}, {Riello}, {De Angeli}, {Carrasco}, {Montegriffo}, {Fabricius}, {Jordi}, {Palaversa}, {Diener}, {Busso}, {Cacciari}, {van Leeuwen}, {Burgess}, {Davidson}, {Harrison}, {Hodgkin}, {Pancino}, {Richards}, {Altavilla}, {Balaguer-N{\'u}{\~n}ez}, {Barstow}, {Bellazzini}, {Brown}, {Castellani}, {Cocozza}, {De Luise}, {Delgado}, {Ducourant}, {Galleti}, {Gilmore}, {Giuffrida}, {Holl}, {Kewley}, {Koposov}, {Marinoni}, {Marrese}, {Osborne}, {Piersimoni}, {Portell}, {Pulone}, {Ragaini}, {Sanna}, {Terrett}, {Walton}, {Wevers}, \& {Wyrzykowski}}]{Evans_gaia_dr2_photom_2018A&A...616A...4E}
{Evans}, D.~W., {Riello}, M., {De Angeli}, F., {et~al.} 2018, \aap, 616, A4

\bibitem[{{Fabricius} {et~al.}(2022){Fabricius}, {Babusiaux}, {Luri}, {Khanna}, {Muraveva}, {Reyl{\'e}}, {Spoto}, {Vallenari}, {Antoja}, {Arenou}, {Balbinot}, {Barache}, {Bauchet}, {Bossini}, {Busonero}, {Cantat-Gaudin}, {Carrasco}, {Diakit{\'e}}, {Figueras}, {Garofalo}, {Garcia-Gutierrez}, {Helmi}, {Jordi}, {Leclerc}, {Licata}, {Masip Vela}, {Mongui{\'o}}, {Ramos}, {Robin}, {Romero-G{\'o}mez}, {Rybizki}, {S{\'a}ez-N{\'u}{\~n}ez}, {Spina}, {Turon}, \& {Weiler}}]{GDR3_documentation_ch14_2022gdr3.reptE..14F}
{Fabricius}, C., {Babusiaux}, C., {Luri}, X., {et~al.} 2022, {Gaia DR3 documentation Chapter 14: Validation}, Gaia DR3 documentation, European Space Agency

\bibitem[{{Fabricius} {et~al.}(2021){Fabricius}, {Luri}, {Arenou}, {Babusiaux}, {Helmi}, {Muraveva}, {Reyl{\'e}}, {Spoto}, {Vallenari}, {Antoja}, {Balbinot}, {Barache}, {Bauchet}, {Bragaglia}, {Busonero}, {Cantat-Gaudin}, {Carrasco}, {Diakit{\'e}}, {Fabrizio}, {Figueras}, {Garcia-Gutierrez}, {Garofalo}, {Jordi}, {Kervella}, {Khanna}, {Leclerc}, {Licata}, {Lambert}, {Marrese}, {Masip}, {Ramos}, {Robichon}, {Robin}, {Romero-G{\'o}mez}, {Rubele}, \& {Weiler}}]{gaiaedr3_validation}
{Fabricius}, C., {Luri}, X., {Arenou}, F., {et~al.} 2021, \aap, 649, A5

\bibitem[{{Ferrario}(2012)}]{Ferrario_2012MNRAS.426.2500F}
{Ferrario}, L. 2012, \mnras, 426, 2500

\bibitem[{{Fleury} {et~al.}(2022){Fleury}, {Caiazzo}, \& {Heyl}}]{Fleury_2022MNRAS.511.5984F}
{Fleury}, L., {Caiazzo}, I., \& {Heyl}, J. 2022, \mnras, 511, 5984

\bibitem[{{Fuhrmann} {et~al.}(2017){Fuhrmann}, {Chini}, {Kaderhandt}, \& {Chen}}]{Fuhrmann_2017ApJ...836..139F}
{Fuhrmann}, K., {Chini}, R., {Kaderhandt}, L., \& {Chen}, Z. 2017, \apj, 836, 139

\bibitem[{{Gaia Collaboration} {et~al.}(2021{\natexlab{a}}){Gaia Collaboration}, {Brown}, {Vallenari}, {Prusti}, {de Bruijne}, {Babusiaux}, {Biermann}, {Creevey}, {Evans}, {Eyer}, {Hutton}, {Jansen}, {Jordi}, {Klioner}, {Lammers}, {Lindegren}, {Luri}, {Mignard}, {Panem}, {Pourbaix}, {Randich}, {Sartoretti}, {Soubiran}, {Walton}, {Arenou}, {Bailer-Jones}, {Bastian}, {Cropper}, {Drimmel}, {Katz}, {Lattanzi}, {van Leeuwen}, {Bakker}, {Cacciari}, {Casta{\~n}eda}, {De Angeli}, {Ducourant}, {Fabricius}, {Fouesneau}, {Fr{\'e}mat}, {Guerra}, {Guerrier}, {Guiraud}, {Jean-Antoine Piccolo}, {Masana}, {Messineo}, {Mowlavi}, {Nicolas}, {Nienartowicz}, {Pailler}, {Panuzzo}, {Riclet}, {Roux}, {Seabroke}, {Sordo}, {Tanga}, {Th{\'e}venin}, {Gracia-Abril}, {Portell}, {Teyssier}, {Altmann}, {Andrae}, {Bellas-Velidis}, {Benson}, {Berthier}, {Blomme}, {Brugaletta}, {Burgess}, {Busso}, {Carry}, {Cellino}, {Cheek}, {Clementini}, {Damerdji}, {Davidson}, {Delchambre}, {Dell'Oro}, {Fern{\'a}ndez-Hern{\'a}ndez}, {Galluccio},
  {Garc{\'\i}a-Lario}, {Garcia-Reinaldos}, {Gonz{\'a}lez-N{\'u}{\~n}ez}, {Gosset}, {Haigron}, {Halbwachs}, {Hambly}, {Harrison}, {Hatzidimitriou}, {Heiter}, {Hern{\'a}ndez}, {Hestroffer}, {Hodgkin}, {Holl}, {Jan{\ss}en}, {Jevardat de Fombelle}, {Jordan}, {Krone-Martins}, {Lanzafame}, {L{\"o}ffler}, {Lorca}, {Manteiga}, {Marchal}, {Marrese}, {Moitinho}, {Mora}, {Muinonen}, {Osborne}, {Pancino}, {Pauwels}, {Petit}, {Recio-Blanco}, {Richards}, {Riello}, {Rimoldini}, {Robin}, {Roegiers}, {Rybizki}, {Sarro}, {Siopis}, {Smith}, {Sozzetti}, {Ulla}, {Utrilla}, {van Leeuwen}, {van Reeven}, {Abbas}, {Abreu Aramburu}, {Accart}, {Aerts}, {Aguado}, {Ajaj}, {Altavilla}, {{\'A}lvarez}, {{\'A}lvarez Cid-Fuentes}, {Alves}, {Anderson}, {Anglada Varela}, {Antoja}, {Audard}, {Baines}, {Baker}, {Balaguer-N{\'u}{\~n}ez}, {Balbinot}, {Balog}, {Barache}, {Barbato}, {Barros}, {Barstow}, {Bartolom{\'e}}, {Bassilana}, {Bauchet}, {Baudesson-Stella}, {Becciani}, {Bellazzini}, {Bernet}, {Bertone}, {Bianchi}, {Blanco-Cuaresma}, {Boch},
  {Bombrun}, {Bossini}, {Bouquillon}, {Bragaglia}, {Bramante}, {Breedt}, {Bressan}, {Brouillet}, {Bucciarelli}, {Burlacu}, {Busonero}, {Butkevich}, {Buzzi}, {Caffau}, {Cancelliere}, {C{\'a}novas}, {Cantat-Gaudin}, {Carballo}, {Carlucci}, {Carnerero}, {Carrasco}, {Casamiquela}, {Castellani}, {Castro-Ginard}, {Castro Sampol}, {Chaoul}, {Charlot}, {Chemin}, {Chiavassa}, {Cioni}, {Comoretto}, {Cooper}, {Cornez}, {Cowell}, {Crifo}, {Crosta}, {Crowley}, {Dafonte}, {Dapergolas}, {David}, {David}, {de Laverny}, {De Luise}, {De March}, {De Ridder}, {de Souza}, {de Teodoro}, {de Torres}, {del Peloso}, {del Pozo}, {Delbo}, {Delgado}, {Delgado}, {Delisle}, {Di Matteo}, {Diakite}, {Diener}, {Distefano}, {Dolding}, {Eappachen}, {Edvardsson}, {Enke}, {Esquej}, {Fabre}, {Fabrizio}, {Faigler}, {Fedorets}, {Fernique}, {Fienga}, {Figueras}, {Fouron}, {Fragkoudi}, {Fraile}, {Franke}, {Gai}, {Garabato}, {Garcia-Gutierrez}, {Garc{\'\i}a-Torres}, {Garofalo}, {Gavras}, {Gerlach}, {Geyer}, {Giacobbe}, {Gilmore}, {Girona},
  {Giuffrida}, {Gomel}, {Gomez}, {Gonzalez-Santamaria}, {Gonz{\'a}lez-Vidal}, {Granvik}, {Guti{\'e}rrez-S{\'a}nchez}, {Guy}, {Hauser}, {Haywood}, {Helmi}, {Hidalgo}, {Hilger}, {H{\l}adczuk}, {Hobbs}, {Holland}, {Huckle}, {Jasniewicz}, {Jonker}, {Juaristi Campillo}, {Julbe}, {Karbevska}, {Kervella}, {Khanna}, {Kochoska}, {Kontizas}, {Kordopatis}, {Korn}, {Kostrzewa-Rutkowska}, {Kruszy{\'n}ska}, {Lambert}, {Lanza}, {Lasne}, {Le Campion}, {Le Fustec}, {Lebreton}, {Lebzelter}, {Leccia}, {Leclerc}, {Lecoeur-Taibi}, {Liao}, {Licata}, {Lindstr{\o}m}, {Lister}, {Livanou}, {Lobel}, {Madrero Pardo}, {Managau}, {Mann}, {Marchant}, {Marconi}, {Marcos Santos}, {Marinoni}, {Marocco}, {Marshall}, {Martin Polo}, {Mart{\'\i}n-Fleitas}, {Masip}, {Massari}, {Mastrobuono-Battisti}, {Mazeh}, {McMillan}, {Messina}, {Michalik}, {Millar}, {Mints}, {Molina}, {Molinaro}, {Moln{\'a}r}, {Montegriffo}, {Mor}, {Morbidelli}, {Morel}, {Morris}, {Mulone}, {Munoz}, {Muraveva}, {Murphy}, {Musella}, {Noval}, {Ord{\'e}novic}, {Orr{\`u}},
  {Osinde}, {Pagani}, {Pagano}, {Palaversa}, {Palicio}, {Panahi}, {Pawlak}, {Pe{\~n}alosa Esteller}, {Penttil{\"a}}, {Piersimoni}, {Pineau}, {Plachy}, {Plum}, {Poggio}, {Poretti}, {Poujoulet}, {Pr{\v{s}}a}, {Pulone}, {Racero}, {Ragaini}, {Rainer}, {Raiteri}, {Rambaux}, {Ramos}, {Ramos-Lerate}, {Re Fiorentin}, {Regibo}, {Reyl{\'e}}, {Ripepi}, {Riva}, {Rixon}, {Robichon}, {Robin}, {Roelens}, {Rohrbasser}, {Romero-G{\'o}mez}, {Rowell}, {Royer}, {Rybicki}, {Sadowski}, {Sagrist{\`a} Sell{\'e}s}, {Sahlmann}, {Salgado}, {Salguero}, {Samaras}, {Sanchez Gimenez}, {Sanna}, {Santove{\~n}a}, {Sarasso}, {Schultheis}, {Sciacca}, {Segol}, {Segovia}, {S{\'e}gransan}, {Semeux}, {Shahaf}, {Siddiqui}, {Siebert}, {Siltala}, {Slezak}, {Smart}, {Solano}, {Solitro}, {Souami}, {Souchay}, {Spagna}, {Spoto}, {Steele}, {Steidelm{\"u}ller}, {Stephenson}, {S{\"u}veges}, {Szabados}, {Szegedi-Elek}, {Taris}, {Tauran}, {Taylor}, {Teixeira}, {Thuillot}, {Tonello}, {Torra}, {Torra}, {Turon}, {Unger}, {Vaillant}, {van Dillen}, {Vanel},
  {Vecchiato}, {Viala}, {Vicente}, {Voutsinas}, {Weiler}, {Wevers}, {Wyrzykowski}, {Yoldas}, {Yvard}, {Zhao}, {Zorec}, {Zucker}, {Zurbach}, \& {Zwitter}}]{gaiaedr3_summary}
{Gaia Collaboration}, {Brown}, A.~G.~A., {Vallenari}, A., {et~al.} 2021{\natexlab{a}}, \aap, 649, A1

\bibitem[{{Gaia Collaboration} {et~al.}(2023){Gaia Collaboration}, {Montegriffo}, {Bellazzini}, {De Angeli}, {Andrae}, {Barstow}, {Bossini}, {Bragaglia}, {Burgess}, {Cacciari}, {Carrasco}, {Chornay}, {Delchambre}, {Evans}, {Fouesneau}, {Fr{\'e}mat}, {Garabato}, {Jordi}, {Manteiga}, {Massari}, {Palaversa}, {Pancino}, {Riello}, {Ruz Mieres}, {Sanna}, {Santove{\~n}a}, {Sordo}, {Vallenari}, {Walton}, {Brown}, {Prusti}, {de Bruijne}, {Arenou}, {Babusiaux}, {Biermann}, {Creevey}, {Ducourant}, {Eyer}, {Guerra}, {Hutton}, {Klioner}, {Lammers}, {Lindegren}, {Luri}, {Mignard}, {Panem}, {Pourbaix}, {Randich}, {Sartoretti}, {Soubiran}, {Tanga}, {Bailer-Jones}, {Bastian}, {Drimmel}, {Jansen}, {Katz}, {Lattanzi}, {van Leeuwen}, {Bakker}, {Casta{\~n}eda}, {Fabricius}, {Galluccio}, {Guerrier}, {Heiter}, {Masana}, {Messineo}, {Mowlavi}, {Nicolas}, {Nienartowicz}, {Pailler}, {Panuzzo}, {Riclet}, {Roux}, {Seabroke}, {Th{\'e}venin}, {Gracia-Abril}, {Portell}, {Teyssier}, {Altmann}, {Audard}, {Bellas-Velidis}, {Benson},
  {Berthier}, {Blomme}, {Busonero}, {Busso}, {C{\'a}novas}, {Carry}, {Cellino}, {Cheek}, {Clementini}, {Damerdji}, {Davidson}, {de Teodoro}, {Nu{\~n}ez Campos}, {Dell'Oro}, {Esquej}, {Fern{\'a}ndez-Hern{\'a}ndez}, {Fraile}, {Garc{\'\i}a-Lario}, {Gosset}, {Haigron}, {Halbwachs}, {Hambly}, {Harrison}, {Hern{\'a}ndez}, {Hestroffer}, {Hodgkin}, {Holl}, {Jan{\ss}en}, {Jevardat de Fombelle}, {Jordan}, {Krone-Martins}, {Lanzafame}, {L{\"o}ffler}, {Marchal}, {Marrese}, {Moitinho}, {Muinonen}, {Osborne}, {Pauwels}, {Recio-Blanco}, {Reyl{\'e}}, {Rimoldini}, {Roegiers}, {Rybizki}, {Sarro}, {Siopis}, {Smith}, {Sozzetti}, {Utrilla}, {van Leeuwen}, {Abbas}, {{\'A}brah{\'a}m}, {Abreu Aramburu}, {Aerts}, {Aguado}, {Ajaj}, {Aldea-Montero}, {Altavilla}, {{\'A}lvarez}, {Alves}, {Anderson}, {Anglada Varela}, {Antoja}, {Baines}, {Baker}, {Balaguer-N{\'u}{\~n}ez}, {Balbinot}, {Balog}, {Barache}, {Barbato}, {Barros}, {Bartolom{\'e}}, {Bassilana}, {Bauchet}, {Becciani}, {Berihuete}, {Bernet}, {Bertone}, {Bianchi}, {Binnenfeld},
  {Blanco-Cuaresma}, {Boch}, {Bombrun}, {Bouquillon}, {Bramante}, {Breedt}, {Bressan}, {Brouillet}, {Brugaletta}, {Bucciarelli}, {Burlacu}, {Butkevich}, {Buzzi}, {Caffau}, {Cancelliere}, {Cantat-Gaudin}, {Carballo}, {Carlucci}, {Carnerero}, {Casamiquela}, {Castellani}, {Castro-Ginard}, {Chaoul}, {Charlot}, {Chemin}, {Chiaramida}, {Chiavassa}, {Comoretto}, {Contursi}, {Cooper}, {Cornez}, {Cowell}, {Crifo}, {Cropper}, {Crosta}, {Crowley}, {Dafonte}, {Dapergolas}, {David}, {de Laverny}, {De Luise}, {De March}, {De Ridder}, {de Souza}, {de Torres}, {del Peloso}, {del Pozo}, {Delbo}, {Delgado}, {Delisle}, {Demouchy}, {Dharmawardena}, {Diakite}, {Diener}, {Distefano}, {Dolding}, {Enke}, {Fabre}, {Fabrizio}, {Faigler}, {Fedorets}, {Fernique}, {Figueras}, {Fournier}, {Fouron}, {Fragkoudi}, {Gai}, {Garcia-Gutierrez}, {Garcia-Reinaldos}, {Garc{\'\i}a-Torres}, {Garofalo}, {Gavel}, {Gavras}, {Gerlach}, {Geyer}, {Giacobbe}, {Gilmore}, {Girona}, {Giuffrida}, {Gomel}, {Gomez}, {Gonz{\'a}lez-N{\'u}{\~n}ez},
  {Gonz{\'a}lez-Santamar{\'\i}a}, {Gonz{\'a}lez-Vidal}, {Granvik}, {Guillout}, {Guiraud}, {Guti{\'e}rrez-S{\'a}nchez}, {Guy}, {Hatzidimitriou}, {Hauser}, {Haywood}, {Helmer}, {Helmi}, {Sarmiento}, {Hidalgo}, {H{\l}adczuk}, {Hobbs}, {Holland}, {Huckle}, {Jardine}, {Jasniewicz}, {Jean-Antoine Piccolo}, {Jim{\'e}nez-Arranz}, {Juaristi Campillo}, {Julbe}, {Karbevska}, {Kervella}, {Khanna}, {Kordopatis}, {Korn}, {K{\'o}sp{\'a}l}, {Kostrzewa-Rutkowska}, {Kruszy{\'n}ska}, {Kun}, {Laizeau}, {Lambert}, {Lanza}, {Lasne}, {Le Campion}, {Lebreton}, {Lebzelter}, {Leccia}, {Leclerc}, {Lecoeur-Taibi}, {Liao}, {Licata}, {Lindstr{\'o}m}, {Lister}, {Livanou}, {Lobel}, {Lorca}, {Loup}, {Madrero Pardo}, {Magdaleno Romeo}, {Managau}, {Mann}, {Marchant}, {Marconi}, {Marcos}, {Marcos Santos}, {Mar{\'\i}n Pina}, {Marinoni}, {Marocco}, {Marshall}, {Martin Polo}, {Mart{\'\i}n-Fleitas}, {Marton}, {Mary}, {Masip}, {Mastrobuono-Battisti}, {Mazeh}, {McMillan}, {Messina}, {Michalik}, {Millar}, {Mints}, {Molina}, {Molinaro}, {Moln{\'a}r},
  {Monari}, {Mongui{\'o}}, {Montero}, {Mor}, {Mora}, {Morbidelli}, {Morel}, {Morris}, {Muraveva}, {Murphy}, {Musella}, {Nagy}, {Noval}, {Oca{\~n}a}, {Ogden}, {Ordenovic}, {Osinde}, {Pagani}, {Pagano}, {Palicio}, {Pallas-Quintela}, {Panahi}, {Payne-Wardenaar}, {Pe{\~n}alosa Esteller}, {Penttil{\"a}}, {Pichon}, {Piersimoni}, {Pineau}, {Plachy}, {Plum}, {Poggio}, {Pr{\v{s}}a}, {Pulone}, {Racero}, {Ragaini}, {Rainer}, {Raiteri}, {Ramos}, {Ramos-Lerate}, {Re Fiorentin}, {Regibo}, {Richards}, {Rios Diaz}, {Ripepi}, {Riva}, {Rix}, {Rixon}, {Robichon}, {Robin}, {Robin}, {Roelens}, {Rogues}, {Rohrbasser}, {Romero-G{\'o}mez}, {Rowell}, {Royer}, {Rybicki}, {Sadowski}, {S{\'a}ez N{\'u}{\~n}ez}, {Sagrist{\`a} Sell{\'e}s}, {Sahlmann}, {Salguero}, {Samaras}, {Sanchez Gimenez}, {Sarasso}, {Schultheis}, {Sciacca}, {Segol}, {Segovia}, {S{\'e}gransan}, {Semeux}, {Shahaf}, {Siddiqui}, {Siebert}, {Siltala}, {Silvelo}, {Slezak}, {Slezak}, {Smart}, {Snaith}, {Solano}, {Solitro}, {Souami}, {Souchay}, {Spagna}, {Spina}, {Spoto},
  {Steele}, {Steidelm{\"u}ller}, {Stephenson}, {S{\"u}veges}, {Surdej}, {Szabados}, {Szegedi-Elek}, {Taris}, {Taylor}, {Teixeira}, {Tolomei}, {Tonello}, {Torra}, {Torra}, {Torralba Elipe}, {Trabucchi}, {Tsounis}, {Turon}, {Ulla}, {Unger}, {Vaillant}, {van Dillen}, {van Reeven}, {Vanel}, {Vecchiato}, {Viala}, {Vicente}, {Voutsinas}, {Wevers}, {Wyrzykowski}, {Yoldas}, {Yvard}, {Zhao}, {Zorec}, {Zucker}, \& {Zwitter}}]{GDR3_Montegriffo_XP_synth_phot_2022arXiv220606215G}
{Gaia Collaboration}, {Montegriffo}, P., {Bellazzini}, M., {et~al.} 2023, \aap, 674, A33

\bibitem[{{Gaia Collaboration} {et~al.}(2021{\natexlab{b}}){Gaia Collaboration}, {Smart}, {Sarro}, {Rybizki}, {Reyl{\'e}}, {Robin}, {Hambly}, {Abbas}, {Barstow}, {de Bruijne}, {Bucciarelli}, {Carrasco}, {Cooper}, {Hodgkin}, {Masana}, {Michalik}, {Sahlmann}, {Sozzetti}, {Brown}, {Vallenari}, {Prusti}, {Babusiaux}, {Biermann}, {Creevey}, {Evans}, {Eyer}, {Hutton}, {Jansen}, {Jordi}, {Klioner}, {Lammers}, {Lindegren}, {Luri}, {Mignard}, {Panem}, {Pourbaix}, {Randich}, {Sartoretti}, {Soubiran}, {Walton}, {Arenou}, {Bailer-Jones}, {Bastian}, {Cropper}, {Drimmel}, {Katz}, {Lattanzi}, {van Leeuwen}, {Bakker}, {Casta{\~n}eda}, {De Angeli}, {Ducourant}, {Fabricius}, {Fouesneau}, {Fr{\'e}mat}, {Guerra}, {Guerrier}, {Guiraud}, {Jean-Antoine Piccolo}, {Messineo}, {Mowlavi}, {Nicolas}, {Nienartowicz}, {Pailler}, {Panuzzo}, {Riclet}, {Roux}, {Seabroke}, {Sordo}, {Tanga}, {Th{\'e}venin}, {Gracia-Abril}, {Portell}, {Teyssier}, {Altmann}, {Andrae}, {Bellas-Velidis}, {Benson}, {Berthier}, {Blomme}, {Brugaletta}, {Burgess},
  {Busso}, {Carry}, {Cellino}, {Cheek}, {Clementini}, {Damerdji}, {Davidson}, {Delchambre}, {Dell'Oro}, {Fern{\'a}ndez-Hern{\'a}ndez}, {Galluccio}, {Garc{\'\i}a-Lario}, {Garcia-Reinaldos}, {Gonz{\'a}lez-N{\'u}{\~n}ez}, {Gosset}, {Haigron}, {Halbwachs}, {Harrison}, {Hatzidimitriou}, {Heiter}, {Hern{\'a}ndez}, {Hestroffer}, {Holl}, {Jan{\ss}en}, {Jevardat de Fombelle}, {Jordan}, {Krone-Martins}, {Lanzafame}, {L{\"o}ffler}, {Lorca}, {Manteiga}, {Marchal}, {Marrese}, {Moitinho}, {Mora}, {Muinonen}, {Osborne}, {Pancino}, {Pauwels}, {Recio-Blanco}, {Richards}, {Riello}, {Rimoldini}, {Roegiers}, {Siopis}, {Smith}, {Ulla}, {Utrilla}, {van Leeuwen}, {van Reeven}, {Abreu Aramburu}, {Accart}, {Aerts}, {Aguado}, {Ajaj}, {Altavilla}, {{\'A}lvarez}, {{\'A}lvarez Cid-Fuentes}, {Alves}, {Anderson}, {Anglada Varela}, {Antoja}, {Audard}, {Baines}, {Baker}, {Balaguer-N{\'u}{\~n}ez}, {Balbinot}, {Balog}, {Barache}, {Barbato}, {Barros}, {Bartolom{\'e}}, {Bassilana}, {Bauchet}, {Baudesson-Stella}, {Becciani}, {Bellazzini},
  {Bernet}, {Bertone}, {Bianchi}, {Blanco-Cuaresma}, {Boch}, {Bombrun}, {Bossini}, {Bouquillon}, {Bragaglia}, {Bramante}, {Breedt}, {Bressan}, {Brouillet}, {Burlacu}, {Busonero}, {Butkevich}, {Buzzi}, {Caffau}, {Cancelliere}, {C{\'a}novas}, {Cantat-Gaudin}, {Carballo}, {Carlucci}, {Carnerero}, {Casamiquela}, {Castellani}, {Castro-Ginard}, {Castro Sampol}, {Chaoul}, {Charlot}, {Chemin}, {Chiavassa}, {Cioni}, {Comoretto}, {Cornez}, {Cowell}, {Crifo}, {Crosta}, {Crowley}, {Dafonte}, {Dapergolas}, {David}, {David}, {de Laverny}, {De Luise}, {De March}, {De Ridder}, {de Souza}, {de Teodoro}, {de Torres}, {del Peloso}, {del Pozo}, {Delgado}, {Delgado}, {Delisle}, {Di Matteo}, {Diakite}, {Diener}, {Distefano}, {Dolding}, {Eappachen}, {Edvardsson}, {Enke}, {Esquej}, {Fabre}, {Fabrizio}, {Faigler}, {Fedorets}, {Fernique}, {Fienga}, {Figueras}, {Fouron}, {Fragkoudi}, {Fraile}, {Franke}, {Gai}, {Garabato}, {Garcia-Gutierrez}, {Garc{\'\i}a-Torres}, {Garofalo}, {Gavras}, {Gerlach}, {Geyer}, {Giacobbe}, {Gilmore},
  {Girona}, {Giuffrida}, {Gomel}, {Gomez}, {Gonzalez-Santamaria}, {Gonz{\'a}lez-Vidal}, {Granvik}, {Guti{\'e}rrez-S{\'a}nchez}, {Guy}, {Hauser}, {Haywood}, {Helmi}, {Hidalgo}, {Hilger}, {H{\l}adczuk}, {Hobbs}, {Holland}, {Huckle}, {Jasniewicz}, {Jonker}, {Juaristi Campillo}, {Julbe}, {Karbevska}, {Kervella}, {Khanna}, {Kochoska}, {Kontizas}, {Kordopatis}, {Korn}, {Kostrzewa-Rutkowska}, {Kruszy{\'n}ska}, {Lambert}, {Lanza}, {Lasne}, {Le Campion}, {Le Fustec}, {Lebreton}, {Lebzelter}, {Leccia}, {Leclerc}, {Lecoeur-Taibi}, {Liao}, {Licata}, {Lindstr{\o}m}, {Lister}, {Livanou}, {Lobel}, {Madrero Pardo}, {Managau}, {Mann}, {Marchant}, {Marconi}, {Marcos Santos}, {Marinoni}, {Marocco}, {Marshall}, {Martin Polo}, {Mart{\'\i}n-Fleitas}, {Masip}, {Massari}, {Mastrobuono-Battisti}, {Mazeh}, {McMillan}, {Messina}, {Millar}, {Mints}, {Molina}, {Molinaro}, {Moln{\'a}r}, {Montegriffo}, {Mor}, {Morbidelli}, {Morel}, {Morris}, {Mulone}, {Munoz}, {Muraveva}, {Murphy}, {Musella}, {Noval}, {Ord{\'e}novic}, {Orr{\`u}}, {Osinde},
  {Pagani}, {Pagano}, {Palaversa}, {Palicio}, {Panahi}, {Pawlak}, {Pe{\~n}alosa Esteller}, {Penttil{\"a}}, {Piersimoni}, {Pineau}, {Plachy}, {Plum}, {Poggio}, {Poretti}, {Poujoulet}, {Pr{\v{s}}a}, {Pulone}, {Racero}, {Ragaini}, {Rainer}, {Raiteri}, {Rambaux}, {Ramos}, {Ramos-Lerate}, {Re Fiorentin}, {Regibo}, {Ripepi}, {Riva}, {Rixon}, {Robichon}, {Robin}, {Roelens}, {Rohrbasser}, {Romero-G{\'o}mez}, {Rowell}, {Royer}, {Rybicki}, {Sadowski}, {Sagrist{\`a} Sell{\'e}s}, {Salgado}, {Salguero}, {Samaras}, {Sanchez Gimenez}, {Sanna}, {Santove{\~n}a}, {Sarasso}, {Schultheis}, {Sciacca}, {Segol}, {Segovia}, {S{\'e}gransan}, {Semeux}, {Shahaf}, {Siddiqui}, {Siebert}, {Siltala}, {Slezak}, {Solano}, {Solitro}, {Souami}, {Souchay}, {Spagna}, {Spoto}, {Steele}, {Steidelm{\"u}ller}, {Stephenson}, {S{\"u}veges}, {Szabados}, {Szegedi-Elek}, {Taris}, {Tauran}, {Taylor}, {Teixeira}, {Thuillot}, {Tonello}, {Torra}, {Torra}, {Turon}, {Unger}, {Vaillant}, {van Dillen}, {Vanel}, {Vecchiato}, {Viala}, {Vicente}, {Voutsinas},
  {Weiler}, {Wevers}, {Wyrzykowski}, {Yoldas}, {Yvard}, {Zhao}, {Zorec}, {Zucker}, {Zurbach}, \& {Zwitter}}]{gaiaedr3_gcns}
{Gaia Collaboration}, {Smart}, R.~L., {Sarro}, L.~M., {et~al.} 2021{\natexlab{b}}, \aap, 649, A6

\bibitem[{{Gentile Fusillo} {et~al.}(2021){Gentile Fusillo}, {Tremblay}, {Cukanovaite}, {Vorontseva}, {Lallement}, {Hollands}, {G{\"a}nsicke}, {Burdge}, {McCleery}, \& {Jordan}}]{gentilefusillo21_2021MNRAS.508.3877G}
{Gentile Fusillo}, N.~P., {Tremblay}, P.~E., {Cukanovaite}, E., {et~al.} 2021, \mnras, 508, 3877

\bibitem[{{Gentile Fusillo} {et~al.}(2019){Gentile Fusillo}, {Tremblay}, {G{\"a}nsicke}, {Manser}, {Cunningham}, {Cukanovaite}, {Hollands}, {Marsh}, {Raddi}, {Jordan}, {Toonen}, {Geier}, {Barstow}, \& {Cummings}}]{gentilefusillo19_2019MNRAS.482.4570G}
{Gentile Fusillo}, N.~P., {Tremblay}, P.-E., {G{\"a}nsicke}, B.~T., {et~al.} 2019, \mnras, 482, 4570

\bibitem[{{Gliese} \& {Jahrei{\ss}}(1991)}]{cns3}
{Gliese}, W. \& {Jahrei{\ss}}, H. 1991, {Preliminary Version of the Third Catalogue of Nearby Stars}, On: The Astronomical Data Center CD-ROM: Selected Astronomical Catalogs

\bibitem[{{Golovin} {et~al.}(2023){Golovin}, {Reffert}, {Just}, {Jordan}, {Vani}, \& {Jahrei{\ss}}}]{golovin23}
{Golovin}, A., {Reffert}, S., {Just}, A., {et~al.} 2023, \aap, 670, A19

\bibitem[{{Halbwachs} {et~al.}(2023){Halbwachs}, {Pourbaix}, {Arenou}, {Galluccio}, {Guillout}, {Bauchet}, {Marchal}, {Sadowski}, \& {Teyssier}}]{GDR3_Halbwachs_2022arXiv220605726H}
{Halbwachs}, J.-L., {Pourbaix}, D., {Arenou}, F., {et~al.} 2023, \aap, 674, A9

\bibitem[{{Hardy} {et~al.}(2023){Hardy}, {Dufour}, \& {Jordan}}]{Hardy_2023MNRAS.520.6111H}
{Hardy}, F., {Dufour}, P., \& {Jordan}, S. 2023, \mnras, 520, 6111

\bibitem[{{Harris} {et~al.}(2006){Harris}, {Munn}, {Kilic}, {Liebert}, {Williams}, {von Hippel}, {Levine}, {Monet}, {Eisenstein}, {Kleinman}, {Metcalfe}, {Nitta}, {Winget}, {Brinkmann}, {Fukugita}, {Knapp}, {Lupton}, {Smith}, \& {Schneider}}]{Harris_2006AJ....131..571H}
{Harris}, H.~C., {Munn}, J.~A., {Kilic}, M., {et~al.} 2006, \aj, 131, 571

\bibitem[{{Holl} {et~al.}(2023{\natexlab{a}}){Holl}, {Fabricius}, {Portell}, {Lindegren}, {Panuzzo}, {Bernet}, {Casta{\~n}eda}, {Jevardat de Fombelle}, {Audard}, {Ducourant}, {Harrison}, {Evans}, {Busso}, {Sozzetti}, {Gosset}, {Arenou}, {De Angeli}, {Riello}, {Eyer}, {Rimoldini}, {Gavras}, {Mowlavi}, {Nienartowicz}, {Lecoeur-Ta{\"\i}bi}, {Garc{\'\i}a-Lario}, \& {Pourbaix}}]{GDR3_Holl_spurious_2022arXiv221211971H}
{Holl}, B., {Fabricius}, C., {Portell}, J., {et~al.} 2023{\natexlab{a}}, \aap, 674, A25

\bibitem[{{Holl} {et~al.}(2023{\natexlab{b}}){Holl}, {Sozzetti}, {Sahlmann}, {Giacobbe}, {S{\'e}gransan}, {Unger}, {Delisle}, {Barbato}, {Lattanzi}, {Morbidelli}, \& {Sosnowska}}]{GDR3_Holl_2022arXiv220605439H}
{Holl}, B., {Sozzetti}, A., {Sahlmann}, J., {et~al.} 2023{\natexlab{b}}, \aap, 674, A10

\bibitem[{{Isern}(2019)}]{Isern_2019ApJ...878L..11I}
{Isern}, J. 2019, \apjl, 878, L11

\bibitem[{{Isern} {et~al.}(2008){Isern}, {Garc{\'\i}a-Berro}, {Torres}, \& {Catal{\'a}n}}]{Isern_2008ApJ...682L.109I}
{Isern}, J., {Garc{\'\i}a-Berro}, E., {Torres}, S., \& {Catal{\'a}n}, S. 2008, \apjl, 682, L109

\bibitem[{{Jim{\'e}nez-Esteban} {et~al.}(2023){Jim{\'e}nez-Esteban}, {Torres}, {Rebassa-Mansergas}, {Cruz}, {Murillo-Ojeda}, {Solano}, {Rodrigo}, \& {Camisassa}}]{JimenezEstebaan_2023MNRAS.518.5106J}
{Jim{\'e}nez-Esteban}, F.~M., {Torres}, S., {Rebassa-Mansergas}, A., {et~al.} 2023, \mnras, 518, 5106

\bibitem[{{Kaltenegger} \& {Faherty}(2021)}]{Kaltenegger_2021Natur.594..505K}
{Kaltenegger}, L. \& {Faherty}, J.~K. 2021, \nat, 594, 505

\bibitem[{{Kilic} {et~al.}(2007){Kilic}, {Stanek}, \& {Pinsonneault}}]{Kilic_2007ApJ...671..761K}
{Kilic}, M., {Stanek}, K.~Z., \& {Pinsonneault}, M.~H. 2007, \apj, 671, 761

\bibitem[{{Kosakowski} {et~al.}(2020){Kosakowski}, {Kilic}, {Brown}, \& {Gianninas}}]{Kosakowski_2020ApJ...894...53K}
{Kosakowski}, A., {Kilic}, M., {Brown}, W.~R., \& {Gianninas}, A. 2020, \apj, 894, 53

\bibitem[{{K{\"u}lebi} {et~al.}(2009){K{\"u}lebi}, {Jordan}, {Euchner}, {G{\"a}nsicke}, \& {Hirsch}}]{Kuelebi_2009A&A...506.1341K}
{K{\"u}lebi}, B., {Jordan}, S., {Euchner}, F., {G{\"a}nsicke}, B.~T., \& {Hirsch}, H. 2009, \aap, 506, 1341

\bibitem[{{Lam} {et~al.}(2022){Lam}, {Yuen}, {Green}, \& {Li}}]{Lam_WDPhotTools_2022RASTI...1...81L}
{Lam}, M.~C., {Yuen}, K.~W., {Green}, M.~J., \& {Li}, W. 2022, RAS Techniques and Instruments, 1, 81

\bibitem[{{Li} {et~al.}(2019){Li}, {Chen}, {Chen}, \& {Han}}]{Li_2019ApJ...871..148L}
{Li}, Z., {Chen}, X., {Chen}, H.-L., \& {Han}, Z. 2019, \apj, 871, 148

\bibitem[{{Liebert} {et~al.}(1979){Liebert}, {Dahn}, {Gresham}, \& {Strittmatter}}]{Liebert_1979ApJ...233..226L}
{Liebert}, J., {Dahn}, C.~C., {Gresham}, M., \& {Strittmatter}, P.~A. 1979, \apj, 233, 226

\bibitem[{{Liebert} {et~al.}(1988){Liebert}, {Dahn}, \& {Monet}}]{Liebert_1988ApJ...332..891L}
{Liebert}, J., {Dahn}, C.~C., \& {Monet}, D.~G. 1988, \apj, 332, 891

\bibitem[{{Lindegren}(2018)}]{lindegren18b}
{Lindegren}, L. 2018, {Gaia Data Processing and Analysis Consortium (DPAC) technical note GAIA-C3-TN-LU-LL-124-01, available at \url{https://www.cosmos.esa.int/web/gaia/public-dpac-documents}}

\bibitem[{{Lindegren} {et~al.}(2018){Lindegren}, {Hern{\'a}ndez}, {Bombrun}, {Klioner}, {Bastian}, {Ramos-Lerate}, {de Torres}, {Steidelm{\"u}ller}, {Stephenson}, {Hobbs}, {Lammers}, {Biermann}, {Geyer}, {Hilger}, {Michalik}, {Stampa}, {McMillan}, {Casta{\~n}eda}, {Clotet}, {Comoretto}, {Davidson}, {Fabricius}, {Gracia}, {Hambly}, {Hutton}, {Mora}, {Portell}, {van Leeuwen}, {Abbas}, {Abreu}, {Altmann}, {Andrei}, {Anglada}, {Balaguer-N{\'u}{\~n}ez}, {Barache}, {Becciani}, {Bertone}, {Bianchi}, {Bouquillon}, {Bourda}, {Br{\"u}semeister}, {Bucciarelli}, {Busonero}, {Buzzi}, {Cancelliere}, {Carlucci}, {Charlot}, {Cheek}, {Crosta}, {Crowley}, {de Bruijne}, {de Felice}, {Drimmel}, {Esquej}, {Fienga}, {Fraile}, {Gai}, {Garralda}, {Gonz{\'a}lez-Vidal}, {Guerra}, {Hauser}, {Hofmann}, {Holl}, {Jordan}, {Lattanzi}, {Lenhardt}, {Liao}, {Licata}, {Lister}, {L{\"o}ffler}, {Marchant}, {Martin-Fleitas}, {Messineo}, {Mignard}, {Morbidelli}, {Poggio}, {Riva}, {Rowell}, {Salguero}, {Sarasso}, {Sciacca}, {Siddiqui}, {Smart},
  {Spagna}, {Steele}, {Taris}, {Torra}, {van Elteren}, {van Reeven}, \& {Vecchiato}}]{lindegren18}
{Lindegren}, L., {Hern{\'a}ndez}, J., {Bombrun}, A., {et~al.} 2018, \aap, 616, A2

\bibitem[{{Lindegren} {et~al.}(2021){Lindegren}, {Klioner}, {Hern{\'a}ndez}, {Bombrun}, {Ramos-Lerate}, {Steidelm{\"u}ller}, {Bastian}, {Biermann}, {de Torres}, {Gerlach}, {Geyer}, {Hilger}, {Hobbs}, {Lammers}, {McMillan}, {Stephenson}, {Casta{\~n}eda}, {Davidson}, {Fabricius}, {Gracia-Abril}, {Portell}, {Rowell}, {Teyssier}, {Torra}, {Bartolom{\'e}}, {Clotet}, {Garralda}, {Gonz{\'a}lez-Vidal}, {Torra}, {Abbas}, {Altmann}, {Anglada Varela}, {Balaguer-N{\'u}{\~n}ez}, {Balog}, {Barache}, {Becciani}, {Bernet}, {Bertone}, {Bianchi}, {Bouquillon}, {Brown}, {Bucciarelli}, {Busonero}, {Butkevich}, {Buzzi}, {Cancelliere}, {Carlucci}, {Charlot}, {Cioni}, {Crosta}, {Crowley}, {del Peloso}, {del Pozo}, {Drimmel}, {Esquej}, {Fienga}, {Fraile}, {Gai}, {Garcia-Reinaldos}, {Guerra}, {Hambly}, {Hauser}, {Jan{\ss}en}, {Jordan}, {Kostrzewa-Rutkowska}, {Lattanzi}, {Liao}, {Licata}, {Lister}, {L{\"o}ffler}, {Marchant}, {Masip}, {Mignard}, {Mints}, {Molina}, {Mora}, {Morbidelli}, {Murphy}, {Pagani}, {Panuzzo}, {Pe{\~n}alosa
  Esteller}, {Poggio}, {Re Fiorentin}, {Riva}, {Sagrist{\`a} Sell{\'e}s}, {Sanchez Gimenez}, {Sarasso}, {Sciacca}, {Siddiqui}, {Smart}, {Souami}, {Spagna}, {Steele}, {Taris}, {Utrilla}, {van Reeven}, \& {Vecchiato}}]{gaiaedr3_astromertry}
{Lindegren}, L., {Klioner}, S.~A., {Hern{\'a}ndez}, J., {et~al.} 2021, \aap, 649, A2

\bibitem[{{Martin} {et~al.}(2005){Martin}, {Fanson}, {Schiminovich}, {Morrissey}, {Friedman}, {Barlow}, {Conrow}, {Grange}, {Jelinsky}, {Milliard}, {Siegmund}, {Bianchi}, {Byun}, {Donas}, {Forster}, {Heckman}, {Lee}, {Madore}, {Malina}, {Neff}, {Rich}, {Small}, {Surber}, {Szalay}, {Welsh}, \& {Wyder}}]{Martin_GALEX_2005ApJ...619L...1M}
{Martin}, D.~C., {Fanson}, J., {Schiminovich}, D., {et~al.} 2005, \apjl, 619, L1

\bibitem[{{McGill} {et~al.}(2023){McGill}, {Anderson}, {Casertano}, {Sahu}, {Bergeron}, {Blouin}, {Dufour}, {Smith}, {Evans}, {Belokurov}, {Smart}, {Bellini}, {Calamida}, {Dominik}, {Kains}, {Kl{\"u}ter}, {Nielsen}, \& {Wambsganss}}]{McGill_2023MNRAS.520..259M}
{McGill}, P., {Anderson}, J., {Casertano}, S., {et~al.} 2023, \mnras, 520, 259

\bibitem[{{McGill} {et~al.}(2018){McGill}, {Smith}, {Evans}, {Belokurov}, \& {Smart}}]{McGill_2018MNRAS.478L..29M}
{McGill}, P., {Smith}, L.~C., {Evans}, N.~W., {Belokurov}, V., \& {Smart}, R.~L. 2018, \mnras, 478, L29

\bibitem[{{Montegriffo} {et~al.}(2023){Montegriffo}, {De Angeli}, {Andrae}, {Riello}, {Pancino}, {Sanna}, {Bellazzini}, {Evans}, {Carrasco}, {Sordo}, {Busso}, {Cacciari}, {Jordi}, {van Leeuwen}, {Vallenari}, {Altavilla}, {Barstow}, {Brown}, {Burgess}, {Castellani}, {Cowell}, {Davidson}, {De Luise}, {Delchambre}, {Diener}, {Fabricius}, {Fr{\'e}mat}, {Fouesneau}, {Gilmore}, {Giuffrida}, {Hambly}, {Harrison}, {Hidalgo}, {Hodgkin}, {Holland}, {Marinoni}, {Osborne}, {Pagani}, {Palaversa}, {Piersimoni}, {Pulone}, {Ragaini}, {Rainer}, {Richards}, {Rowell}, {Ruz-Mieres}, {Sarro}, {Walton}, \& {Yoldas}}]{GDR3_Montegriffo_XP_calibration_2022arXiv220606205M}
{Montegriffo}, P., {De Angeli}, F., {Andrae}, R., {et~al.} 2023, \aap, 674, A3

\bibitem[{{O'Brien} {et~al.}(2023){O'Brien}, {Tremblay}, {Gentile Fusillo}, {Hollands}, {G{\"a}nsicke}, {Koester}, {Pelisoli}, {Cukanovaite}, {Cunningham}, {Doyle}, {Elms}, {Farihi}, {Hermes}, {Holberg}, {Jordan}, {Klein}, {Kleinman}, {Manser}, {De Martino}, {Marsh}, {McCleery}, {Melis}, {Nitta}, {Parsons}, {Raddi}, {Rebassa-Mansergas}, {Schreiber}, {Silvotti}, {Steeghs}, {Toloza}, {Toonen}, {Torres}, {Weinberger}, \& {Zuckerman}}]{OBRien_2023MNRAS.518.3055O}
{O'Brien}, M.~W., {Tremblay}, P.~E., {Gentile Fusillo}, N.~P., {et~al.} 2023, \mnras, 518, 3055

\bibitem[{{Pecaut} \& {Mamajek}(2013)}]{pecaut13}
{Pecaut}, M.~J. \& {Mamajek}, E.~E. 2013, \apjs, 208, 9

\bibitem[{{Pourbaix} {et~al.}(2022){Pourbaix}, {Arenou}, {Gavras}, {Gosset}, {Halbwachs}, {Siopis}, {Sozzetti}, {Bauchet}, {Damerdji}, {Delchambre}, {Delisle}, {Giacobbe}, {Holl}, {Jorissen}, {Lattanzi}, {Leclerc}, {Morel}, {Sadowski}, {Sahlmann}, \& {Segransan}}]{GDR3_documentation_ch7_2022gdr3.reptE...7P}
{Pourbaix}, D., {Arenou}, F., {Gavras}, P., {et~al.} 2022, {Gaia DR3 documentation Chapter 7: Non-single stars}, Gaia DR3 documentation, European Space Agency

\bibitem[{{Pourbaix} {et~al.}(2004){Pourbaix}, {Tokovinin}, {Batten}, {Fekel}, {Hartkopf}, {Levato}, {Morrell}, {Torres}, \& {Udry}}]{Pourbaix_SB9_2004A&A...424..727P}
{Pourbaix}, D., {Tokovinin}, A.~A., {Batten}, A.~H., {et~al.} 2004, \aap, 424, 727

\bibitem[{{Raghavan} {et~al.}(2010){Raghavan}, {McAlister}, {Henry}, {Latham}, {Marcy}, {Mason}, {Gies}, {White}, \& {ten Brummelaar}}]{Raghavan_2010ApJS..190....1R}
{Raghavan}, D., {McAlister}, H.~A., {Henry}, T.~J., {et~al.} 2010, \apjs, 190, 1

\bibitem[{{Reyl{\'e}} {et~al.}(2023){Reyl{\'e}}, {Jardine}, {Fouqu{\'e}}, {Caballero}, {Smart}, \& {Sozzetti}}]{Reyle_2023arXiv230202810R}
{Reyl{\'e}}, C., {Jardine}, K., {Fouqu{\'e}}, P., {et~al.} 2023, in The 21st Cambridge workshop on Cool Stars, Stellar Systems, and the Sun, Toulouse, France, 2022, ed. A.~S. {Brun}, J.~{Bouvier}, \& P.~{Petit}, arXiv:2302.02810

\bibitem[{{Riello} {et~al.}(2021){Riello}, {De Angeli}, {Evans}, {Montegriffo}, {Carrasco}, {Busso}, {Palaversa}, {Burgess}, {Diener}, {Davidson}, {Rowell}, {Fabricius}, {Jordi}, {Bellazzini}, {Pancino}, {Harrison}, {Cacciari}, {van Leeuwen}, {Hambly}, {Hodgkin}, {Osborne}, {Altavilla}, {Barstow}, {Brown}, {Castellani}, {Cowell}, {De Luise}, {Gilmore}, {Giuffrida}, {Hidalgo}, {Holland}, {Marinoni}, {Pagani}, {Piersimoni}, {Pulone}, {Ragaini}, {Rainer}, {Richards}, {Sanna}, {Walton}, {Weiler}, \& {Yoldas}}]{gaiaedr3_photometry}
{Riello}, M., {De Angeli}, F., {Evans}, D.~W., {et~al.} 2021, \aap, 649, A3

\bibitem[{{Rowell}(2013)}]{Rowell_2013MNRAS.434.1549R}
{Rowell}, N. 2013, \mnras, 434, 1549

\bibitem[{{Rybizki} {et~al.}(2022){Rybizki}, {Green}, {Rix}, {El-Badry}, {Demleitner}, {Zari}, {Udalski}, {Smart}, \& {Gould}}]{Rybizki_fidelity_2021arXiv210111641R}
{Rybizki}, J., {Green}, G.~M., {Rix}, H.-W., {et~al.} 2022, \mnras, 510, 2597

\bibitem[{{Sarna} {et~al.}(2000){Sarna}, {Ergma}, \& {Ger{\v{s}}kevit{\v{s}}-Antipova}}]{Sarna_2000MNRAS.316...84S}
{Sarna}, M.~J., {Ergma}, E., \& {Ger{\v{s}}kevit{\v{s}}-Antipova}, J. 2000, \mnras, 316, 84

\bibitem[{{Schmidt}(1959)}]{Schmidt_1959ApJ...129..243S}
{Schmidt}, M. 1959, \apj, 129, 243

\bibitem[{{Scholz}(2020)}]{Scholz_2020A&A...637A..45S}
{Scholz}, R.~D. 2020, \aap, 637, A45

\bibitem[{{Spaeth} {et~al.}(2023){Spaeth}, {Reffert}, {Trifonov}, \& {Golovin}}]{Spaeth_2023RNAAS...7...12S}
{Spaeth}, D., {Reffert}, S., {Trifonov}, T., \& {Golovin}, A. 2023, Research Notes of the American Astronomical Society, 7, 12

\bibitem[{{Steinmetz} {et~al.}(2020){Steinmetz}, {Matijevi{\v{c}}}, {Enke}, {Zwitter}, {Guiglion}, {McMillan}, {Kordopatis}, {Valentini}, {Chiappini}, {Casagrande}, {Wojno}, {Anguiano}, {Bienaym{\'e}}, {Bijaoui}, {Binney}, {Burton}, {Cass}, {de Laverny}, {Fiegert}, {Freeman}, {Fulbright}, {Gibson}, {Gilmore}, {Grebel}, {Helmi}, {Kunder}, {Munari}, {Navarro}, {Parker}, {Ruchti}, {Recio-Blanco}, {Reid}, {Seabroke}, {Siviero}, {Siebert}, {Stupar}, {Watson}, {Williams}, {Wyse}, {Anders}, {Antoja}, {Birko}, {Bland-Hawthorn}, {Bossini}, {Garc{\'\i}a}, {Carrillo}, {Chaplin}, {Elsworth}, {Famaey}, {Gerhard}, {Jofre}, {Just}, {Mathur}, {Miglio}, {Minchev}, {Monari}, {Mosser}, {Ritter}, {Rodrigues}, {Scholz}, {Sharma}, {Sysoliatina}, \& {RAVE Collaboration}}]{RAVE_2020AJ....160...82S}
{Steinmetz}, M., {Matijevi{\v{c}}}, G., {Enke}, H., {et~al.} 2020, \aj, 160, 82

\bibitem[{{Taylor}(2005)}]{topcat_2005ASPC..347...29T}
{Taylor}, M.~B. 2005, in Astronomical Society of the Pacific Conference Series, Vol. 347, Astronomical Data Analysis Software and Systems XIV, ed. P.~{Shopbell}, M.~{Britton}, \& R.~{Ebert}, 29

\bibitem[{{Taylor}(2019)}]{topcat_2019ASPC..523...43T}
{Taylor}, M.~B. 2019, in Astronomical Society of the Pacific Conference Series, Vol. 523, Astronomical Data Analysis Software and Systems XXVII, ed. P.~J. {Teuben}, M.~W. {Pound}, B.~A. {Thomas}, \& E.~M. {Warner}, 43

\bibitem[{{Torres} {et~al.}(2022){Torres}, {Canals}, {Jim{\'e}nez-Esteban}, {Rebassa-Mansergas}, \& {Solano}}]{Torres_2022MNRAS.511.5462T}
{Torres}, S., {Canals}, P., {Jim{\'e}nez-Esteban}, F.~M., {Rebassa-Mansergas}, A., \& {Solano}, E. 2022, \mnras, 511, 5462

\bibitem[{{Torres} {et~al.}(2023){Torres}, {Cruz}, {Murillo-Ojeda}, {Jim{\'e}nez-Esteban}, {Rebassa-Mansergas}, {Solano}, {Camisassa}, {Raddi}, \& {Doliguez Le Lourec}}]{Torres_2023A&A...677A.159T}
{Torres}, S., {Cruz}, P., {Murillo-Ojeda}, R., {et~al.} 2023, \aap, 677, A159

\bibitem[{{Ulla} {et~al.}(2022){Ulla}, {Creevey}, {{\'A}lvarez}, {Andrae}, {Bailer-Jones}, {Bellas-Velidis}, {Brugaletta}, {Carballo}, {Dafonte}, {Delchambre}, {Dharmawardena}, {Drimmel}, {Fouesneau}, {Fr{\'e}mat}, {Garabato}, {Hatzidimitriou}, {Heiter}, {Kordopatis}, {Korn}, {Lanzafame}, {Lobel}, {Manteiga}, {Marshall}, {Pailler}, {Pallas-Quintela}, {Recio-Blanco}, {Rybizki}, {Sarro Baro}, {Schultheis}, {Sordo}, {Soubiran}, {Th{\'e}venin}, \& {Vallenari}}]{GDR3_documentation_ch11_2022gdr3.reptE..11U}
{Ulla}, A., {Creevey}, O.~L., {{\'A}lvarez}, M.~A., {et~al.} 2022, {Gaia DR3 documentation Chapter 11: Astrophysical parameters}, Gaia DR3 documentation, European Space Agency

\bibitem[{{Vincent} {et~al.}(2023){Vincent}, {Bergeron}, \& {Dufour}}]{Vincent_2023MNRAS.521..760V}
{Vincent}, O., {Bergeron}, P., \& {Dufour}, P. 2023, \mnras, 521, 760

\bibitem[{{Virtanen} {et~al.}(2020){Virtanen}, {Gommers}, {Oliphant}, {Haberland}, {Reddy}, {Cournapeau}, {Burovski}, {Peterson}, {Weckesser}, {Bright}, {van der Walt}, {Brett}, {Wilson}, {Millman}, {Mayorov}, {Nelson}, {Jones}, {Kern}, {Larson}, {Carey}, {Polat}, {Feng}, {Moore}, {VanderPlas}, {Laxalde}, {Perktold}, {Cimrman}, {Henriksen}, {Quintero}, {Harris}, {Archibald}, {Ribeiro}, {Pedregosa}, {van Mulbregt}, \& {SciPy 1. 0 Contributors}}]{scipy_2020NatMe..17..261V}
{Virtanen}, P., {Gommers}, R., {Oliphant}, T.~E., {et~al.} 2020, Nature Methods, 17, 261

\bibitem[{{Yanny} {et~al.}(2009){Yanny}, {Rockosi}, {Newberg}, {Knapp}, {Adelman-McCarthy}, {Alcorn}, {Allam}, {Allende Prieto}, {An}, {Anderson}, {Anderson}, {Bailer-Jones}, {Bastian}, {Beers}, {Bell}, {Belokurov}, {Bizyaev}, {Blythe}, {Bochanski}, {Boroski}, {Brinchmann}, {Brinkmann}, {Brewington}, {Carey}, {Cudworth}, {Evans}, {Evans}, {Gates}, {G{\"a}nsicke}, {Gillespie}, {Gilmore}, {Nebot Gomez-Moran}, {Grebel}, {Greenwell}, {Gunn}, {Jordan}, {Jordan}, {Harding}, {Harris}, {Hendry}, {Holder}, {Ivans}, {Ivezi{\v{c}}}, {Jester}, {Johnson}, {Kent}, {Kleinman}, {Kniazev}, {Krzesinski}, {Kron}, {Kuropatkin}, {Lebedeva}, {Lee}, {French Leger}, {L{\'e}pine}, {Levine}, {Lin}, {Long}, {Loomis}, {Lupton}, {Malanushenko}, {Malanushenko}, {Margon}, {Martinez-Delgado}, {McGehee}, {Monet}, {Morrison}, {Munn}, {Neilsen}, {Nitta}, {Norris}, {Oravetz}, {Owen}, {Padmanabhan}, {Pan}, {Peterson}, {Pier}, {Platson}, {Re Fiorentin}, {Richards}, {Rix}, {Schlegel}, {Schneider}, {Schreiber}, {Schwope}, {Sibley}, {Simmons},
  {Snedden}, {Allyn Smith}, {Stark}, {Stauffer}, {Steinmetz}, {Stoughton}, {SubbaRao}, {Szalay}, {Szkody}, {Thakar}, {Sivarani}, {Tucker}, {Uomoto}, {Vanden Berk}, {Vidrih}, {Wadadekar}, {Watters}, {Wilhelm}, {Wyse}, {Yarger}, \& {Zucker}}]{SEGUE_2009AJ....137.4377Y}
{Yanny}, B., {Rockosi}, C., {Newberg}, H.~J., {et~al.} 2009, \aj, 137, 4377

\bibitem[{{Yuan}(1992)}]{Yuan_1992A&A...261..105Y}
{Yuan}, J.~W. 1992, \aap, 261, 105

\end{thebibliography}

\appendix

\section{Discrete source classifier (DSC) in \gdrthree{}}
\label{sec:dsc}

One of the key data products within \gdrthree{} is the \texttt{astrophysical\_parameters} table, which provides, among other information, the probabilistic classification of sources into several classes, including white dwarfs. Specifically for white dwarfs, the DSC uses two classifiers: \texttt{specmod}, which provides probabilities derived from XP spectra, and \texttt{combmod}, which combines spectroscopic probabilities with those obtained from astrometry and photometry.

All of the white dwarfs identified in this study have posterior class probabilities below 0.5. The probability values range from $2\times10^{-12}$ to 0.46 in \texttt{combmod}, and from 0 to 0.04 in \texttt{specmod}. In addition, one white dwarf (\object{\gdrthree{}~1938960722332184704}) is excluded from the classifier.

\begin{figure*}
    \centering
    \includegraphics[width=0.49\textwidth]{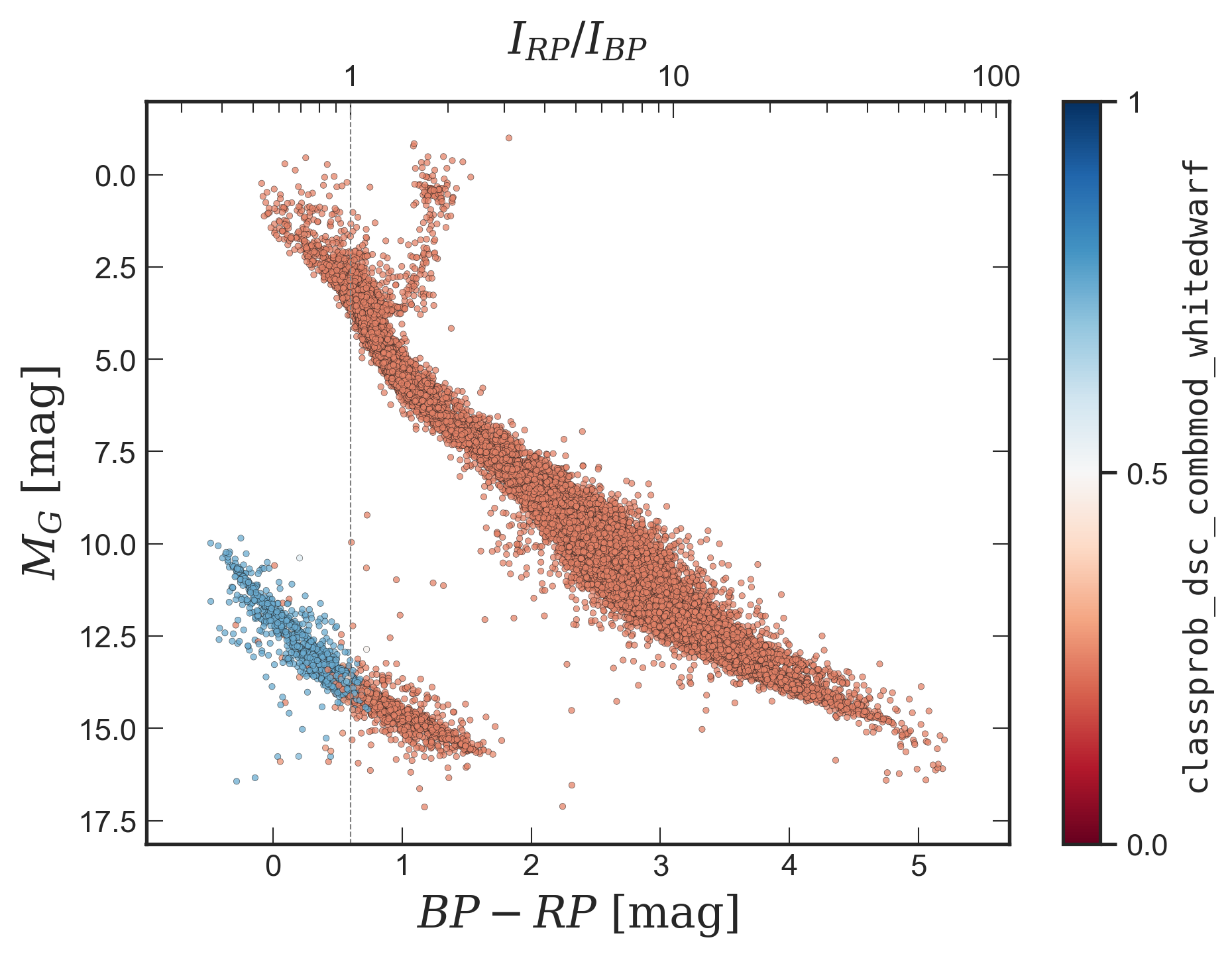}
    \includegraphics[width=0.49\textwidth]{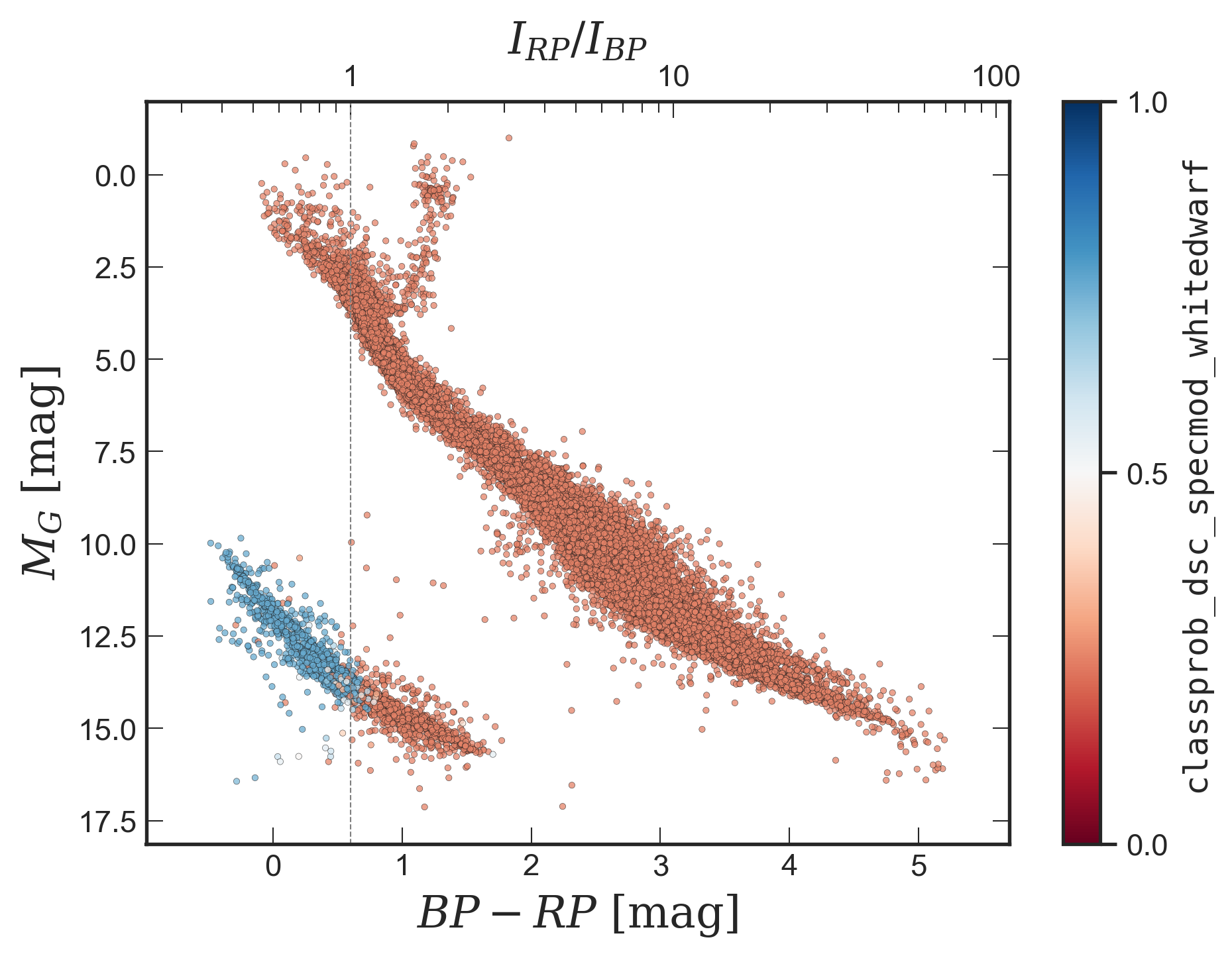}
    \caption{CMD for the 50~pc sample colour-coded with the probability of being a white dwarf as listed in \gdrthree{}. The dashed line indicates the colour at which the $BP$ and $RP$ fluxes are equal. \textit{Left:} probability derived from XP-spectroscopy, photometry, and astrometry (DSC-Combmod classifier). \textit{Right:} probability derived from XP-spectroscopy only (DSC-Specmod classifier).}
    \label{fig:hrd_classprob}
\end{figure*}

Previously it has been shown that the DSC probabilities for white dwarfs and binaries may be poorly calibrated and consequently due to the poor performance of the classifier (with a purity of only about 25\%) the use of these probabilities to construct samples of such objects is not recommended \citep{GDR3_Creevey_ApsisI_2022arXiv220605864C, GDR3_documentation_ch11_2022gdr3.reptE..11U}.
This limitation becomes apparent when analysing the CMD colour-coded with white dwarf probabilities (see \figref{fig:hrd_classprob}). In particular, in most cases, only objects in the upper part of the white dwarf sequence have probabilities greater than 0.5.
What is striking is that the sharp decrease in the derived probabilities occurs at about $BP-RP = 0.6$, corresponding to the colour where the $RP$ flux starts to exceed the $BP$ flux.

\end{document}